\documentclass[a4paper,fleqn,usenatbib]{mnras}

\usepackage[T1]{fontenc}
\usepackage{ae,aecompl}

\usepackage{graphicx}	% Including figure files
\usepackage{amsmath}	% Advanced maths commands
\usepackage{amssymb}	% Extra maths symbols

\usepackage{threeparttable}

\hypersetup
{
    pdfauthor={Kokubo},
    pdftitle={Kokubo 2022},
}

\title[Extragalactic giant eruption LBV SDSS1133]{Multiple giant eruptions and X-ray emission in the recoiling AGN/LBV candidate SDSS1133}

\author[M.~Kokubo]{
Mitsuru Kokubo $^{1, 2}$
\thanks{E-mail: mkokubo@astro.princeton.edu}
\thanks{JSPS Fellow}
\\
$^{1}$ Department of Astrophysical Sciences, Princeton University, Princeton, New Jersey 08544, USA\\
$^{2}$ Astronomical Institute, Tohoku University, 6-3 Aramaki-Aza-Aoba, Aoba-ku, Sendai, Miyagi 980-8578, Japan\\
}

\date{Accepted 2022 June 14. Received 2022 June 14; in original form 2021 January 19}

\pubyear{2022}

\begin{document}
\label{firstpage}
\pagerange{\pageref{firstpage}--\pageref{lastpage}}
\maketitle

\begin{abstract}

We present a comprehensive analysis of $20$~years worth of multi-color photometric light curves, multi-epoch optical spectra, and X-ray data of an off-nuclear variable object SDSS1133 in Mrk~177 at $z=0.0079$.
The UV-optical light curves reveal that SDSS1133 experienced four outbursts in 2001, 2014, 2019, and 2021.
The persistent UV-optical luminosity in the non-outbursting state is $\sim 10^{41}$~erg~s${}^{-1}$ with small-scale flux variations, and peak luminosities during the outbursts reach $\sim 10^{42}$~erg~s${}^{-1}$.
The optical spectra exhibit enduring broad hydrogen Balmer P-Cygni profiles with the absorption minimum at $\sim -2,000~\text{km}~\text{s}^{-1}$, indicating the presence of fast moving ejecta.
{\it Chandra} detected weak X-ray emission at a $0.3-10$~keV luminosity of $L_{X} = 4 \times 10^{38}$~erg~s${}^{-1}$ after the 2019 outburst.
These lines of evidence suggests that SDSS1133 is an extreme luminous blue variable (LBV) star experiencing multiple giant eruptions with interactions of the ejected shell with different shells and/or circumstellar medium (CSM), and disfavors the recoiling Active Galactic Nuclei (AGN) scenario suggested in the literature.
We suggest that pulsational pair-instability may provide a viable explanation for the multiple energetic eruptions in SDSS1133.
If the current activity of SDSS1133 is a precursor of a supernova explosion, we may be able to observe a few additional giant eruptions and then the terminal supernova explosion or collapse to a massive black hole in future observations.

\end{abstract}


%%%%%%%%%%%%%%%%%%%%%%%%%%%%%%%%%%%%%%%%%%%%%%%%%%

%%%%%%%%%%%%%%%%% BODY OF PAPER %%%%%%%%%%%%%%%%%%

%-------------------------------------------------------------------

\section{Introduction}
\label{sec:intro}

\begin{figure}
\center{
\includegraphics[clip, width=2.0in]{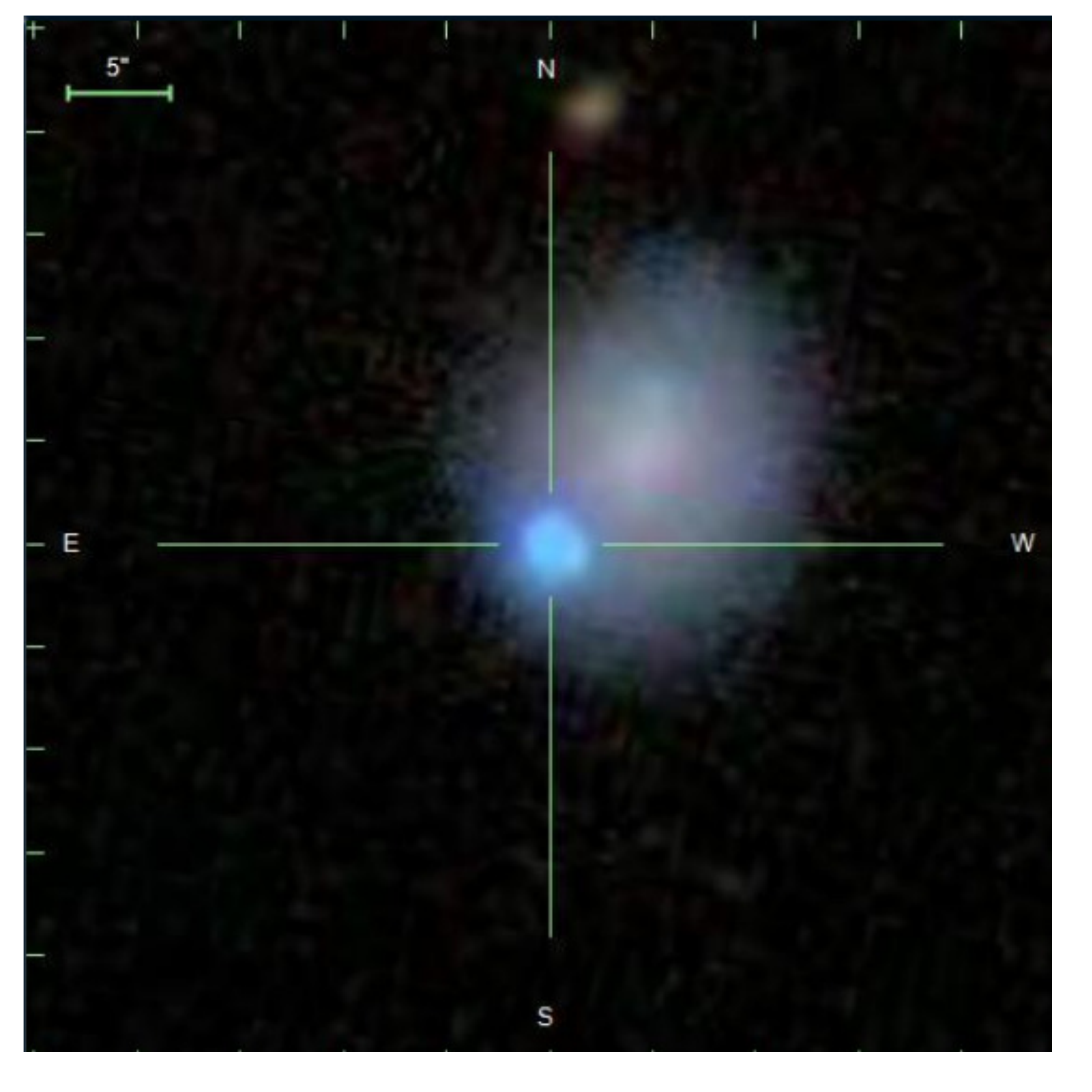}
}
 \caption{The SDSS DR7 $g$, $r$, and $i$ band color image centered on SDSS1133. The original images are obtained on 2002 April 1. The host galaxy is Mrk~177 at $z=0.0079$.}
 \label{fig:sdss1133}
\end{figure}

SDSS J113323.97+550415.8 (hereafter SDSS1133) has long been recognised as an unusually persistent extragalactic variable object \citep[see also][]{lop06,zho06,kee12,rei13,koss14,bur20,war21}.
SDSS1133 is a point source with a quasar-like optical color, located at the outskirt of a blue compact dwarf galaxy Mrk~177 (UGCA 239) at the galactocentric distance of $5''.8 = 0.81$~kpc (Figure~\ref{fig:sdss1133}).
By analysing historic photometry data of SDSS1133, \cite{koss14} show that SDSS1133 is optically detected over 63~yr since 1950.
The absolute magnitude of SDSS1133 is in between $M_{g} \sim -13$~mag and $-14$~mag in its non-outbursting phase (before $1999$ and after $2003$), and it got two orders of magnitude brighter ($M_{g} \sim -16$~mag) in the outbursting phase in $2001-2002$.
The optical spectrum obtained in 2003 by the Sloan Digital Sky Survey (SDSS) reveals complex hydrogen Balmer line profiles, where the broad emission component has a full-width at half maximum (FWHM) greater than $\sim 2,000$~km~s${}^{-1}$, and the SDSS pipeline classifies SDSS1133 as a quasi-stellar object (QSO).
The narrow line redshift is consistent with that of the Mrk~177's nucleus ($z=0.0079$; see Section~\ref{sec:specdata}).
Mrk~177 exhibits a disturbed morphology, suggesting that it is a post-merger galaxy \citep{koss14}.

It is suggested that SDSS1133 could be a recoiling type~1 Active Galactic Nuclei (AGN) with extreme variability \citep{koss14}.
Under the recoiling AGN scenario, SDSS1133 is explained as an offset AGN ejected from the center of recently-merged galaxy Mrk~177, kicked by a gravitational wave recoil after the coalescence of the supermassive black hole binary (SMBHB) \citep[e.g.,][and references therein]{kom12}.
The absolute magnitude of SDSS1133 is consistent with that of the low-luminosity AGNs in the local Universe \citep[e.g.,][]{ho97b,gre07,rei13}.
The large variability amplitude of $\sim$2 mag is rarely seen in AGNs, but can be achieved by a rare population of extremely variable AGNs \citep[e.g.,][]{mac10,mac12,gra17}.
The broad emission line profile can be interpreted as complex virial motions of broad line region clouds surrounding the AGN.
The optical and near-infrared colors remain roughly constant over the observed periods and are redder than the stellar locus, being consistent with the recoiling AGN interpretation \citep{koss14}.

However, other observational properties suggest that SDSS1133 is a kind of stellar activity-related variables.
The absolute magnitude and temporal behavior of SDSS1133 are more consistent with gap transients or SN impostors (namely giant eruptions of extragalactic luminous blue variable stars; hereafter LBV), or faint supernovae (SN) \citep[e.g.,][]{van00,smi11,smi17,pas19,per20}.
Also, the broad Balmer line profile of SDSS1133 exhibits blue-shifted P-Cygni-like absorption features up to $\sim -5,000$ km~s${}^{-1}$, which is reminiscent of those observed in SNe~II and some giant eruption LBVs \citep[e.g.,][]{smi08,pas10,pas13,mau13,koss14,war21}.
SDSS1133 exhibits narrow/intermediate width Fe~II and Ca~II emission lines \citep{koss14}, which are rarely seen in AGNs while sometimes observed in SNe and LBVs \citep[e.g.,][]{smi10,fol11,tad13,koss14,smi18,war21}.

Recently, \cite{war21} report another outburst event of SDSS1133 detected by the Zwicly Transient Facility (reported as a transient ZTF19aafmjfw, also known as Gaia19bwn) beginning on 2019 April 7, peaked on 2019 June 5, and faded again by 2019 December 1 \citep[see also][]{sta19,pur19b,pur19}.
Moreover, the {\it Gaia} all-sky photometry reveals that SDSS1133 experienced another outburst peaked on 2021 June 21.
Although \cite{koss14} and \cite{bur20} suggested a scenario that SDSS1133 is a LBV erupting for decades since 1950 and then followed by a terminal explosion observed as a SN~IIn after 2001, the detection of these new outburst events rules out this possibility of terminal explosion and any other one-off event scenarios for SDSS1133 such as SNe, tidal disruption events, or binary mergers.
Based on the detection of the recurrent outbursts and its peculiar enduring spectroscopic properties, \cite{war21} suggest that SDSS1133 is likely to be an erupting LBV with the progenitor still alive, but the AGN scenario cannot be completely ruled out.

Although a lot of observational data are available and discussed in the literature, there is still no consensus on the nature of SDSS1133.
The purpose of this paper is to present a comprehensive analysis of the multi-wavelength data of SDSS1133 to conclude the true nature of this enigmatic object.
In Section~\ref{sec:data}, we present analysis for multi-epoch imaging and spectroscopic data of SDSS1133, including already published and unpublished X-ray, ultraviolet (UV), and optical data.
In Section~\ref{sec:discussion}, we compare the observational properties of SDSS1133 with known LBVs as well as AGNs and SNe, and we conclude that SDSS1133 is one of the brightest extragalactic giant eruption LBV, akin to $\eta$ Carinae's Great Eruption and known SN progenitor LBV outbursts.
We suggest that the X-ray-to-radio emission of SDSS1133 is mediated by interactions of an ejected shell with different shells and/or circumstellar medium (CSM), where the CSM is formed by enduring stellar wind and the ejecta are produced via multiple episodes of non-terminal explosions (e.g., pulsational pair-instability).
Conclusions are summarized in Section~\ref{sec:conclusion}.

\section{Data analysis}
\label{sec:data}

\begin{figure*}
\center{
\includegraphics[clip, width=6.8in]{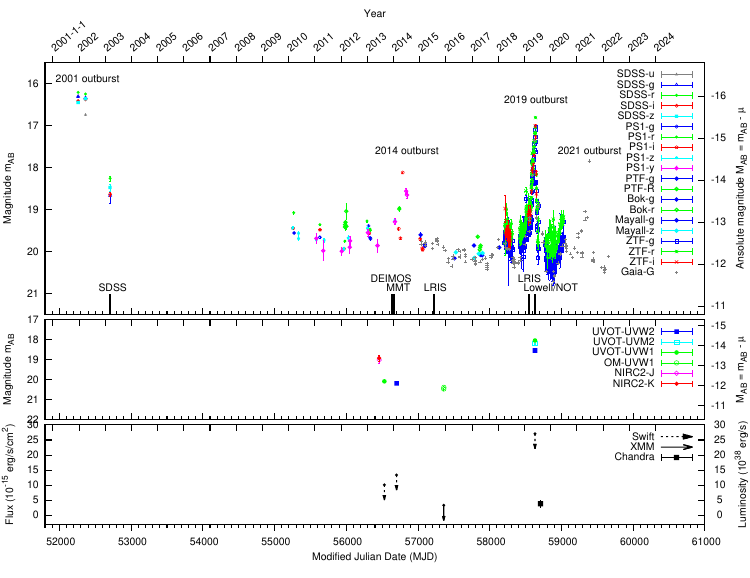}
}
 \caption{
 Top: the optical light curves of SDSS1133. The PSF magnitudes obtained by the GALFIT modelling are shown, except for the synthesized SDSS magnitudes from the SDSS spectrum in 2003 and $Gaia$ $G$ band light curve. Galactic extinction is corrected. Four outbursts observed in 2001, 2014, 2019, and 2021 are labelled. Epochs of optical spectroscopic observations (Section~\ref{sec:specdata}) are indicated by vertical bars.
 Middle: UV and NIR photometric measurements. The Keck/NIRC2 measurements are taken from \citet{koss14}, and the {\it Swift}/UVOT and {\it XMM-Newton}/OM PSF measurements are analysed in this work. 
 Bottom: the unabsorbed $0.3-10$~keV X-ray flux light curve. The arrows indicate the 90\% upper limits on the X-ray flux.
 }
 \label{fig:ztf_lc}
\end{figure*}

\begin{figure*}
\center{
\includegraphics[clip, width=6.8in]{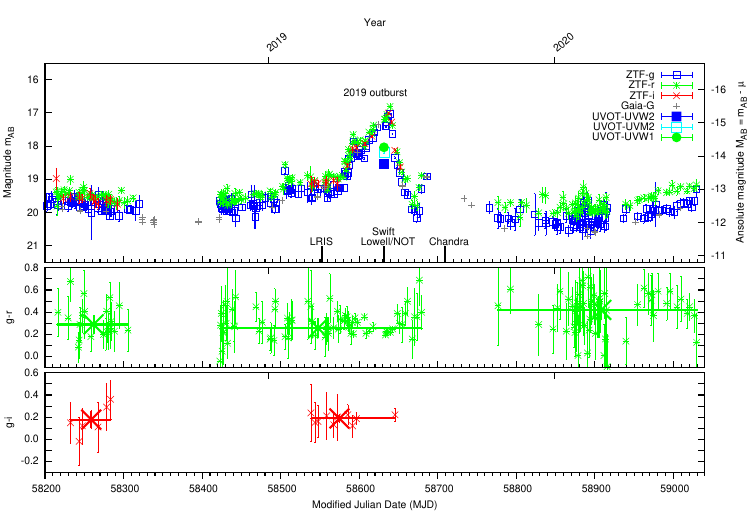}
}
 \caption{Top: the same as Figure~\ref{fig:ztf_lc}, but the plot range is restricted from 2018 March 23 to 2020 July 10 (58200 < MJD < 59040), where the densely-sampled ZTF data are available. The optical and UV data points are shown simultaneously.
 Epochs of optical spectroscopy and X-ray observations ({\it Swift} and {\it Chandra}) are indicated by vertical bars.
 Middle: the $g-r$ colors at the epochs when quasi-simultaneous ZTF $g$ and $r$ band measurements are available. The large symbols are the mean colors of each of the observing period indicated by the horizontal lines ($g-r=0.29$, 0.26, and 0.42~mag in 2018, 2019, and 2020, respectively). Bottom: the $g-i$ colors. The mean colors are $g-i=0.18$ and 0.19~mag in 2018 and 2019, respectively.
 }
 \label{fig:ztf_lc_zoom}
\end{figure*}

\begin{table*}
\centering
\caption{UV-optical PSF photometry for SDSS1133 obtained by the {\tt GALFIT} modelling.}
\label{tab:photometry_table}
\begin{tabular}{cccccc}
\hline
MJD & Date & mag. & error in mag. & Telescope & Filter \\
    & (YYYY-MM-DD) & & & & \\
\hline
56701.96 & 2014-02-13 & 20.26 & 0.03 & Swift & UVW2 \\
58631.43 & 2019-05-28 & 18.61 & 0.03 & Swift & UVW2 \\
\hline
58631.44 & 2019-05-28 & 18.26 & 0.03 & Swift & UVM2 \\
\hline
56531.44 & 2013-08-27 & 20.14 & 0.02 & Swift & UVW1 \\
57358.55 & 2015-12-02 & 20.52 & 0.05 & XMM & UVW1 \\
57360.53 & 2015-12-04 & 20.44 & 0.04 & XMM & UVW1 \\
58631.44 & 2019-05-28 & 18.10 & 0.02 & Swift & UVW1 \\
\hline
52261.37 & 2001-12-18 & 16.35 & 0.01 & SDSS & $u$ \\
52365.14 & 2002-04-01 & 16.79 & 0.01 & SDSS & $u$ \\
\hline
52261.37 & 2001-12-18 & 16.34 & 0.01 & SDSS & $g$ \\
\dots & \dots & \dots & \dots & \dots & \dots \\
\hline
\end{tabular}
%\bigskip
   \begin{tablenotes}
     \item [1] MJD indicates the mean epoch of each observation run. Galactic extinction is uncorrected. All magnitudes are listed as AB magnitude. The statistical photometric errors are reported (systematic errors are not included). The full table is only available in electronic form.
   \end{tablenotes}
\end{table*}

\begin{table*}
	\centering
	\caption{{\it Swift}/XRT, {\it XMM-Newton}/EPIC, and {\it Chandra}/ACIS unabsorbed $0.3-10$~keV X-ray flux and luminosity ($L_{0.3-10~\text{keV}}$) of SDSS1133.}
	\label{tab:xray_table}
	\begin{tabular}{cccccc} % four columns, alignment for each
		\hline
		 MJD & Date & Flux  & Luminosity $L_{0.3-10~\text{keV}}$ & Luminosity $L_{0.3-10~\text{keV}}$ & Telescope\\
		     &  &  &  &  (SMC, LMC,  MW) & \\
		     & (YYYY/MM/DD) &  ($10^{-14}$~erg~s${}^{-1}$~cm${}^{-2}$) & ($10^{39}$~erg~s${}^{-1}$) & ($10^{39}$~erg~s${}^{-1}$) & \\
		\hline
		56531.4 & 2013/08/26-28 & $<1.018$  & $<1.018$ & $<2.756$, 2.602, 1.891 & Swift\\
		\hline
		56702.0 & 2014/02/06-25 & $<1.340$  & $<1.339$ & $<3.626$, 3.423, 2.488 & Swift\\
		\hline
		57359.6 & 2015/12/02-04 & $<0.337$  & $<0.337$ & $<1.098$, 0.949, 0.671 & XMM \\
		\hline
		58631.4 & 2019/05/27-29 & $<2.706$  & $<2.705$ & $<7.319$, 6.910, 5.024 & Swift\\
		\hline
		58708.9 & 2019/08/13-14 & $0.393_{-0.116}^{+0.116}$  & $0.393_{-0.116}^{+0.116}$ & $ 0.669_{-0.198}^{+0.198}$, $ 0.644_{-0.190}^{+0.190}$, $ 0.522_{-0.154}^{+0.154}$ & Chandra\\
		\hline
	\end{tabular}
%\bigskip
   \begin{tablenotes}
     \item [1] MJD indicates the mean epoch of each observation run.  An absorbed power-law model with $\Gamma=2$ and $N_{H}=2 \times 10^{20}$~cm${}^{-2}$ is assumed. {\it Swift}/XRT and {\it XMM-Newton}/EPIC data are non-detection, thus the 90\% upper limits are shown. The reported uncertainty of the {\it Chandra}/ACIS luminosity is $\pm 1\sigma$.
     \item [2] The 4th column is the $0.3-10$~keV X-ray luminosity corrected for the host galaxy X-ray absorption assuming (SMC, LMC, MW)-like extinction, where the additional gas column of $N_{H, \text{host}} = (2.84, 1.46, 0.31) \times 10^{22}$~cm${}^{-2}$ in the host galaxy is included in the X-ray spectral modelling, respectively (Section~\ref{sec:multiwavelength_sed}).
   \end{tablenotes}
\end{table*}

In this Section, we describe details of the analysis of the multi-wavelength data of SDSS1133 retrieved from various publicly-available data archives.
The X-ray, UV, and optical imaging data from the SDSS, PanSTARRS1 (PS1), ZTF, Palomar Transient Factory (PTF), Mayall and Bok Legacy Surveys, {\it Swift}, {\it XMM-Newton}, and {\it Chandra} data archives are described in Sections~\ref{sec:sdss}$-$\ref{sec:chandra}, respectively.
A part of the {\it Swift} and optical imaging data (SDSS and PS1) have already been published by \cite{koss14}, but here we (re)analyse all the data in a consistent manner using {\tt GALFIT} version 3.0.5 \citep{pen11} to derive host-subtracted Point Spread Function (PSF) magnitude light curves of SDSS1133 (see Section~\ref{sec:structural_decomposition} for details).
Other historic measurements for SDSS1133 reported in the literature are summarised in Sections~\ref{sec:gaia} and \ref{sec:koss_photometry}.
The optical and X-ray broad-band photometries of SDSS1133 derived in this work are tabulated respectively in Tables~\ref{tab:photometry_table} and \ref{tab:xray_table}.
Figures~\ref{fig:ztf_lc} and \ref{fig:ztf_lc_zoom} show the broad-band light curves.
The data analysis of the SDSS spectrum and unpublished Keck/LRIS spectra retrieved from the Keck Observatory Archive (KOA) is described in Section~\ref{sec:specdata}.

%Both of SDSS1133 and Mrk~177 were spectroscopically observed by Sloan Digital Sky Survey (SDSS) in 2003, and their redshifts are evaluated by the SDSS spectroscopic pipeline as $z=0.00906$ and $0.00785$, respectively.
%The difference of the recession velocity is $\sim 363$~km~s${}^{-1}$.

%The host galaxy,Mrk 177, is at a distance of 28.9 Mpc (distance modulus 32.3 mag;Tully1994). We use this redshift-independent distance indicator forthe subsequent analysis of SDSS1133 and Mrk 177. At this distance,1 arcsec corresponds to 140 pc. Galactic foreground extinction is very low, AV=0.03 mag (Schlafly & Finkbeiner2011)

Following \cite{koss14}, we use $d_{L} = 28.9$~Mpc (distance modulus $\mu = 32.30$~mag) as a luminosity distance to Mrk~177 \citep[taken from NASA/IPAC Extragalactic Database (NED);][]{tul88,tul94} to calculate luminosities and absolute magnitudes of SDSS1133\footnote{NED reports an uncertainty on the distance as $d_{L} = 28.9 \pm 8.9$~Mpc ($\mu = 32.30 \pm 0.80$~mag). Throughout this paper we do not include the uncertainty on the distance, but it should be kept in mind that the derived luminosity is uncertain by $0.27$~dex due to this distance uncertainty.}.
Since SDSS1133 is a low-redshift object ($z=0.0079$), $K$-correction is negligible in most cases.
The magnitudes are reported in the AB magnitude unit.
The Galactic extinction toward the direction of SDSS1133 is very low, $E(B-V) = 0.010$~mag \citep{fit99,sch11}.
The Galactic extinction coefficients for the SDSS and PS1 systems are taken from NED, and those for other photometric bands are derived by assuming a $T=10,000$~K black body spectrum at $z=0.0079$ and using the \cite{fit99}'s extinction curve and filter transmission curves.

\if0

python
import numpy as np
from astropy import units as u
from astropy import constants as const

d = 28.9*u.Mpc
ed = 8.9*u.Mpc

mu = 5.0*np.log10(np.random.normal(28.9,8.9,1000000)/(10.0e-6))
np.std(mu[~np.isnan(mu)])

% 0.80

\fi

\if0

 /media/mkokubo/analysis_201809/
 SDSS1133/images/lightcurve_all/calc_extinction/Band_Alambda.dat

A_bessell_U 0.0448609704106
A_bessell_B 0.0379093048372
A_bessell_V 0.0286994726074
A_bessell_R 0.022634925698
A_bessell_I 0.0161131034307
A_u 0.045330017243
A_g 0.0354992311447
A_r 0.0243531225141
A_i 0.0180515276596
A_z 0.0134990304651
A_J 0.00767609881494
A_H 0.00484598548251
A_K 0.00326297773424
A_W1 0.00190013038838
A_W2 0.00131504216792
A_uvot_UVW2 0.0744748026233
A_uvot_UVM2 0.0777201667409
A_uvot_UVW1 0.0620553348786
A_om_UVW1 0.0537606521947
A_gaia_G 0.0270169262578
A_ptf_R 0.0222151681269
A_ztf_g 0.0345388189746
A_ztf_r 0.0230894037501
A_ztf_i 0.0164205448295
A_bok_g 0.0348206012371
A_bok_r 0.0231686267083
A_mayall_g 0.0349759116379
A_mayall_z 0.0127676016809
A_galex_NUV 0.0725694222762
A_dss_103aO 0.0405403128585
A_dss_IIIaJ 0.0344434156011
A_dss_IIIaF 0.0220667764497

\fi

\subsection{SDSS imaging (2001-2002)}
\label{sec:sdss}

SDSS $u$, $g$, $r$, $i$, and $z$ band images were obtained on 2001 December 18 and 2002 April 1 (MJD=52261.4 and 52365.1, respectively; Figure~\ref{fig:sdss1133}).
%Full width at half maximums (FWHMs) of the the point-spread function (PSF) of the SDSS images are in a range of $1''.1-1''.5$, and the point-like SDSS1133 is clearly separated from the nucleus of the host galaxy Mrk117 with the galactocentric distance of $5''.8$ (0.81~kpc) \citep{koss14}.
%\cite{koss14} have already presented the SDSS photometry of SDSS1133, but we re-analysed the SDSS images and derived the PSF magnitudes at the epochs of the SDSS observations.
We downloaded the SDSS corrected frames and calibration files from the SDSS Data Release (DR) 7 Data Archive Server.
Photometric zero points were calculated from global zero points and airmass values \footnote{\href{http://classic.sdss.org/dr7/algorithms/fluxcal.html}{http://classic.sdss.org/dr7/algorithms/fluxcal.html}}, and PSF model images were reconstructed using the calibration files\footnote{\href{https://classic.sdss.org/dr7/products/images/read_atlas.html}{https://classic.sdss.org/dr7/products/images/read\_atlas.html}}.

The background variance maps were calculated by using {\tt SExtractor} version 2.25.0 \citep{ber96,ber11}, and variance maps were created by adding the background variance and objects' Poisson variance.
The PSF and the variance maps were used for the {\tt GALFIT} modelling to perform the PSF photometry of SDSS1133, as described in Section~\ref{sec:structural_decomposition}.

\subsection{PanSTARRS1 (2010-2015)}
\label{sec:panstarrs}

PanSTARRS1 (PS1) $3\pi$ survey started observing the entire northern sky since 2010 \citep{chambers16}.
The PS1 data obtained before 2015 are now publicly available as PS1 DR2.
We downloaded PS1 $g$, $r$, $i$, $z$, and $y$ band $6000 \times 6000$~pixels ($25'.8 \times  25'.8$) single-epoch warp and variance images at the position around SDSS1133 ({\tt skycell}.2375.070 of the PS1 sky tessellation) from the PS1 DR2 Image Cutout Service.
The publicly available single-epoch images as of DR2 are spanning from 2010 February 27 to 2015 February 12.
Several $g$ and $i$ band images were visually identified to be affected by chip and readout cells gaps or telescope tracking errors, and were excluded from the analysis.
PSF models of the single-epoch images at the position of SDSS1133 were created using {\tt SExtractor} and {\tt PSFEx} version 3.21.1 \citep{ber96,ber11}.
Photometric zero points calculated by the PS1 data analysis pipeline (written in the image headers) were directly used \citep[][]{wat20,magnier20}. 
% calculated by using scaling factors taken from the image headers 

% stack and warp calibration is not identical in PV3. 
% warp is ubercalibrated, while stack is using normal calibration 
% because ubercalibration is not available at the time of stacking.
% Section 6 of wat20.

\subsection{PTF (2012-2015)}
\label{sec:ptf}

Palomer Transient Factory (PTF) was a wide-field optical time-domain survey using the Samuel Oschin 48-inch Schmidt telescope at Palomar Observatory \citep{law09}.
PTF observed SDSS1133 in $g$ and Mould-$R$ bands in 2012, 2014, and, 2015.
Calibrated PTF images were downloaded from the IPAC database \citep{lah14}.
As in Sections~\ref{sec:sdss} and \ref{sec:panstarrs}, image background variance maps were calculated by using {\tt SExtractor}, and PSF models were created by using {\tt PSFEx}.
photometric zero points defined by a 8~pixel diameter aperture taken from the image header \citep[][]{lah14} were rescaled by using the PSF models to represent infinite-aperture zero point magnitudes.

\subsection{ZTF (2018-2020)}
\label{sec:ztf}

\begin{figure}
\center{
\includegraphics[clip, width=2.8in]{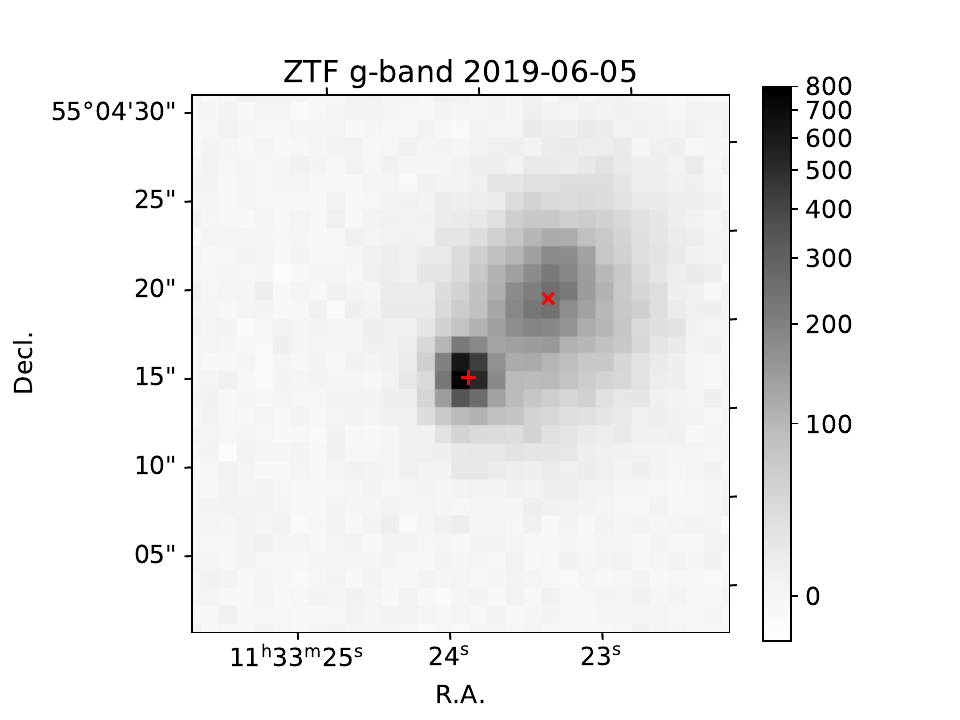}
\includegraphics[clip, width=2.8in]{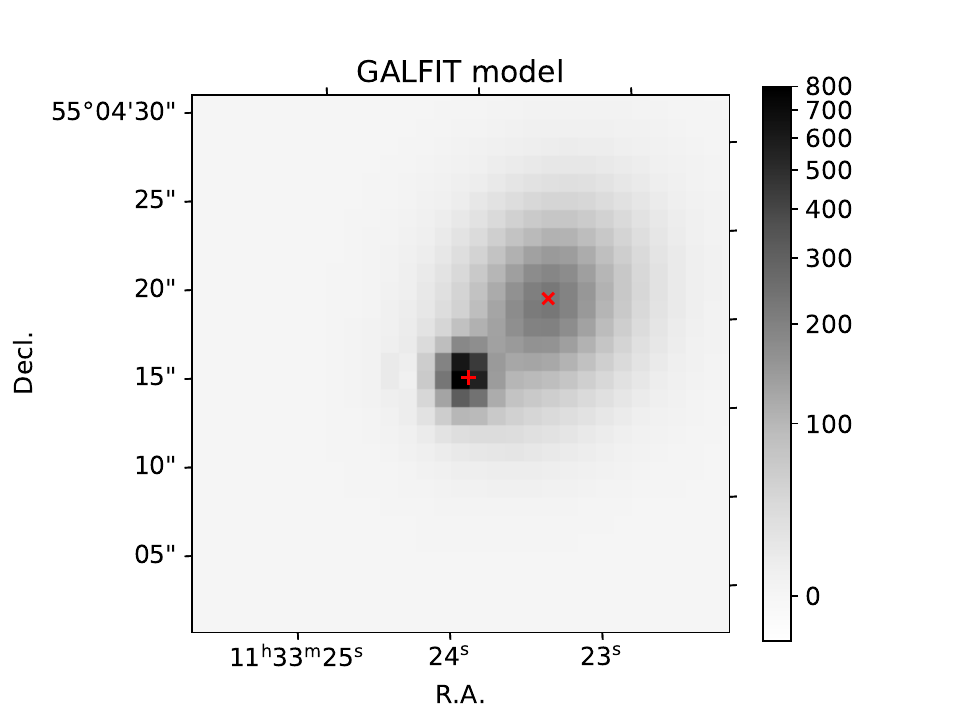}
\includegraphics[clip, width=2.8in]{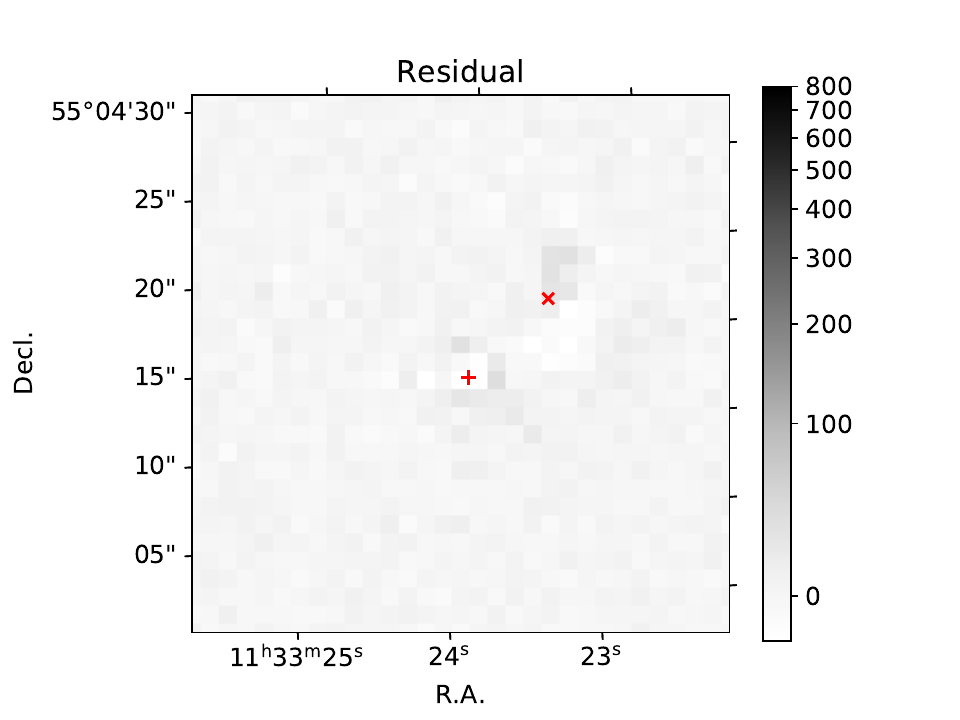}
}
 \caption{ZTF $g$ band image on 2019 June 5 (SDSS1133's bright phase; top), GALFIT best-fit model (middle), and residual map (bottom). The images have a size of $\sim 30'' \times 30''$ and are centered on SDSS1133. The images are shown in asinh scale for visualization purposes. The SDSS coordinates of SDSS1133 and Mrk~177's nucleus are indicated by $+$ and $\times$ symbols, respectively.
 }
 \label{fig:ztf_image_g}
\end{figure}

Zwicky Transient Facility (ZTF) has been conducting a time-domain survey using a wide-field camera with 47 square degree field of view mounted on the Samuel Oschin 48-inch Schmidt telescope at Palomar Observatory since 2017.
Northern Sky Public Survey is a three-day cadence survey of the Northern Sky with ZTF-$g$ and ZTF-$r$ bands, and other private programs are also obtaining $g$, $r$, and $i$ band images over a fraction of the sky \citep{gra19}.
As described by \cite{war21}, ZTF detected the 2019 outburst event of SDSS1133 and reported a transient detection alert on 2019 April 23 (MJD=58596), and recorded the transient as ZTF19aafmjfw \citep[see][for details of the ZTF alert system]{pat19} \footnote{\href{https://antares.noao.edu/alerts/data/14589650}{https://antares.noao.edu/alerts/data/14589650}; \href{https://lasair.roe.ac.uk/object/ZTF19aafmjfw/}{https://lasair.roe.ac.uk/object/ZTF19aafmjfw/}}.

The pipeline-processed data both from public and private surveys are made public via the ZTF Public Data Release.
%ZTF Data Release 3 contains reference-subtracted images, where the reference images were coadded ZTF images taken during the period of 2018 February 17 - June 13, 2018 February 11 - April 18, and 2018 April 15 - 2019 March 16 for $g$, $r$, and $i$ band, respectively.
ZTF DR4 pipeline-processed $g$, $r$, and $i$ band single-epoch images (sciimg.fits; primary science image of database field = 788) and associated PSF model images (sciimgdaopsfcent.fits; PSF estimate at science image center) were downloaded from the Science Data System at IPAC \citep{mas19,bel19}.
The ZTF DR4 includes images taken during the period of 2018 March 25 - 2020 June 29, 2018 April 6 - 2020 June 29, and 2018 April 6 - 2019 June 18 in the $g$, $r$, and $i$ band, respectively.
%Each image has $3072 \times 3080$  pixels with a pixel scale of 1''.012~pixel${}^{-1}$, and was obtained with an exposure time of 30 second.
{\tt INFOBITS} in the image header was used to remove bad-quality flagged images from the analysis below.
% 159, 160, 13  --->  155, 153, 13
%At the position of SDSS1133, 286, 281, and 29 unflagged images in the $g$, $r$, and $i$ band are available in the ZTF DR4.
The magnitude zero-points estimated by the ZTF pipeline (written in the image header) were used to calibrate the instrument magnitudes to AB magnitudes \citep[specifically, the PS1 system;][]{mas19}.
Color terms \citep{mas20} were neglected, which do not affect the final results since SDSS1133's colors, $g-r$ or $r-i$, are close to 0 (Figure~\ref{fig:ztf_lc_zoom}).

% g-band
%TOTEXPT =    480.00            / total effective exposure time [sec]            
%JDSTART = 2458166.882269       / Obs. JD of earliest image used [days]          
%JDEND   = 2458282.750093       / Obs. JD of latest image used [days]        
%2018 February 17 (MJD=58166) to June 13 (MJD=58282)

% r-band
%TOTEXPT =    450.00            / total effective exposure time [sec]            
%JDSTART = 2458160.811088       / Obs. JD of earliest image used [days]          
%JDEND   = 2458226.762164       / Obs. JD of latest image used [days]    
%2018 February 11 (MJD=58160) to April 18 (MJD=58226)

% i-band
%TOTEXPT =    450.00            / total effective exposure time [sec]            
%JDSTART = 2458223.733542       / Obs. JD of earliest image used [days]          
%JDEND   = 2458558.812454       / Obs. JD of latest image used [days]     
%2018 April 15 (MJD=58223) to 2019 March 16 (MJD=58558)

\subsection{Mayall and Bok Legacy Imaging Surveys (2015-2018)}
\label{sec:desi}

The sky region of SDSS1133 was imaged by the DESI Legacy Imaging Surveys \citep{zou19,dey19}, specifically by Mayall $z$ band Legacy Survey and by Bok 90Prime $g$ and $r$ band Survey projects, during the period between 2016 and 2018.
The Mayall Survey is using a mosaic camera (Mosaic-3) mounted on the 4-m Mayall telescope, and Bok 90Prime survey is using the 90Prime camera at the prime focus of the Bok 2.3-m telescope.
Also, since the sky region of SDSS1133 is located inside the spring Hobby-Eberly Telescope Dark Energy Experiment (HETDEX) field, two Mayall $g$ band images obtained on 2015 March 15-16 during the course of an imaging survey for the HETDEX (NOAO program ID = 2015A-0075; PI: Robin B. Ciardullo) are available.

%# Bok 90Prime g 5 images and r 4 images
%# Mayall      g 7 images and z 4 images

Mayall and Bok $g$, $r$, and $z$ band imaging data \citep[community pipeline (CP) {\tt ooi\_v1} products before being processed by {\tt Tractor};][]{dey19} containing SDSS1133 and Mrk~177 were downloaded from the NOAO Science Archive\footnote{\href{http://archive1.dm.noao.edu/}{http://archive1.dm.noao.edu/}}.
We used {\tt SExtractor} and {\tt PSFEx} to estimate PSF models of the images.
The zero-point AB magnitudes were evaluated by comparing instrumental PSF magnitudes of three field stars around SDSS1133 to the PS1 PSF magnitudes taken from the PS1 DR2 {\tt MeanObject} View table \citep{chambers16}, with color transformations given in \cite{dey19}.

\subsection{{\it Swift} (2013-2019)}
\label{sec:swift}

{\it Swift} \citep{geh04} X-ray Telescope (XRT) and UV-Optical Telescope (UVOT) have targeted SDSS1133 several times.
\cite{koss14} carried out Target-of-Opportunity XRT and UVOT UVW1 observations for SDSS1133 on 2013 August 26, 27, and 28.
Also, unpublished Swift UVW2 data obtained on 2014 February 6-25 are available (target ID = 00032905).
Then, several follow-up observations were triggered after the 2019 outburst; {\it Swift} observed SDSS1133 on 2019 May 27 (MJD=58630.1) and 29 (MJD=58632.8) for 2.4 and 2.0 ks, respectively, with the XRT and UVOT in the UVW2, UVM2, and UVW1 filters (target ID = 00011418).
We performed the Swift data analysis with HEAsoft software packages \citep[Version~6.25;][]{nasa14}, using calibration datasets in the High Energy Astrophysics Science Archive Research Center (HEASARC)'s calibration database (CALDB; accessed 2020 December 17).
%https://www.swift.ac.uk/analysis/uvot/
Hereafter each of the {\it Swift} data in 2013, 2014, and 2019 were separately binned into a single measurement.

%The {\it Swift} data were binned into three epochs, where the mean epochs are 
%MJD=56531.4 (2013 August 26-28; UVW1 + XRT), 
%56702.0 (2014 February 6-25; UVW2 + XRT), and 
%58631.4 (2019 May 27-29; UVW2, UVM2, UVW1 + XRT).

\subsubsection{Swift/XRT}
\label{sec:swift_xrt}

We downloaded pipeline-processed {\it Swift} XRT Level 1 products from the HEASARC Archive.
Calibrated and cleaned event files in the XRT photon counting mode were produced using {\tt xrtpipeline} with standard screening criteria.

%We defined three XRT visits as 2013 August 26-28 (MJD=56531.4; 18.6~ks), 2014 February 6-25 (MJD=56702.0; 13.8~ks), and 2019 May 27-29 (MJD=58631.4; 4.3~ks), and each of the XRT datasets was coadded separately.
Each of the 2013, 2014, and 2019 XRT datasets (18.6~ks, 13.8~ks, and 4.3~ks, respectively) was coadded separately.
Spectral extraction and coadding of the event files were performed by using {\tt xselect}.
A circular aperture of $47''$ in radius centered on the SDSS1133's optical coordinate was used for the source count extraction, and an annular aperture with the inner and outer radii of $100''$ and $200''$ centered on the SDSS1133's optical coordinates was used for the background count extraction.
As shown in Section~\ref{sec:chandra}, not only SDSS1133 but also the Mrk~177's nuclear X-ray emission contribute to the X-ray counts in the $r = 47''$ circular aperture.
Thus, precisely speaking, the luminosity upper limits obtained above should be interpreted as the limits for the total luminosity of SDSS1133 and Mrk~177, giving a conservative upper limits on the SDSS1133's luminosity.

{\tt xrtmkarf} was used to create aperture-corrected (uncorrected) auxiliary response files (ARF) for the source (background) spectra, and an exposure time-weighted mean response file for each visit was created by using {\tt addarf}.
A common response matrix file (RMF; swxpc0to12s6\_20130101v014.rmf) taken from CALDB was used for the subsequent analysis.
% https://darts.isas.jaxa.jp/astro/asca/hajimete_ver700/node63.html#addrmf
%Vignetting-corrected summed exposure maps for the three visits were created by using {\tt ximage}.

We found that the X-ray counts of SDSS1133 are quite low, thus the spectral bins were grouped into a single full band ($0.3-10$~keV) by using {\tt XSPEC/grppha}.
The back-ground-subtracted full band net count rates were calculated to be 
$-0.991 (\pm 1.941) \times 10^{-4}$, 
$-1.312 (\pm 2.141) \times 10^{-4}$, and 
$ 4.383 (\pm 5.268) \times 10^{-4}$~count~s${}^{-1}$ 
in 2013, 2014, and 2019, respectively.
%at 
%MJD=
%56531.4, 
%56702.0, and 
%58631.4, respectively.
We conclude that all of these measurements were non-detection.
Under the assumption of Gaussian statistics with zero mean, 90\% upper limits on the count rates are
$2.488 \times 10^{-4}$, 
$2.745 \times 10^{-4}$, and 
$6.754 \times 10^{-4}$~count~s${}^{-1}$, respectively.
The corresponding 90\% upper limits on the $0.3-10$~keV band flux and luminosity were evaluated by assuming an absorbed power-law model ({\tt phabs*powerlaw} in {\tt XSPEC}) with a fixed power-law photon index of $\Gamma=2$ (in the form of $f_{\nu} \propto \nu^{1-\Gamma}$) and a Galactic HI column density of $N_{H}=2 \times 10^{20}$~cm$^{-2}$, and the results are summarized in Table~\ref{tab:xray_table}.
The {\it Swift} non-detection suggests that the X-ray luminosity of SDSS1133 is less than $\sim 3\times 10^{39}$~erg~s$^{-1}$ even at around the peak epoch of the 2019 outburst (Figure~\ref{fig:ztf_lc_zoom}).

Our re-analysis of the {\it Swift}/XRT data shows that the ``marginal'' {\it Swift}/XRT X-ray detection on 2013 August 26-28 reported by \cite{koss14} was actually non-detection.
As described in Section~\ref{sec:chandra}, this conclusion is further confirmed by the {\it Chandra}/ACIS data which indicate that the X-ray luminosity of SDSS1133 is just below the {\it Swift}/XRT's detection limit.

\if0

# 00000000001-xrt
('2013-08-26T20:02:56', 'sw00032905001xpcw3po_cl.evt')
('2013-08-27T11:41:52', 'sw00032905002xpcw3po_cl.evt')
('2013-08-28T00:30:52', 'sw00032905003xpcw3po_cl.evt')

# 00000000002-xrt
('2014-02-06T14:05:50', 'sw00032905004xpcw3po_cl.evt')
('2014-02-07T01:14:50', 'sw00032905005xpcw3po_cl.evt')
('2014-02-08T01:07:51', 'sw00032905006xpcw3po_cl.evt')
('2014-02-13T21:36:45', 'sw00032905007xpcw3po_cl.evt')
('2014-02-15T01:05:49', 'sw00032905008xpcw3po_cl.evt')
('2014-02-21T07:33:49', 'sw00032905009xpcw3po_cl.evt')
('2014-02-25T18:40:51', 'sw00032905010xpcw3po_cl.evt')

# 00000000001-xrt
('2019-05-27T02:21:01', 'sw00011418002xpcw3po_cl.evt')
('2019-05-27T02:20:57', 'sw00011418002xwtw2po_cl.evt')
('2019-05-29T18:27:53', 'sw00011418003xpcw3po_cl.evt')
('2019-05-29T18:27:49', 'sw00011418003xwtw2po_cl.evt')

\fi

\subsubsection{Swift/UVOT}
\label{sec:swift_uvot}

{\it Swift}/UVOT observed SDSS1133 through UVW1 (2600\AA), UVM2 (2246\AA), and UVW2 (1928\AA) bands.
We downloaded pipeline-processed Swift UVOT Level 2 products from the HEASARC Archive.
%The PSF photometry on the {\it Swift} UVOT images was performed as follows.
%The UVM2 band data obtained on 2019 May 27 and 29 (1.4~ks) are combined into a single image using {\tt uvotimsum}.
%The UVW2 band data obtained on 2014 February 6-25 (total exposure is 13.6~ks) and 2019 May 27-29 (1.4~ks) are coadded separately, and the UVW1 band data obtained on 2013 August 26-28 (19.8~ks) and 2019 May 27-29 (1.4~ks) are coadded separately.
For each filter, each of the 2013, 2014, and 2019 data was combined into a single image using {\tt uvotimsum}.

Time dependent sensitivity corrections (several tens percent)\footnote{The most up-to-date UVOT sensitivity calibration file (included in CALDB ver.20201215) was used.} and small deadtime and coincidence loss corrections (at most 2\%) were calculated by using {\tt uvotsource} and applied for the measured fluxes \citep{poo08,bre11}.
The zero point AB magnitudes for the source count rates for the standard aperture are 18.95, 18.54, and 19.11~mag for UVW1, UVM2, and UVW2 band, respectively\footnote{These photometric zero-points convert count rates into AB magnitudes, where the count rates are corrected for deadtime, coincidence-loss, aperture corrections, large-scale sensitivity variations, and detector sensitivity variations.}.
Using tables of encircled energy curves for the UVOT PSFs in the CALDB, we estimated zero point AB magnitudes for the infinite aperture (defined to be $30''$ in radius) to be used for the PSF photometry as 19.11, 18.72, and 19.24~mag for UVW1, UVM2, and UVW2 band, respectively.
The PSF models of the UVOT images were created with {\tt PSFEx}.
%The total exposure times are long enough to approximate the statistical photometric error by Poisson statistics \citep{kui08,poo08}\footnote{The Swift UVOT frame exposure time is 0.0110322~sec.}.
When reporting the {\it Swift}/UVOT photometric measurements, the errors on the magnitudes only include statistical errors.
We should note that, in addition to the statistical errors, the reported magnitudes can be affected by various systematic errors; e.g., the {\it Swift}/UVOT photometric zero-point estimates induce $\sim 0.03$~mag errors on the magnitudes.

\subsection{{\it XMM-Newton} (2015 December)}
\label{sec:xmmnewton}

{\it XMM-Newton} \citep{jan01} X-ray and UV imaging data were obtained on 2015 December 2 and 4 (Obs. ID=0762960201 and 0762960301; PI: Michael Koss), which is unpublished in the literature.
We downloaded the two {\it XMM-Newton} datasets (Observation Data File; ODF) from the {\it XMM-Newton} Science Archive.
The Current Calibration Files (CCF) were downloaded from the CCF ftp server (accessed 2020 October 28).
The data analysis was performed by using the {\it XMM-Newton} Science Analysis System \citep[{\tt SAS} version 18.0.0;][]{gab04}.

\subsubsection{$XMM$-$Newton$/EPIC}
\label{sec:xmmnewton_epic}

\if0

# Src.reg
# Region file format: DS9 version 4.1
# Filename: ./0762960301/reprocess/PNimage_EPICclean_2928_0762960301_EPN_S003_ImagingEvts.ds
global color=green dashlist=8 3 width=1 font="helvetica 10 normal roman" select=1 highlite=1 dash=0 fixed=0 edit=1 move=1 delete=1 include=1 source=1
fk5
circle(11:33:23.978,+55:04:15.90,32")

# Bkg.reg
# Region file format: DS9 version 4.1
# Filename: ./0762960301/reprocess/PNimage_EPICclean_2928_0762960301_EPN_S003_ImagingEvts.ds
global color=green dashlist=8 3 width=1 font="helvetica 10 normal roman" select=1 highlite=1 dash=0 fixed=0 edit=1 move=1 delete=1 include=1 source=1
fk5
circle(11:33:32.834,+55:06:27.06,72")

\fi

Filtered photon counting event files of European Photon Imaging Camera (EPIC) MOS and PN cameras were generated by using standard {\tt SAS} tasks {\tt emproc}, {\tt epproc}, and {\tt evselect}.
The exposure times of the MOS1, MOS2, and PN data are 8.2, 8.4, and 4.1~ks on 2015 December 2, and 11.3, 11.5, 7.2~ks on 2015 December 4, respectively.
MOS and PN spectra of the two datasets were extracted and then combined into a single source spectrum, background spectrum, and response matrix (RSP) by using {\tt SAS/multiespecget}.
Here the combined RSP file was defined by summing up RSP files (=RMF$\times$ARF) of all of the MOS and PN data, and the effective exposure time of the combined dataset was calculated to be 8.5~ks.
A circular aperture of $32''$ in radius centered on the optical position of SDSS1133 was used for the source spectral extraction, and an offset circular aperture of $72''$ in radius was used for the background spectral extraction.
As described in Section~\ref{sec:swift_xrt}, the source extraction aperture of $32''$ contains not only SDSS1133 but also the Mrk~177's nuclear X-ray emission (see Section~\ref{sec:chandra}), thus the luminosity upper limit obtained here only provide a conservative upper limit on the SDSS1133's luminosity.

The same analysis as Section~\ref{sec:swift_xrt} was applied for the {\it XMM-Newton} combined spectral data.
The effective background-subtracted $0.3-10$~keV net count rate was calculated to be
$2.981~(\pm 2.266) \times 10^{-3}$ count~s${}^{-1}$, 
thus we conclude that SDSS1133 is undetected in the {\it XMM-Newton} data.
The 90\% upper limit on the effective count rate is 
$2.905 \times 10^{-3}$ count~s${}^{-1}$.
Assuming the same absorbed power-law spectral model ({\tt phabs*powerlaw} in {\tt XSPEC}), we obtained an 90\% upper limit on the unabsorbed X-ray flux in the $0.3-10$ keV band as
$3.370\times 10^{-15}$~erg~cm${}^{-2}$~s${}^{-1}$, 
corresponding to an 90\% upper limit on  the $0.3-10$ keV luminosity of 
$L_{0.3-10~\text{keV}} = 3.367 \times 10^{38}$~erg~s${}^{-1}$, as summarized in Table~\ref{tab:xray_table}.

\subsubsection{$XMM$-$Newton$/OM}
\label{sec:xmmnewton_om}

Both of the two {\it XMM-Newton}/Optical Monitor \citep[OM;][]{mas01} imaging mode datasets were obtained through the UVW1 band (2910~\AA) with the pre-defined EPIC Imaging mode configuration, where the total on-source exposure times were 10.9 and 11.8~ks, respectively.
Astrometrically-aligned mosaiced sky images ($\sim 16' \times 16'$, $2 \times 2$ binning, $0".953$~pixel${}^{-1}$) of the two datasets were separately generated from the ODFs by using {\tt SAS/omichain}.
Since SDSS1133 is faint, the coincidence-loss is negligible.
The dead time fraction ($0.017$), sensitivity degradation correction factor ($1.126$), and large-aperture zero point in AB system ($18.566$~mag)\footnote{The UVW1-filter zero point is defined by an aperture of 35 unbinned pixels ($16''.7$) radius, and for count rates corrected for the coincidence-loss, dead time, and sensitivity degradation.} were directly taken from the CCF.
% cts_corr = cts_obs / (1-deadfrac) * 1.126
We applied the same analysis as Section~\ref{sec:swift_uvot} to the {\it XMM-Newton}/OM data; {\tt SExtractor} and {\tt PSFEx} were respectively used to subtract global background from the images and create PSF models.

\if0

#
# coincidence-loss and deadtime loss correction
# 
# T   = Tft + Tnetexp ~ 10ms
# Tft = 0.1740ms
# deadfrac = Tft/T
#
# From Poisson (Mason+2001), 
# cts_obs * T = 1 - exp( -[ cts_corr * T * (1-deadfrac) ] )
# This is equivalent to
# cts_corr = [ ln( 1 - cts_obs * T ) ] / (Tft - T)
# This is the correction formula.
#
# When cts_corr is small, 
# cts_obs * T = cts_corr * T * (1-deadfrac)
# --> cts_obs     = cts_corr * (1-deadfrac)
# --> cts_corr    =  cts_obs / (1-deadfrac)
#
#

python
import numpy as np
from astropy import units as u
from astropy import constants as const

# cts_detected is r=35 unbinned = 17.5 binned aperture centered on SDSS1133
# cts_detected = [ 6.98745, 6.96354 ] / u.second

#
# cts_detected is r=11".2 unbinned = 5".6 binned aperture centered on SDSS1133
# cts_detected = [ 1.39973, 1.37238	 ] / u.second
#https://www.mssl.ucl.ac.uk/www_astro/XMM-OM-SUSS/DataProcessingPhotometry.shtml

Tft          = 0.1740 * u.millisecond
T            = [ 9.8908,  10.3756 ] * u.millisecond
cts_detected = [ 1.39973, 1.37238	 ] / u.second
phin = ( np.log(1.0 - cts_detected*T)/(Tft - T) ).to(1/u.second)
print( cts_detected, phin )
print( phin/cts_detected )

\fi

\subsection{{\it Chandra} (2019 August)}
\label{sec:chandra}

\begin{figure}
\center{
\includegraphics[clip, width=3.9in]{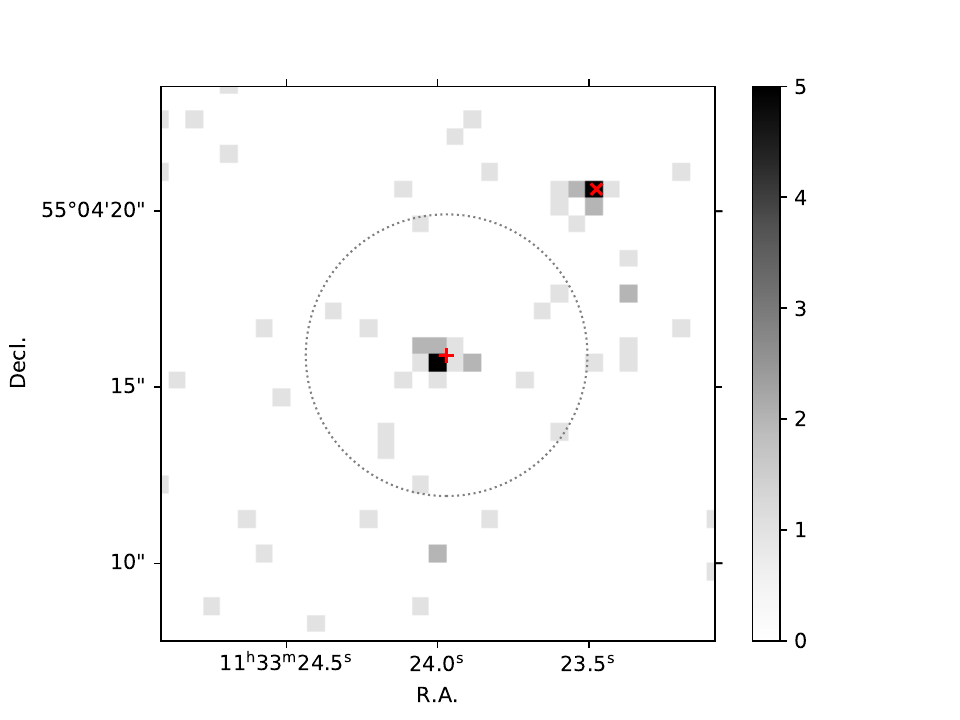}
}
 \caption{{\it Chandra}/ACIS $0.5-8$~keV co-added count image centered on SDSS1133 ($\sim 15'' \times 15''$, 0.492~arcsec~pixel${}^{-1}$) obtained on 2019 August 13 and 14 (74.2~ks; Obs.~ID = 21434 and 22684). The colorbar indicates the X-ray counts. North is up and East is to the left.
 The source extraction aperture ($4''$ in radius) centered on SDSS1133 is shown as a dotted line. 
 The X-ray source located $\sim 6''$ North West from SDSS1133 corresponds to the nucleus of the host galaxy Mrk~177. The SDSS optical coordinates of SDSS1133 and Mrk~177's nucleus are indicated by $+$ and $\times$ symbols, respectively.
 }
 \label{fig:chandra}
\end{figure}

{\it Chandra} \citep{wei02} observed SDSS1133 on 2019 August 13 (MJD=58708.4, 16~ks) and 14 (MJD=58709.3, 58~ks) with the Advanced CCD Imaging Spectrometer, ACIS (Obs. ID = 21434 and 22684; PI: David Pooley), which is unpublished in the literature.
Since the optical light curves show that SDSS1133 dimmed by $\sim 2$~mag compared to the peak of the 2019 outburst by the end of 2019 June \citep[Figure~\ref{fig:ztf_lc_zoom};][]{pur19}, the {\it Chandra} observations put a constraint on the X-ray emission of the SDSS1133 in its post-outburst phase.

We used {\tt CIAO} v4.12.1 \citep{fru06} to download and analyse the archival ACIS data.
Figure~\ref{fig:chandra} shows a $0.5-8$~keV co-added clipped count image centered on SDSS1133 ($\sim 15'' \times 15''$) created by using {\tt CIAO/merge\_obs}.
SDSS1133 is clearly detected in this image, and also the nucleus of the host galaxy Mrk~177 is detected at a similar count rate.

The event files, ARFs, and RMFs, reprocessed with {\tt CIAO/chandra\_repro} using CALDB 4.9.2.1, were input to {\tt CIAO/specextract} to extract count rate spectra, where a circular aperture of $4''$ in radius and annular aperture with $10''$ inner and $40''$ outer radii were used for the source and background extraction, respectively.
The optical position of SDSS1133 was used to extract the spectra.
Aperture corrections were applied for the ARFs for the source.
Then, the spectra from the two exposures were summed by using {\tt CIAO/combine\_spectra} to create a 74.2~ks spectrum.

The background-subtracted full-band ($0.5-8$~keV) net count rate was calculated to be $2.327~(\pm 0.688) \times 10^{-4}$ count~s${}^{-1}$ with the signal-to-noise ratio of $S/N = 3.380$, under the assumption of the Gaussian statistics.
The number of the observed X-ray counts is too small for the spectral analysis (see Figure~\ref{fig:chandra}), thus here we report integrated quantities in a $0.5-8$~keV band.
By using {\tt CIAO/modelflux} with the ARFs and RMFs, and assuming a power-law spectral model with a fixed power-law photon index of $\Gamma=2$ ({\tt xspowerlaw}) and a Galactic HI column density of $N_{H}=2\times 10^{20}$~cm${}^{-2}$ ({\tt xsphabs}), we obtained an unabsorbed X-ray flux in the $0.5-8$ keV band as 
$3.109~(\pm 0.919) \times 10^{-15}$~erg~cm${}^{-2}$~s${}^{-1}$, or 
$0.5-8$ keV luminosity of 
$L_{0.5-8~\text{keV}} = 3.107~(\pm 0.918) \times 10^{38}$~erg~s${}^{-1}$.
The corresponding $0.3-10$ keV luminosity is 
$3.929 \times 10^{38}$~erg~s${}^{-1}$ (Table~\ref{tab:xray_table}).
% which is about 4 times lower than that reported by \cite{koss14} using Swift/XRT data ($L_{0.3-10~\text{keV}} = 1.5 \times 10^{39}$~erg~s${}^{-1}$).
As already mentioned in Section~\ref{sec:swift_xrt}, this faint X-ray luminosity suggests that the {\it Swift}/XRT observations were not sensitive enough to detect X-ray emission from SDSS1133.

Under the assumption that SDSS1133 is a LBV, an optically-thin thermal plasma model may be a more adequate reference model \citep[e.g.,][]{naz12}.
By using a thin thermal plasma emission model {\tt xsapec} \citep{smi01} in {\tt CIAO/modelflux}, and fixing the temperature to $k_{B}T = 0.6$ keV \citep[following][]{naz12}, Galactic HI column density to $N_{H}=2\times 10^{20}$~cm${}^{-2}$, and metal abundance to $Z_{\odot}$,  we obtained an unabsorbed X-ray flux in the $0.5-8$ keV band as $3.425~(\pm 1.013) \times 10^{-15}$~erg~cm${}^{-2}$~s${}^{-1}$, or $0.5-8$ keV luminosity of 
$L_{0.5-8~\text{keV}} = 3.423~(\pm 1.012) \times 10^{38}$~erg~s${}^{-1}$.
The difference of the estimated luminosity between the power-law model and thermal plasma model is small, suggesting that the luminosity estimates are not very sensitive to the choice of the spectrum models.
In the subsequent sections, we only refer to the X-ray luminosities obtained under the assumption of the absorbed power-law model summarised in Table~\ref{tab:xray_table}.

By applying the same analysis on the nucleus of Mrk~177, the count rate was evaluated as $1.770~(\pm 0.634) \times 10^{-4}$ count~s${}^{-1}$.
We obtained an unabsorbed X-ray flux in the $0.5-8$ keV band as $2.339~(\pm 0.838) \times 10^{-15}$~erg~cm${}^{-2}$~s${}^{-1}$, or $0.5-8$ keV luminosity of 
$L_{0.5-8~\text{keV}} = 2.337~(\pm 0.837) \times 10^{38}$~erg~s${}^{-1}$.
The corresponding $0.3-10$ keV luminosity is $2.957 \times 10^{38}$~erg~s${}^{-1}$.
Coincidentally, this luminosity is very similar to that of SDSS1133.
Since the separation between SDSS1133 and Mrk~177's nucleus is $5''.8$, the low spatial resolution X-ray telescopes ({\it Swift} and {\it XMM-Newton}) cannot resolve the two X-ray components, as already pointed out in Sections~\ref{sec:swift_xrt} and Sections~\ref{sec:xmmnewton_epic}.

\if0

/home/mkokubo/anaconda3/bin/python
from astropy import units as u
from astropy import constants as const
import numpy as np

# SDSS1133

# power-law

scale = 3.109e-15/2.327e-4
error = 0.688e-4 * scale
print(error)

dl = 28.9*u.Mpc
L  = ( np.array([ 3.109, 0.919 ])*1.0e-15 * u.erg/u.s/u.cm**2 * 4.0*np.pi*dl**2.0 ).to(u.erg/u.s)
print(L)

L  = ( np.array([ 3.9318 ])*1.0e-15 * u.erg/u.s/u.cm**2 * 4.0*np.pi*dl**2.0 ).to(u.erg/u.s)
print(L)

# thin thermal plasma

scale = 3.425e-15/2.327e-4
error = 0.688e-4 * scale
print(error)

dl = 28.9*u.Mpc
L  = ( np.array([ 3.425, 1.013 ])*1.0e-15 * u.erg/u.s/u.cm**2 * 4.0*np.pi*dl**2.0 ).to(u.erg/u.s)
print(L)

# Mrk117

# power-law

scale = 2.339e-15/1.770e-4
error = 0.634e-4 * scale
print(error)

dl = 28.9*u.Mpc
L  = ( np.array([ 2.339, 0.838 ])*1.0e-15 * u.erg/u.s/u.cm**2 * 4.0*np.pi*dl**2.0 ).to(u.erg/u.s)
print(L)

L  = ( np.array([ 2.9585 ])*1.0e-15 * u.erg/u.s/u.cm**2 * 4.0*np.pi*dl**2.0 ).to(u.erg/u.s)
print(L)

\fi

\subsection{UV-optical structural decomposition of SDSS1133 and Mrk 177 with {\tt GALFIT}}
\label{sec:structural_decomposition}

\begin{table}
\centering
\caption{The mean structural S\'{e}rsic parameters of Mrk~177 obtained by the {\tt GALFIT} modelling for each UV-optical dataset.}
\label{tab:structural_decomposition_table}
\begin{tabular}{ccccccc}
\hline
mag. & ${R}_{e}$ & ${n}$ & ${b/a}$ & PA        & Telescope & Filter \\
     & ($''$)    &       &         & ($\circ$) &           &        \\
\hline
17.43 & 5.12 & 1.17 & 0.77 & 172.69 & Swift  & UVW2 \\
17.35 & 7.54 & 8.11 & 0.78 & 180.22 & Swift  & UVM2 \\
17.28 & 5.40 & 1.26 & 0.78 & 164.26 & Swift  & UVW1 \\
17.02 & 5.01 & 0.83 & 0.78 & 156.41 & XMM    & UVW1 \\
16.62 & 5.41 & 1.09 & 0.74 & 156.09 & SDSS   & $u$ \\
15.69 & 4.72 & 0.80 & 0.78 & 154.68 & SDSS   & $g$ \\
15.71 & 4.74 & 0.81 & 0.77 & 152.72 & PS1    & $g$ \\
15.92 & 4.62 & 0.84 & 0.79 & 158.44 & PTF    & $g$ \\
15.69 & 4.59 & 0.84 & 0.78 & 157.23 & ZTF    & $g$ \\
15.66 & 4.76 & 0.87 & 0.77 & 155.25 & Bok    & $g$ \\
15.64 & 4.83 & 0.90 & 0.76 & 156.43 & Mayall & $g$ \\
15.34 & 4.57 & 0.89 & 0.80 & 155.90 & SDSS   & $r$ \\
15.37 & 4.57 & 0.90 & 0.78 & 152.86 & PS1    & $r$ \\
15.27 & 4.46 & 0.93 & 0.79 & 157.77 & PTF    & $r$ \\
15.32 & 4.40 & 0.91 & 0.79 & 157.92 & ZTF    & $r$ \\
15.30 & 4.57 & 0.96 & 0.79 & 155.90 & Bok    & $r$ \\
15.15 & 4.54 & 0.99 & 0.81 & 155.37 & SDSS   & $i$ \\
15.25 & 4.52 & 0.95 & 0.81 & 153.86 & PS1    & $i$ \\
15.19 & 4.27 & 0.97 & 0.81 & 156.97 & ZTF    & $i$ \\
15.10 & 4.17 & 1.03 & 0.81 & 154.58 & SDSS   & $z$ \\
15.11 & 4.36 & 1.01 & 0.80 & 153.01 & PS1    & $z$ \\
15.05 & 4.45 & 1.11 & 0.80 & 155.16 & Mayall & $z$ \\
15.10 & 4.29 & 0.95 & 0.80 & 153.70 & PS1    & $y$ \\
\hline
\end{tabular}
%\bigskip
   \begin{tablenotes}
     \item [1] PA is the position angle of the semi-major axis from the North to the East.
   \end{tablenotes}
\end{table}

For each of the UV-optical imaging datasets (SDSS, PS1, PTF, ZTF, Mayall, Bok, {\it Swift}/UVOT, and XMM-Newton/OM), we performed structural decomposition of SDSS1133 and Mrk177 using {\tt GALFIT} to obtain host-subtracted PSF magnitudes of SDSS1133.
A 2-component model was adopted, where SDSS1133 was modelled by a PSF and Mrk~177 was modelled by a S\'{e}rsic profile (see Figure~\ref{fig:ztf_image_g} for an example of the {\tt GALFIT} modelling for a ZTF image).
Specifically, {\tt GALFIT} $\chi^2$-minimization was performed leaving 3 ($X_{\text{PSF}}$, $Y_{\text{PSF}}$, mag${}_{\text{PSF}}$) and 7 ($X_{\text{Sersic}}$, $Y_{\text{Sersic}}$, mag${}_{\text{Sersic}}$, ${R}_{e,\text{Sersic}}$, ${n}_{\text{Sersic}}$, ${b/a}_{\text{Sersic}}$, PA${}_{\text{Sersic}}$) parameters were fitted as free parameters for the point-source and S\'{e}rsic component, respectively.
The fitting was performed on $\sim 80'' \times 80''$ cutout images centered on SDSS1133.

PanSTARRS, PTF, ZTF, Mayall, and Bok datasets have multiple images.
To reduce internal scatter in the PSF light curves of each dataset, we performed a two-step fitting procedure as follows.
For each dataset of each filter, first {\tt GALFIT} was performed for each image leaving all the 10 parameters as free parameters, and then {\tt GALFIT} was performed again with the 4 S\'{e}rsic parameters (${R}_{e,\text{Sersic}}$, ${n}_{\text{Sersic}}$, ${b/a}_{\text{Sersic}}$, PA${}_{\text{Sersic}}$) being fixed to mean values of the fitted parameters.
After the fitting, multiple measurements obtained on the same day were combined into a single measurement by taking a weighted mean.
In total, 
the PanSTARRS data have 5, 7, 11, 9, and 10 data points for the $g$, $r$, $i$, $z$, and $y$ band, 
the PTF data have 1 and 7 data points for the $g$, and $R$ band, 
the ZTF data have 223, 210, and 28 data points for the $g$, $r$, and $i$ band, 
the Mayall data have 2 and 4 data points for $g$ and $z$, and 
the Bok data have 5 and 4 data points for $g$ and $z$, band,
respectively.

The results of the PSF photometry for SDSS1133 are summarized in Table~\ref{tab:photometry_table}.
For the SDSS data, the PSF magnitudes derived here are consistent with the values reported in \cite{koss14}, who derived host-subtracted PSF magnitudes of SDSS1133 by fitting a PSF+linear background model for $5'' \times 5''$ images centered on SDSS1133.
The UVW1 magnitude of SDSS1133 on 2013 August 27 reported by \cite{koss14} was $21.41 \pm 0.30$~mag, which differs from our result ($20.14$~mag) by $\sim 1.27$~mag.
We speculate that the difference may be due to overestimation of the host galaxy flux contribution in \cite{koss14}'s analysis, where the a 5'' region around SDSS1133 was fitted with a PSF model and a linear background model.
For reference, we measure the $3''$ radius aperture flux of SDSS1133 on 2013 August 27 (with aperture correction applied by using {\tt uvotsource} and without host galaxy subtraction) as $19.29 \pm 0.05$ (stat.) $\pm 0.03$ (sys.)~mag, which is close to our PSF magnitude estimate ($20.14$~mag).
This indicates that the host galaxy contamination is small in the UV bands.

Table~\ref{tab:structural_decomposition_table} shows the mean structural parameters of Mrk~177 obtained by the {\tt GALFIT} modelling.
We can see that the derived S\'{e}rsic parameters are consistent among the various survey datasets, except for the {\it Swift} UVM2 band (specifically the S\'{e}rsic index $n$); this deviation is probably because the single S\'{e}rsic component does not provide a good fitting to the patchy appearance of the galaxy in the shallow UV image.
Since the accurate structural modelling is beyond the scope of this paper and it does not necessarily improve the PSF photometry of SDSS1133 in the UV band, we did not perform further fine-tuned refit.
Actually, we confirmed that the UVM2 band PSF magnitude of SDSS1133 changes only by $0.15$~mag when fixing the S\'{e}rsic index to be $n=1$.

\subsection{{\it Gaia} (2015-2022)}
\label{sec:gaia}

The $Gaia$ mission \citep{gai16} has been observing SDSS1133 in $Gaia$'s $G$ band since 2015 January 18, including the period of the second flare event in 2019.
The $Gaia$ $G$ band is a white-light filter covering entire optical wavelengths \citep{wei18}.
We downloaded $Gaia$ light curve data of SDSS1133 \citep[recorded as Gaia19bwn;][]{pur19b} from the $Gaia$ Photometric Science Alerts Database webpage (accessed 2022 May 24), excluding `null' and `untrusted' data points \footnote{\href{http://gsaweb.ast.cam.ac.uk/alerts/tableinfo}{http://gsaweb.ast.cam.ac.uk/alerts/tableinfo}.}.
In total 141 data points are available between 2015 January 18 and 2022 March 14 as of 2022 May.
The $Gaia$ light curve is plotted in Figure~\ref{fig:ztf_lc}, but we should note that  the $Gaia$ photometry is host-unsubtracted.

\subsection{Other UV-optical, NIR, and radio measurements}
\label{sec:koss_photometry}

Historic photometric measurements for SDSS1133 based on Photographic Digital Sky Survey (DSS) plates from the POSS I and II surveys are reported by \cite{koss14}.
Although the detections of SDSS1133 are not significant \citep[see Figure~1 of][]{koss14}, the visual magnitudes on 1950 March 20, 1994 April 14, and 1999 April 25 are measured as
$18.6 \pm 0.7$, 
$18.4 \pm 0.4$, and
$18.8 \pm 0.5$~mag, where the observations were carried out with 103aO ($u$ and $g$ band), GG395 ($g$ band), and IIIaF emulsion ($i$ band), respectively.

In addition, a $g$ band upper limit is reported by \cite{zho06} on January 2005; the observation using the 2.16-m telescope of the Beijing Observatory failed to detect the source within 3~mag of the 2002 SDSS observation, providing a rough limit on the $g$ band brightness as $g > 19.4$~mag \citep{koss14}.

\cite{koss14} carried out $J$ and $K_{p}$ band Adaptive Optics assisted imaging observations for SDSS1133 with Keck/the Near Infrared Camera 2 (NIRC2) on 2013 June 16 (MJD=56459.3)
These observations put constraints on the size of the SDSS1133's emission region to be $<$ 12~pc, and $J$- and $K_{p}$ band AB magnitudes are estimated as $19.02 \pm 0.18$ and $18.92 \pm 0.16$~mag, respectively.

$GALEX$ All-sky Imaging survey provides a NUV band ($1750-2800$~\AA; $\lambda_{\text{eff}}=2267$~\AA) magnitude on 2004 March 6 as $21.62 \pm 0.40$~mag \citep{koss14}, though the detection significance is not high and it can be a false detection.

At a radio band, \cite{perez15} report a non-detection by 5.0 GHz electronic European VLBI Network (eEVN) radio observations for SDSS1133 on 2014 December 12 provides a 3$\sigma$ upper limit of 150 microJy~beam${}^{-1}$, corresponding to a 5.0 GHz luminosity of $5.2 \times 10^{24}$ erg~s${}^{-1}$~Hz${}^{-1}$.
Archival FIRST radio observations at 1.4 GHz also provides a 3$\sigma$ upper limit of about 450 microJy~beam${}^{-1}$, corresponding to a luminosity of $1.6 \times 10^{25}$ erg~s${}^{-1}$~Hz${}^{-1}$.

\subsection{Optical spectra}
\label{sec:specdata}

\begin{figure*}
\center{
\includegraphics[clip, width=6.2in]{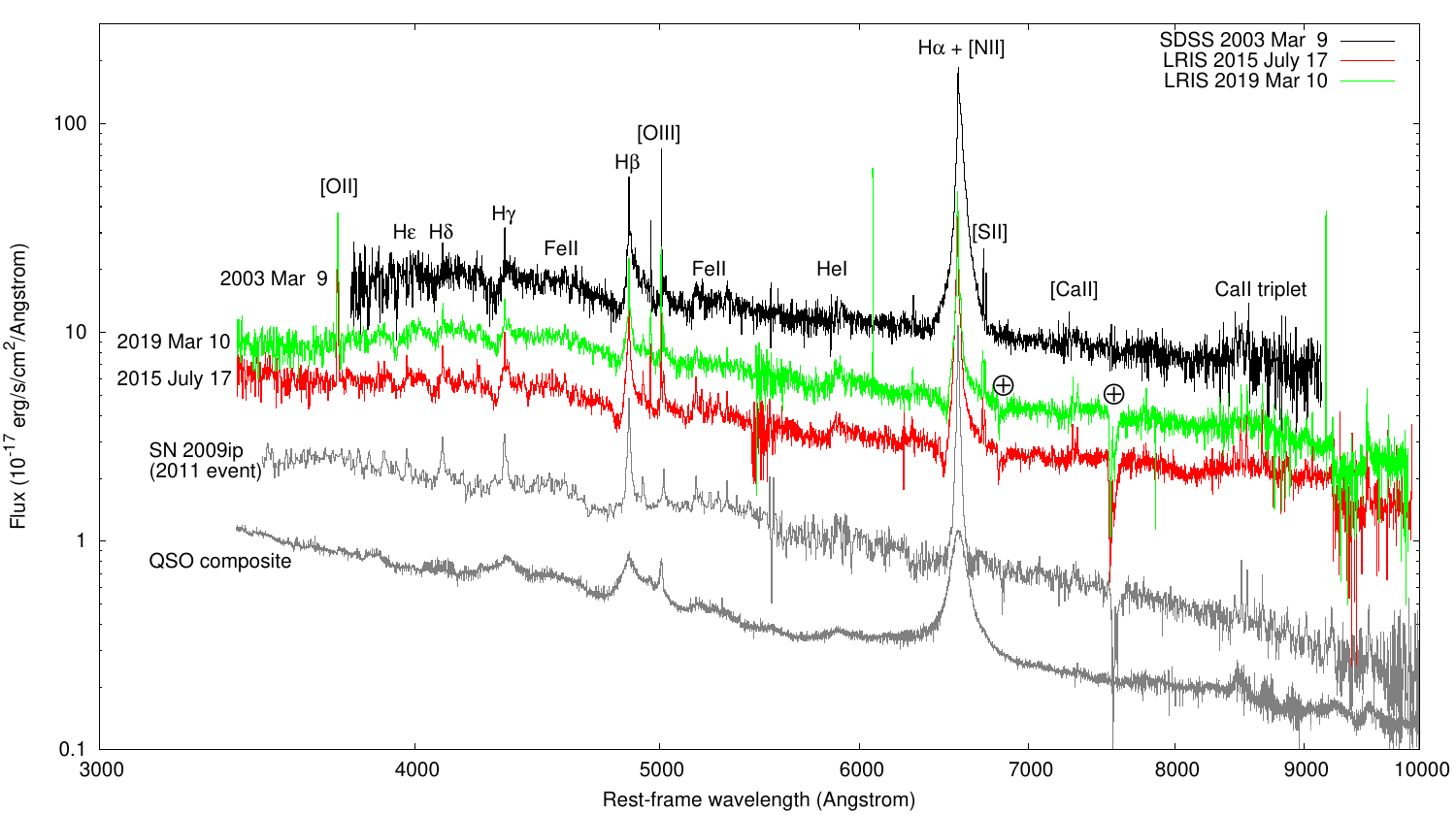}
}
 \caption{Keck/LRIS and SDSS spectra of SDSS1133. 
 Several emission/absorption features are labelled.
 Hydrogen Balmer series shows broad P-Cygni-like features (see Section~\ref{sec:specdata}). Galactic extinction is corrected. The Keck/LRIS spectra are not corrected for the telluric absorption (indicated by $\oplus$). The Galactic-extinction corrected $g$ band magnitudes at the epochs of the SDSS and LRIS observations were estimated as $18.56$, $19.95$, and $19.38$~mag, respectively. 
 For comparison, a VLT/X-Shooter optical spectrum of a pre-SN eruption of SN~2009ip \citep[2011 event on 2011 September 24;][]{pas13} and a SDSS QSO composite spectrum \citep[$1<z<2.1$;][]{sel16} are also plotted.
 }
 \label{fig:keck_spectra}
\end{figure*}

\begin{figure*}
\center{
\includegraphics[clip, width=6.2in]{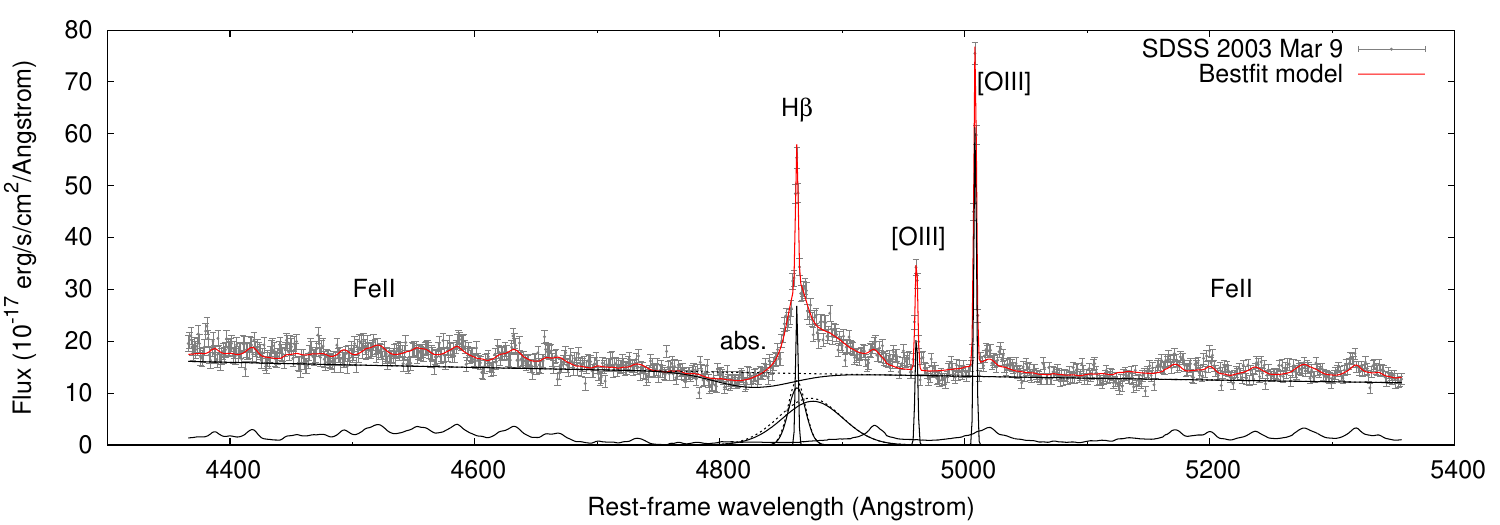}
\includegraphics[clip, width=6.2in]{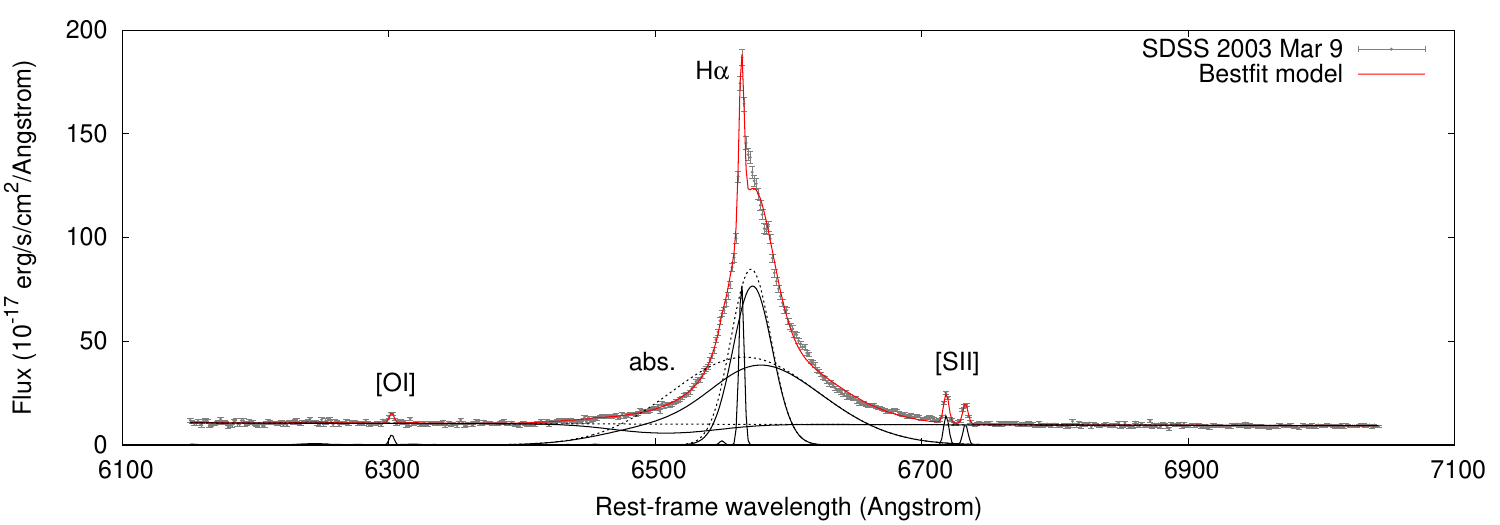}
}
 \caption{Spectral decomposition of the SDSS spectrum of SDSS1133 at the wavelength ranges of H$\beta$ (top) and H$\alpha$ (bottom). 
 The red line represents the summed model spectra. 
 A power-law continuum, iron (mostly Fe~II) pseudo-continuum template, broad, very broad, and narrow Balmer lines, and several narrow emission lines are included in the model (black lines).
 The power-law continuum and broad and very broad Balmer emission lines are assumed to be affected by a broad P-Cygni-like absorption; solid and dotted lines are absorbed and unabsorbed model components, respectively (see Section~\ref{sec:spectral_decomposition} for details).
 }
 \label{fig:sdss_decomposition}
\end{figure*}

\begin{figure*}
\center{
\includegraphics[clip, width=2.8in]{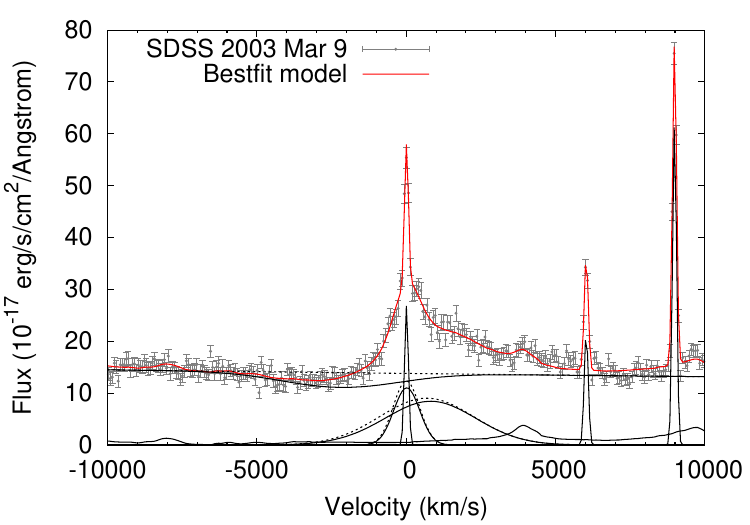}
\includegraphics[clip, width=2.8in]{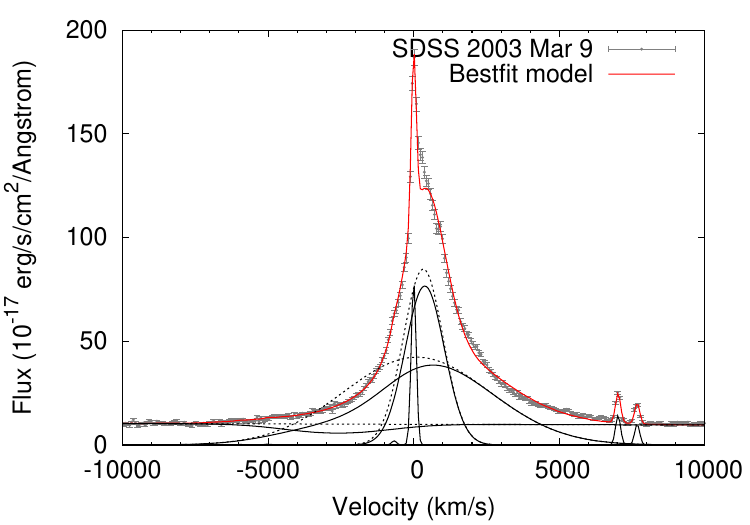}
}
\center{
\includegraphics[clip, width=2.8in]{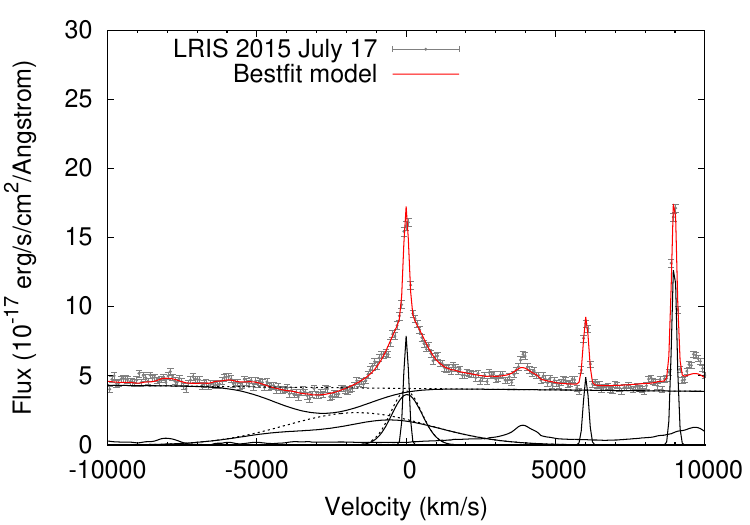}
\includegraphics[clip, width=2.8in]{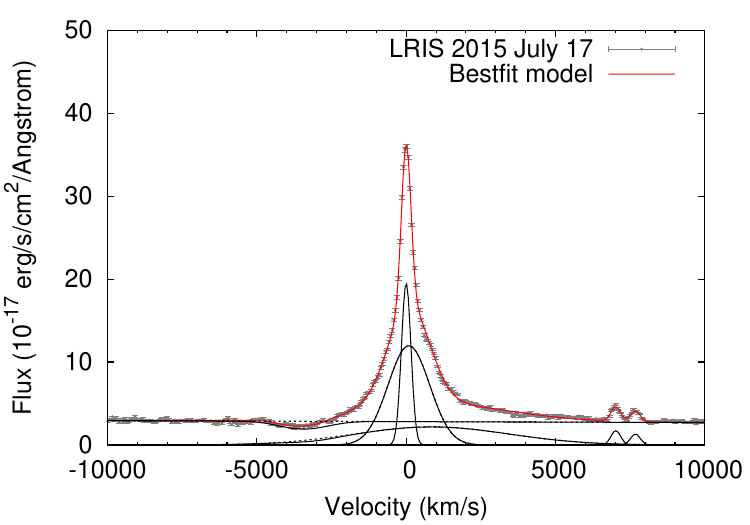}
}
\center{
\includegraphics[clip, width=2.8in]{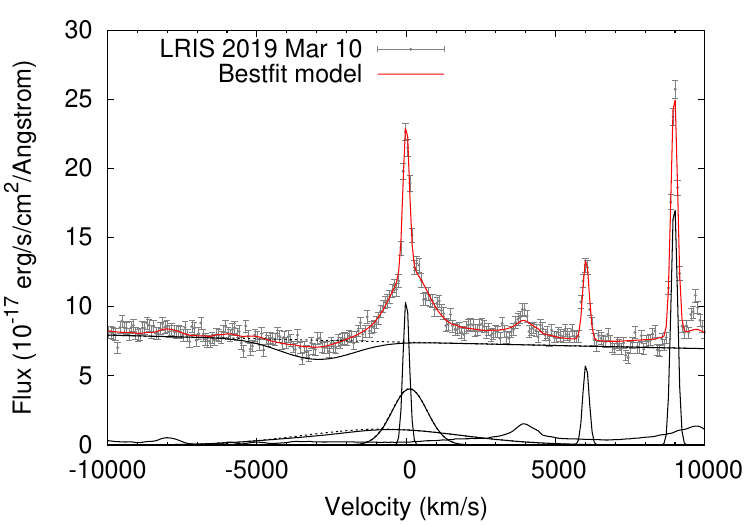}
\includegraphics[clip, width=2.8in]{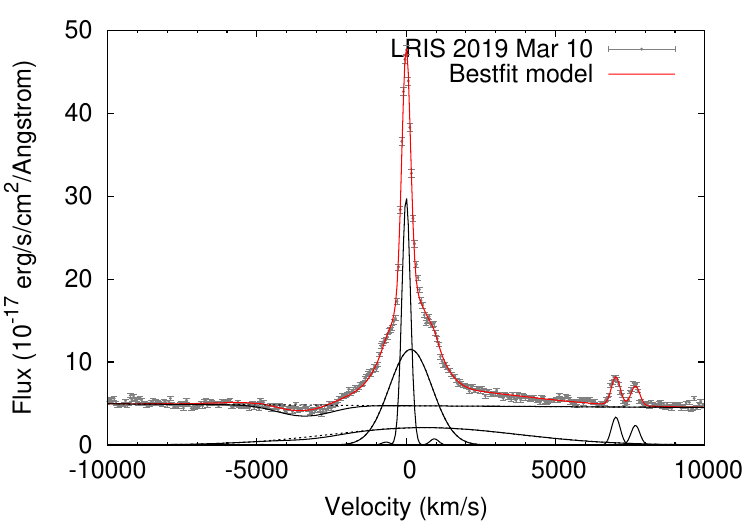}
}
 \caption{The same as Figure~\ref{fig:sdss_decomposition}, but the SDSS and Keck/LRIS spectra are shown as a function of velocity relative to the narrow H$\beta$ line center. 
 %The Galactic extinction-corrected $g$ band magnitudes are $g = 18.66$, 19.95, and 19.38~mag at the epochs of the SDSS and the two LRIS observations, respectively.
 }
 \label{fig:sdss_decomposition_velocity}
\end{figure*}

\begin{figure}
\center{
\includegraphics[clip, width=3.2in]{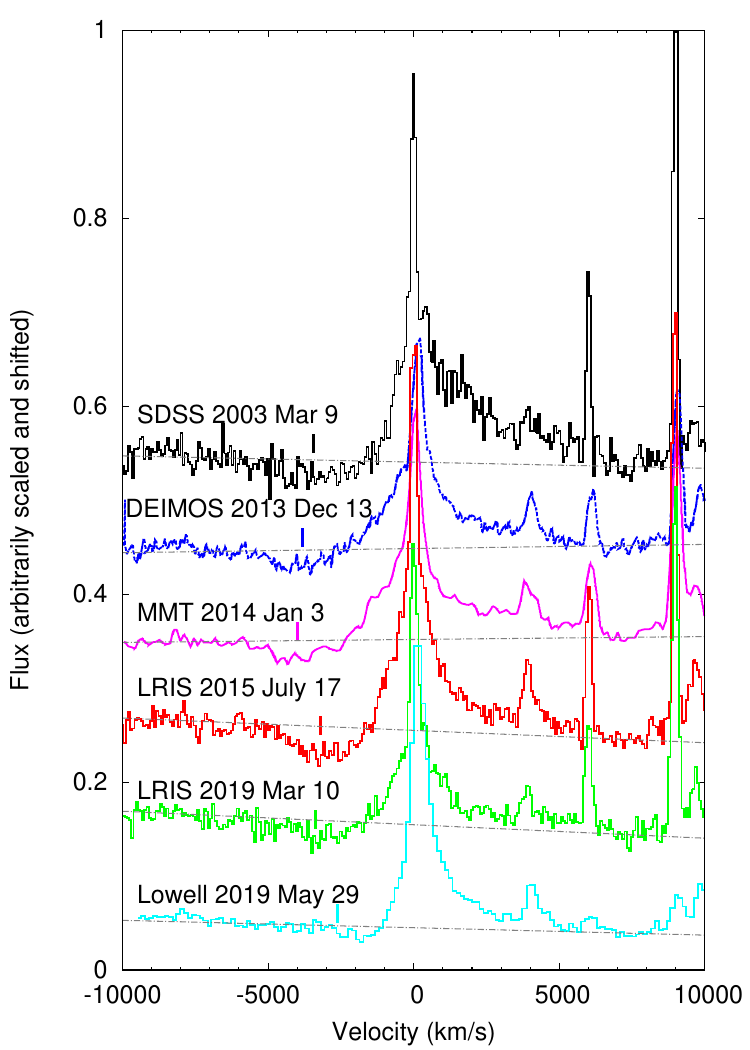}
}
 \caption{Long-term time-series of the H$\beta$ velocity spectra of SDSS1133.
 Keck/DEIMOS and MMT spectra are taken from Fig.~9 of \citet{koss14}, and Lowell Discovery Telescope/DeVeny spectrum is taken from Fig.~14 of \citet{war21}.
 The spectra are horizontally shifted so that the narrow H$\beta$ peaks correcpond to 0~km~s${}^{-1}$, and the flux is arbitrarily shifted and scaled for the purpose of clarity.
 Dotted lines indicate continuum levels estimated by linear regression for relatively iron-free spectral windows at $-7,000 \pm 200$~km~s${}^{-1}$ and  $+7,000 \pm 200$~km~s${}^{-1}$ (see Figure~\ref{fig:sdss_decomposition_velocity}).
 Persistent blue-shifted broad absorption is clearly seen, with the maximum absorption velocity at $ \sim -5,000$~km~s${}^{-1}$.
 The vertical bars indicate the flux-weighted mean absorption velocity measured using the flux values below the linear regression lines in the velocity range between $-5,000$~km~s${}^{-1}$ and $0$~km~s${}^{-1}$.
 }
 \label{fig:spectra}
\end{figure}

\begin{figure}
\center{
\includegraphics[clip, width=3.4in]{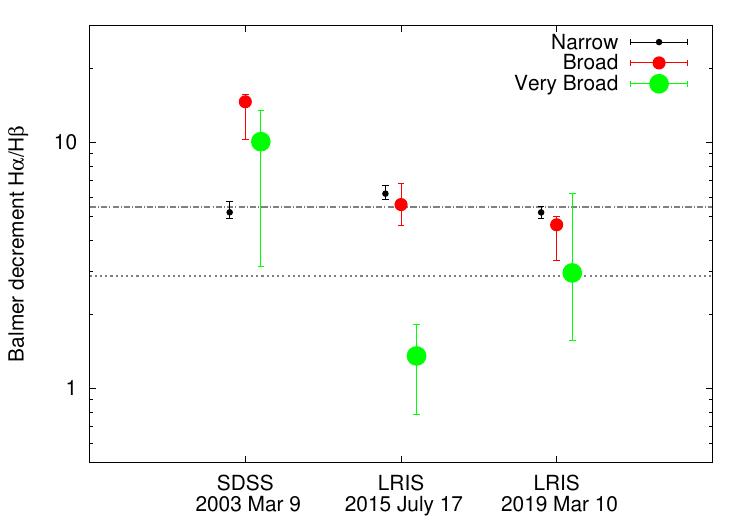}
}
 \caption{The Balmer decrement for the narrow, broad, and very broad emission line components of SDSS1133. The median Balmer decrement for the narrow component, H$\alpha$/H$\beta = 5.49$, is indicated by the dot-dashed line. Theoretically expected value of H$\alpha$/H$\beta = 2.86$ for Case B recombination H~II region with $T=10^{4}$~K and $n_{e}=10^{2}~$cm${}^{-3}$ is indicated by the dotted line.
 }
 \label{fig:balmer_decrement}
\end{figure}

\cite{koss14} present multi-epoch spectra of SDSS1133; the SDSS spectrum on 2003 March 9 (MJD=52707), Keck/DEIMOS spectrum on 2013 December 13 (MJD=56639), and the  Multiple Mirror Telescope (MMT) spectrum on 2014 January 3 (MJD=56660).
\cite{pur19b,pur19} conducted optical spectroscopic follow-up observations for the 2019 outburst event of SDSS1133 (refereed to as Gaia19bwn in their Fig.~2) with ALFOSC on Nordic Optical Telescope (NOT) on 2019 May 29 (MJD=58632), near the light curve maximum epoch (Figure~\ref{fig:ztf_lc_zoom}).
\cite{war21} present an optical spectrum taken on 2019 May 29 with the Lowell Discovery Telescope, the same date as the \cite{pur19b}'s observation.

Other than these already published data, there are unpublished Keck/LRIS spectroscopic data of SDSS1133 obtained on 2015 July 17 (MJD=57220.2; 560~s $\times$ 2) and 2019 March 10 (MJD=58552.5; 560~s) in the KOA.
The second Keck/LRIS was obtained just before the 2019 outburst event, probing the pre-outburst gradually rising phase (Figure~\ref{fig:ztf_lc_zoom}).
Below we describe the data reduction of the Keck/LRIS spectra (Section~\ref{sec:keck_lris}), and spectral decomposition analysis of the Keck/LRIS and SDSS spectra (Section~\ref{sec:spectral_decomposition}).
The comparisons of the multi-epoch spectra are presented in Section~\ref{sec:historical_spectra}.

\subsubsection{Keck/LRIS spectra (2015, 2019)}
\label{sec:keck_lris}

We downloaded the raw Keck/LRIS spectral data of SDSS1133, spectrophotometric star (HZ44), and associated calibration data from KOA (PI: Fiona Harrison, program ID: C280LA, C300).
All the LRIS data were obtained with a $1''.0$-width long-slit, 600/4000 (blue arm) and 400/8500 (red arm) grisms, and on-chip $2$~pixel binning in the slit direction was applied (the pixel scale is $0''.27$~pixel${}^{-1}$).
The observations were carried out under non-photometric (variable seeing and atmospheric extinction) conditions.
The PSF FWHM were evaluated as $\sim 8$~pixel ($2''.2$) on 2015 July 17 and $\sim 11$~pixel ($3''.0$) on 2019 March 10 from the standard star's spatial profiles on the dispersed images.

Standard {\tt IRAF}\footnote{IRAF is distributed by the National Optical Astronomy Observatory, which is operated by the Association of Universities for Research in Astronomy (AURA) under a cooperative agreement with the National Science Foundation.} data reduction was performed on the LRIS data, where overscan, split-illumination flat, wavelength calibration, and image distortion corrections were applied.
Pixels affected by cosmic-rays were identified and masked by using {\tt lacos\_spec} \citep{dok01}. 
The object spectra were extracted by using {\tt IRAF/apall}'s aperture extraction task with a 8~pixel ($2''.16$) aperture.
The host galaxy component was not subtracted, and telluric absorption correction was not applied.
The Galactic extinction was corrected by using \cite{fit99}'s extinction curve.
The relative flux offset between the blue arm and red arm spectra of each epoch was corrected by requiring the fluxes at around $\lambda_{\text{obs}}=5500$~\AA\  to be consistent with each other.
Then, the absolute fluxes of the spectra were roughly calibrated by scaling the spectra to match the $g$ band photometric magnitudes at the epochs of the spectroscopic observations, where the $g$ band photometric magnitudes were evaluated by linearly interpolating the Galactic extinction-corrected $g$ band light curve in Figure~\ref{fig:ztf_lc}.
The Galactic-extinction corrected $g$ band magnitudes were estimated as $19.95$ and $19.38$~mag on 2015 July 17 and 2019 March 10, respectively, which are respectively $1.39$ and $0.82$~mag fainter than the $g = 18.56$~mag estimated from the SDSS spectrum on 2003 March 9.

The Keck/LRIS spectra and publicly available SDSS spectrum of SDSS1133 are shown in Figure~\ref{fig:keck_spectra}.
We can see that overall spectral properties of SDSS1133 are unchanged during the periods of these spectroscopic observations.
The P-Cygni profile is detected in all of the hydrogen Balmer series and possibly in He~I $\lambda 5877$, with the velocity width of about a few 1000 km~s${}^{-1}$ \citep[see also][]{koss14,war21}.
Many narrow-to-intermediate width metal emission lines (e.g., Fe~II, Ca~II, and O~I) are clearly detected especially in the Keck/LRIS spectra.

%57220.249 19.9529  LBLR_OBJ_21554_chip3_bbcfcw.ms
%57220.256 19.9529  LBLR_OBJ_22197_chip3_bbcfcw.ms
%58552.515 19.3780  LBLR_OBJ_44548_chip3_bbcfcw.ms

By applying a supernova spectral classification software GELATO \citep[GEneric cLAssification TOol v.2.0;][]{harutyunyan08}\footnote{\href{https://gelato.tng.iac.es}{https://gelato.tng.iac.es}} to the SDSS and LRIS spectra, we find that SDSS1133 is reminiscent of interacting transients; the SDSS and LRIS 2015 spectra are consistently best-fitted by a spectrum of a pre-SN eruption of a type IIn SN~2009ip obtained on 2011 September 24 \citep[][]{pas13}, and the LRIS 2019 spectrum is best-fitted by a late-phase (+664.5~d) spectrum of a type IIn SN~1995G obtained on 1996 December 19 \citep{pas02}.
In Figure~\ref{fig:keck_spectra}, a VLT/X-Shooter optical spectrum of the 2011 event of SN~2009ip downloaded from the ESO Science Archive Facility Phase 3 (boxcar smoothed by 5~\AA\ and arbitrarily scaled) is plotted for comparison.
A more detailed analysis of the spectral properties is given below.

\subsubsection{Spectral decomposition}
\label{sec:spectral_decomposition}

\begin{table*}
\centering
\caption{Best-fit model parameters and 68\% percentile range derived from the spectral fitting.}
\label{tab:decomposition}
\begin{tabular}{lrrr}
\hline
Parameters & SDSS 2003 March 9 & LRIS 2015 July 17 & LRIS 2019 March 10 \\
\hline
\hline
\multicolumn{4}{|c|}{H$\beta$ region ($\lambda_{\text{obs}}=4400-5400$~\AA)} \\
\hline
$\lambda_{\text{H}\beta, \text{n,obs}} ({\text{\AA}})$                               & 	$4901.25_{-0.03}^{+0.03}$&$4899.72_{-0.03}^{+0.02}$&$4900.43_{-0.04}^{+0.04}$	\\
$\sigma_{\text{H}\beta, \text{n,rest}} (\text{km}~\text{s}^{-1})$                    & 	$91.90_{-1.93}^{+1.86}$&$121.10_{-1.92}^{+1.63}$&$146.41_{-3.33}^{+2.65}$	\\
$f_{\text{H}\beta, \text{n}} (10^{-17}~\text{erg}~\text{s}^{-1}~\text{cm}^{-2})$                    & 	$102.09_{-7.40}^{+7.04}$&$39.43_{-2.87}^{+1.98}$&$64.55_{-3.43}^{+3.98}$	\\
$\lambda_{\text{H}\beta, \text{b,obs}} ({\text{\AA}})$                               & 	$4901.16_{-0.73}^{+0.29}$&$4899.72_{-0.49}^{+0.41}$&$4902.23_{-0.81}^{+0.42}$	\\
$\sigma_{\text{H}\beta, \text{b,rest}} (\text{km}~\text{s}^{-1})$                    & 	$609.44_{-40.77}^{+58.97}$&$721.74_{-145.16}^{+102.82}$&$825.22_{-73.17}^{+181.56}$	\\
$f_{\text{H}\beta, \text{b}} (10^{-17}~\text{erg}~\text{s}^{-1}~\text{cm}^{-2})$                    & 	$311.25_{-18.50}^{+127.56}$&$116.47_{-21.47}^{+24.40}$&$139.13_{-10.76}^{+54.26}$	\\
$\lambda_{\text{H}\beta, \text{vb,obs}} ({\text{\AA}})$                              & 	$4912.29_{-26.02}^{+7.32}$&$4872.95_{-4.87}^{+3.30}$&$4887.42_{-20.95}^{+37.80}$	\\
$\sigma_{\text{H}\beta, \text{vb,rest}} (\text{km}~\text{s}^{-1})$                   & 	$2197.98_{-397.34}^{+867.36}$&$3410.30_{-585.59}^{+594.06}$&$3706.91_{-1092.25}^{+7030.83}$	\\
$f_{\text{H}\beta, \text{vb}} (10^{-17}~\text{erg}~\text{s}^{-1}~\text{cm}^{-2})$                   & 	$822.47_{-246.33}^{+1461.77}$&$325.40_{-79.46}^{+239.41}$&$174.84_{-89.74}^{+157.04}$	\\
$f_{\text{[OIII]4959}} (10^{-17}~\text{erg}~\text{s}^{-1}~\text{cm}^{-2})$                          & 	$84.92_{-4.60}^{+4.27}$&$25.05_{-1.01}^{+0.90}$&$36.18_{-2.39}^{+1.64}$	\\
$f_{\text{[OIII]5007}} (10^{-17}~\text{erg}~\text{s}^{-1}~\text{cm}^{-2})$                          & 	$242.94_{-6.06}^{+6.02}$&$68.74_{-1.20}^{+1.08}$&$110.44_{-3.00}^{+2.11}$	\\
$\alpha_{\lambda}$                                                                   & 	$-1.44_{-0.05}^{+0.05}$&$-1.61_{-0.04}^{+0.04}$&$-1.96_{-0.04}^{+0.03}$	\\
$f_{\lambda, 5100(1+z)}(10^{-17}~\text{erg}~\text{s}^{-1}~\text{cm}^{-2}~{\text{\AA}}^{-1})$        & 	$12.89_{-0.10}^{+0.10}$&$3.79_{-0.03}^{+0.02}$&$6.77_{-0.11}^{+0.03}$	\\
$\lambda_{\text{H}\beta, \text{abs,obs}} ({\text{\AA}})$                             & 	$4865.53_{-18.67}^{+2.45}$&$4853.99_{-1.68}^{+2.49}$&$4851.87_{-5.03}^{+1.80}$	\\
$\sigma_{\text{H}\beta, \text{abs,rest}} (\text{km}~\text{s}^{-1})$                  & 	$2671.89_{-637.50}^{+247.49}$&$1941.97_{-141.13}^{+234.57}$&$1697.22_{-755.87}^{+663.85}$	\\
$\tau_{\text{H}\beta, 0}$                                                            & 	$24.77_{-11.11}^{+66.36}$&$46.82_{-11.30}^{+30.75}$&$13.91_{-8.51}^{+15.58}$	\\
$\text{scale}_{\text{Fe}}$                                                           & 	$1.34_{-0.07}^{+0.07}$&$0.51_{-0.01}^{+0.02}$&$0.54_{-0.02}^{+0.05}$	\\
$z_{\text{Fe}}/z_{\text{H}\beta, \text{n}}$                                          & 	$1.06_{-0.01}^{+0.01}$&$1.06_{-0.01}^{+0.00}$&$1.07_{-0.01}^{+0.01}$	\\
\hline
\multicolumn{4}{|c|}{H$\alpha$ region ($\lambda_{\text{obs}}=6200-7100$~\AA)} \\
\hline
$\lambda_{\text{H}\alpha, \text{n,obs}} ({\text{\AA}})$                        & 	$6617.12_{-0.05}^{+0.06}$&$6610.89_{-0.03}^{+0.03}$&$6613.03_{-0.04}^{+0.03}$	\\
$\sigma_{\text{H}\alpha, \text{n,rest}} (\text{km}~\text{s}^{-1})$             & 	$120.27_{-3.22}^{+4.45}$&$224.89_{-2.49}^{+2.43}$&$201.89_{-2.25}^{+2.01}$	\\
$f_{\text{H}\alpha, \text{n}} (10^{-17}~\text{erg}~\text{s}^{-1}~\text{cm}^{-2})$             & 	$531.08_{-15.75}^{+35.25}$&$244.11_{-3.60}^{+3.51}$&$335.58_{-4.80}^{+4.45}$	\\
$\lambda_{\text{H}\alpha, \text{b,obs}} ({\text{\AA}})$                        & 	$6624.34_{-0.17}^{+1.67}$&$6612.87_{-0.07}^{+0.09}$&$6616.36_{-0.19}^{+0.25}$	\\
$\sigma_{\text{H}\alpha, \text{b,rest}} (\text{km}~\text{s}^{-1})$             & 	$961.50_{-18.10}^{+45.58}$&$978.14_{-13.63}^{+10.21}$&$1002.05_{-25.75}^{+18.32}$	\\
$f_{\text{H}\alpha, \text{b}} (10^{-17}~\text{erg}~\text{s}^{-1}~\text{cm}^{-2})$             & 	$4556.67_{-183.62}^{+161.09}$&$651.09_{-10.64}^{+6.54}$&$643.73_{-22.08}^{+15.43}$	\\
$\lambda_{\text{H}\alpha, \text{vb,obs}} ({\text{\AA}})$                       & 	$6619.45_{-1.65}^{+17.01}$&$6629.68_{-2.49}^{+1.34}$&$6626.35_{-4.39}^{+2.54}$	\\
$\sigma_{\text{H}\alpha, \text{vb,rest}} (\text{km}~\text{s}^{-1})$            & 	$3509.66_{-246.65}^{+46.31}$&$3617.45_{-48.69}^{+85.79}$&$4397.68_{-141.26}^{+172.28}$	\\
$f_{\text{H}\alpha, \text{vb}} (10^{-17}~\text{erg}~\text{s}^{-1}~\text{cm}^{-2})$            & 	$8289.15_{-2552.35}^{+315.46}$&$441.12_{-10.12}^{+20.19}$&$515.36_{-17.64}^{+35.50}$	\\
$f_{\text{[NII]6548}} (10^{-17}~\text{erg}~\text{s}^{-1}~\text{cm}^{-2})$                     &	$13.48_{-6.18}^{+12.92}$&$0.00_{-0.00}^{+0.21}$&$3.96_{-2.32}^{+3.16}$	\\
$f_{\text{[NII]6584}} (10^{-17}~\text{erg}~\text{s}^{-1}~\text{cm}^{-2})$                     &	$0.18_{-0.16}^{+26.74}$&$0.00_{-0.00}^{+0.01}$&$8.39_{-2.05}^{+2.32}$	\\
$f_{\text{[SII]6717}} (10^{-17}~\text{erg}~\text{s}^{-1}~\text{cm}^{-2})$                     &	$97.78_{-4.20}^{+3.22}$&$21.99_{-0.71}^{+0.68}$&$38.64_{-1.28}^{+1.34}$	\\
$f_{\text{[SII]6731}} (10^{-17}~\text{erg}~\text{s}^{-1}~\text{cm}^{-2})$                     &	$68.79_{-3.92}^{+2.86}$&$16.68_{-0.66}^{+0.64}$&$27.28_{-1.18}^{+1.21}$	\\
$f_{\text{[OI]6300}} (10^{-17}~\text{erg}~\text{s}^{-1}~\text{cm}^{-2})$                      &	$31.86_{-3.30}^{+2.85}$&$9.71_{-0.67}^{+0.66}$&$16.61_{-1.29}^{+1.25}$	\\
$\alpha_{\lambda}$                                                             & 	$-0.99_{-0.20}^{+0.07}$&$-1.23_{-0.04}^{+0.06}$&$-1.37_{-0.06}^{+0.06}$	\\
$f_{\lambda, 5100(1+z)} (10^{-17}~\text{erg}~\text{s}^{-1}~\text{cm}^{-2}~\text{\AA}^{-1})$   & 	$12.82_{-0.24}^{+0.82}$&$3.84_{-0.07}^{+0.04}$&$6.69_{-0.11}^{+0.10}$	\\
$\lambda_{\text{H}\alpha, \text{abs,obs}} ({\text{\AA}})$                      & 	$6559.69_{-1.60}^{+11.80}$&$6535.14_{-0.56}^{+0.60}$&$6538.05_{-1.20}^{+1.01}$	\\
$\sigma_{\text{H}\alpha, \text{abs,rest}} (\text{km}~\text{s}^{-1})$           & 	$2274.90_{-1110.01}^{+100.42}$&$1079.75_{-39.16}^{+78.32}$&$1239.13_{-55.37}^{+103.47}$	\\
$\tau_{\text{H}\alpha, 0}$                                                     & 	$70.37_{-67.64}^{+9.46}$&$23.45_{-1.17}^{+2.67}$&$21.65_{-1.41}^{+3.10}$	\\
$\text{scale}_{\text{Fe}}$                                                     & 	$0.57_{-0.47}^{+0.15}$&$0.10_{-0.03}^{+0.04}$&$0.10_{-0.06}^{+0.06}$	\\
$z_{\text{Fe}}/z_{\text{H}\alpha, \text{n}}$                                   & 	$0.97_{-0.06}^{+0.11}$&$1.09_{-0.00}^{+0.03}$&$1.05_{-0.01}^{+0.03}$	\\
\hline
\end{tabular}
%\bigskip
   \begin{tablenotes}
     \item [1] The spectral fitting was performed at the H$\beta$ and H$\alpha$ spectral ranges independently. The Galactic extinction is corrected, but not the host galaxy extinction. $\lambda_{\text{obs}}$, $\sigma_{\text{rest}}$, and $f$ indicate the observed-frame central wavelength, rest-frame velocity standard deviation, and integrated flux of the narrow (n), broad (b), and very broad (vb) Gaussian line components. The narrow lines are only marginally spectrally-resolved. The continuum is assumed to have a power-law form of $f_{\lambda} =  f_{\lambda, 5100(1+z)}[\lambda/((1+z)5100\text{\AA})]^{\alpha_{\lambda}}$, and the P-Cygni absorption strength is modelled as $e^{-\tau_{\lambda}}$ where $\tau_{\lambda} = \tau_{0} (2\pi \sigma_{\text{abs}, \text{obs}}^2)^{-1/2} e^{-((\lambda-\lambda_{\text{abs}, \text{obs}})/\sigma_{\text{abs}, \text{obs}})^2}$.
     $\text{scale}_{\text{Fe}}$ and $z_{\text{Fe}}/{z_{\text{H}}}$ indicate the scaling factor to the \cite{bor92} Fe template spectrum and relative redshift of the Fe template to the hydrogen Balmer narrow line.
     The details of the fitted spectral model is described in Section~\ref{sec:spectral_decomposition}. 
   \end{tablenotes}
\end{table*}

The most remarkable feature of the optical spectra of SDSS1133 is the broad Balmer P-Cygni profile.
The spectral decomposition including the broad P-Cygni absorption feature is not performed by \cite{koss14}, thus we conducted spectral decomposition analysis of the H$\beta$ and H$\alpha$ P-Cygni profile for the SDSS and Keck/LRIS spectra.

Each spectrum at the wavelength range of $\lambda_{\text{obs}}=4400-5400$~\AA\ was fitted with a power-law continuum, broad and very broad Gaussian H$\beta$ lines, and narrow Gaussian H$\beta$ and OIII emission lines, where the power-law continuum and broad and very broad H$\beta$ were absorbed by a factor of $e^{-\tau_{\lambda}}$ where $\tau_{\lambda}$ was modelled as a Gaussian.
Since the optical spectra of SDSS1133 clearly show Fe~II line forest \citep{koss14}, we also included a Fe~II pseudo-continuum template spectrum in the model.
Since no template spectrum for SNe/stellar objects is available in the literature, we adopted the Fe emission template spectrum of \cite{bor92} which is created from a spectrum of the narrow line Seyfert 1 galaxy PG0050+124 (I Zw 1).
The Fe emission template spectrum has the intrinsic FWHM of I Zw 1 of 900~km~s${}^{-1}$, and we directly used the Fe~II spectrum without further Gaussian convolution.
The Fe template spectrum has a flux density of $1.38 \times 10^{-15}$~erg~s${}^{-1}$~\AA${}^{-1}$ at $\lambda_{\text{rest}}=5000$~\AA, and the flux scaling factor to the Fe template spectrum and its redshift are fitting parameters.

The same model fitting was also performed at the wavelength range of $\lambda_{\text{obs}}=6200-7100$~\AA\, where the [NII], [SII], and [OI] narrow emission lines were included in the model.
%The Galactic extinction was corrected assuming \cite{fit99}'s extinction curve.
The narrow emission lines were assumed to have the same redshift and line velocity width.

$\chi^2$ minimization model fitting was performed by using the Levenberg-Marquardt algorithm implemented as {\tt optimize.least\_squares} in {\tt SciPy}.
The fitting was performed in the observed-frame.
Uncertainties on the best-fitting parameters were evaluated by 10,000 trials of Monte Carlo resampling, where 10,000 mock spectra were generated by adding Gaussian flux noise to the original spectrum using the calculated standard deviations of the flux density.

The fitting parameters are tabulated in Table~\ref{tab:decomposition}, and the best-fitting model for the SDSS spectrum is shown in Figure~\ref{fig:sdss_decomposition}.
The narrow lines are only marginally spectrally-resolved, thus the narrow velocity widths reflect the instrumental broadening.
The narrow emission line redshift is measured to be $z = 0.008$, being consistent with the redshift of Mrk~177.
The Fe~II emission template spectrum, though its velocity width does not match that of SDSS1133, well explains the pseudo-continuum bumps observed at the redward and blueward spectral regions of the H$\beta$ emission line, while the Fe~II emission at the H$\alpha$ wavelength range is very weak and hard to be seen.
The very broad component is not well constrained especially at the H$\beta$ wavelength range, but is required to explain the extended wing of the H$\alpha$ profile.

The asymmetric P-Cygni-like Balmer line profile can largely be explained by the combination of the variable broad blue-shifted absorption on the very broad symmetric Balmer line, and the broad blue-shifted absorption also account for the continuum absorption observed at the blueward spectral region of the Balmer lines.
%The broad absorption feature has a similar velocity width with the broad H$\beta$ emission line, suggesting that the broad line emission and absorption are originated from the same region.
The best-fitting models for the SDSS and LRIS spectra are shown in Figure~\ref{fig:sdss_decomposition_velocity} as a function of rest-frame velocity from $-10,000$~km~s${}^{-1}$ to 10,000~km~s${}^{-1}$ relative to the narrow Balmer emission lines.
The best-fitting absorption model suggests that the absorption peaks at $\sim -2,000$~km~s${}^{-1}$ and the highest velocity component is $\sim -5,000$~km~s${}^{-1}$.

\subsubsection{Time series of the H$\beta$ line profile}
\label{sec:historical_spectra}

Figure~\ref{fig:spectra} summarises the multi-epoch H$\beta$ spectra of SDSS1133 described at the beginning of Section~\ref{sec:specdata}.
The $g$ band magnitudes at the epochs of the 
SDSS, 
DEIMOS, 
MMT, 
LRIS 2015,
LRIS 2019, and 
Lowell 
observations are evaluated by linearly interpolating the $g$ band light curve in Figure~\ref{fig:ztf_lc};
$g=18.66$ (SDSS 2003),
19.65 (DEIMOS 2013), 
19.65 (MMT 2014), 
19.95 (LRIS 2015), 
19.38 (LRIS 2019), and 
17.36~mag (Lowell 2019), 
respectively.

Figure~\ref{fig:spectra} reveals that the P-Cygni absorption features of SDSS1133 have been persistent at least for 16 years since 2003, even during the outburst phase in 2019 at the epoch of the Lowell observation when SDSS1133 was $\sim$ 2 mag brighter than the non-outbursting phase (Figure~\ref{fig:ztf_lc_zoom}).
The absorption minimum varies from $-2000$ to $-4000$~km~s${}^{-1}$, while the maximum velocity component is roughly kept constant at $\sim -5,000$~km~s${}^{-1}$.
The persistent P-Cygni absorption feature suggests that the high velocity ejecta producing the P-Cygni absorption is not due to an one-off transient event but is due to continuous or multiple ejections (see Section~\ref{sec:broad_pcygni} for details).

\subsubsection{Balmer decrement and dust extinction in the host galaxy}
\label{sec:balmer_decrement}

The dust extinction toward SDSS1133 may be inferred by using the Balmer decrement, specifically the flux ratio of H$\alpha$ to H$\beta$.
Figure~\ref{fig:balmer_decrement} presents the H$\alpha$/H$\beta$ Balmer decrement for the narrow, broad, and very broad emission line components of SDSS1133.
While the Balmer decrements of the broad and very broad components are not well constrained due to the line profile modelling uncertainties, the Balmer decrements of the narrow component constrain H$\alpha$/H$\beta$ to be $\sim 5-6$.
By combining the three measurements of narrow Balmer decrements, the median and 68\% percentile range is H$\alpha$/H$\beta = 5.49_{-0.51}^{+0.80}$.

The theoretically expected value of the Balmer decrement is 2.86 for Case B recombination H~II region with $T=10^{4}$~K and $n_{e}=10^{2}~$cm${}^{-3}$ \citep[e.g.,][]{ost89,dom13}.
This means that the median value of the relative extinction at the H$\beta$ and H$\alpha$ wavelengths is 
$A_{\text{H}\beta} - A_{\text{H}\alpha} = 2.5\log_{10}(5.49/2.86) = 0.71$~mag.
Assuming \cite{pei92}'s Small Magellanic Cloud (SMC)-like, Large Magellanic Cloud (LMC)-like, and Milky Way (MW)-like dust extinction curves ($R_{V}=A_{V}/E(B-V) = 2.93$, $3.16$, and $3.08$, respectively), the Balmer decrement of the narrow component in SDSS1133 corresponds to
\begin{eqnarray}
E(B-V)_{\text{host}} &=& 0.63_{-0.09}^{+0.13}~\text{(SMC)}, \nonumber \\
& & 0.61_{-0.09}^{+0.13}~\text{(LMC)}, \nonumber \\
& & 0.65_{-0.10}^{+0.14}~\text{(MW)},
\label{eqn:eb_v_balmer_decrement}
\end{eqnarray}
in the unit of magnitude, respectively (Figure~\ref{fig:balmer_decrement}).
Here $E(B-V)_{\text{host}} \equiv A_{B, {\text{host}}} - A_{V, {\text{host}}}$ indicates the color excess due to the dust extinction in the SDSS1133's host environment.
The hydrogen column density corresponding to the derived value of $E(B-V)_{\text{host}}$ is $N_{H, \text{host}}$ = $( 2.84_{-0.42}^{+0.60}$, $1.46_{-0.22}^{+0.31}$, $0.31_{-0.05}^{+0.07})$ $\times 10^{22}$~cm${}^{-2}$ for the SMC-like, LMC-like, and MW-like dust-to-gas ratios, respectively \citep{pei92}.
The inferred large dust extinction ($A_{V}=R_{V}E(B-V)=1.8-2$~mag) may be attributed to dust grains either in the CSM like $\eta$ Carinae's dusty Homunculus nebula \citep[e.g.,][]{smi12,morris17}, or interstellar materials located along our line of sight towards SDSS1133.

%For comparison, the visual extinction toward $\eta$ Carinae is estimated to be $A_{V} = 1.4$~mag \citep[e.g.,][]{smi11b}, which is comparable to the above estimate for SDSS1133.

\if0
lmc 2.93
Narrow BD= 5.49 - 0.51 + 0.8
Narrow EB_V= 0.628 - 0.094 + 0.132
Narrow  A_V= 1.839 - 0.275 + 0.385
Narrow  N_H[10^22]= 2.84 - 0.42 + 0.6
Narrow  A_5100= 1.965 - 0.293 + 0.412
lmc 3.16
Narrow BD= 5.49 - 0.51 + 0.8
Narrow EB_V= 0.608 - 0.091 + 0.127
Narrow  A_V= 1.922 - 0.287 + 0.403
Narrow  N_H[10^22]= 1.46 - 0.22 + 0.31
Narrow  A_5100= 2.107 - 0.315 + 0.442
mw 3.08
Narrow BD= 5.49 - 0.51 + 0.8
Narrow EB_V= 0.65 - 0.097 + 0.136
Narrow  A_V= 2.003 - 0.299 + 0.42
Narrow  N_H[10^22]= 0.31 - 0.05 + 0.07
Narrow  A_5100= 2.141 - 0.32 + 0.449
\fi

$E(B-V)_{\text{host}}$ derived above assumes that the hydrogen Balmer line-emitting region is in the Case B recombination condition.
The intrinsic Balmer decrements can be much larger than 2.86 when the collisional excitation becomes important and/or the line-emitting region is optically-thick to the hydrogen Balmer lines \citep[e.g.,][]{ost89,kok19}.
In SDSS1133, the narrow Balmer lines may possibly have multiple origins, and the Balmer decrement measurements can be affected by the flux contaminations from the dense circumstellar photo-ionizing region where the intrinsic Balmer decrements are larger than 2.86.

Also, the assumptions of the dust extinction curves and $R_{V}$ directly affect the estimates of the dust extinction.
Since the SDSS1133's host galaxy Mrk~177 is a low-mass, low-metallicity galaxy (Section~\ref{sec:hostgalaxy}), the SMC-like dust extinction with $R_{V} = 2.93$ may provide a good approximation to the interstellar dust extinction in Mrk~177.
However, we know little about the size distribution, dust composition, and dust-to-gas ratio of the circumstellar dust grains possibly surrounding SDSS1133.
For example, the intense radiation from SDSS1133 may preferentially remove small dust grains and result in a flat extinction curve \citep[e.g.,][]{taz20}.
Detailed discussion about the properties of the circumstellar dust grain of the SDSS1133 is beyond the scope of this paper, and will be discussed elsewhere.

\section{Discussion}
\label{sec:discussion}

\begin{table*}
\centering
\caption{Source properties discussed in this work and possible scenarios}
\label{tbl:scenario}
\begin{tabular}{lccc}
\hline
Source Property & LBV & recoiling AGN & Sections \\
\hline
\hline
Offset from the galactic center       & YES     & Unusual & \ref{sec:hostgalaxy} \\
Large-amplitude UV-optical outbursts  & YES     & Unusual & \ref{sec:outbursts} \\
Narrow line variability?              & YES     & NO & \ref{sec:metal_lines} \\
Narrow line BPT diagnostics           & YES     & Unusual & \ref{sec:metal_lines} \\
Ca~II emission lines                  & YES     & Unusual & \ref{sec:ca_lines} \\
Broad P-Cygni feature                 & YES(Unusual) & NO & \ref{sec:broad_pcygni}, \ref{sec:broadlines} \\
UV-optical luminosity                 & YES(Unusual) & YES &  \ref{sec:BBmodel_fitting}, \ref{sec:hrd} \\
Small X-ray/optical luminosity ratio  & Unusual & Unusual & \ref{sec:xray_emission}, \ref{sec:multiwavelength_sed} \\
\hline
\end{tabular}
%\bigskip
   \begin{tablenotes}
     \item [1] YES indicates that the observational feature can be readily explained by the scenario. NO indicates it cannot be explained. Unusual indicates it can be explained but it is unusual of observed systems. The last column indicates the section numbers where the corresponding source properties are discussed.
   \end{tablenotes}
\end{table*}

In this Section we discuss the nature of SDSS1133 inferred from the multi-epoch, multi-band photometric and spectroscopic measurements described in the previous Section.
As mentioned in Section~\ref{sec:intro}, SDSS1133 is suggested to be either an recoiling (offset) AGN or extragalactic outbursting LBV star (=SN impostor).
The multiple outbursts and persistent emission rule out the possibility of an one-off transient event.

Following \cite{koss14} and \cite{bur20}, we summarize the major observational properties of SDSS1133 discussed in this section and evaluate each scenario in Table~\ref{tbl:scenario}.
As summarized in Table~\ref{tbl:scenario}, narrow line BPT diagnostics, detection of the narrow line variability, presence of the broad P-Cygni absorption features in the hydrogen Balmer lines, and small X-ray/optical luminosity ratio observed in SDSS1133 all suggest that SDSS1133 is presumably an extragalactic LBV instead of a recoiling AGN.
Although the high X-ray/UV/optical luminosity and high-velocity P-Cygni absorption features are unusual for a LBV, we point out several well-known giant eruption LBVs (e.g., $\eta$~Carinae and SN~2009ip) share the similar properties with SDSS1133.
We conclude that SDSS1133 belongs to an extreme population of LBVs experiencing multiple non-terminal explosions and subsequent strong interactions of the ejected shell with different shells and/or CSM.
Details of these source properties and comparisons with other giant eruption LBVs and SN impostors are discussed below.

\subsection{Host galaxy properties}
\label{sec:hostgalaxy}

The SDSS1133's host galaxy Mrk~177 is a dwarf galaxy with the $g$ band absolute magnitude of $M_{g} = -16.6$~mag, which is comparable to the SMC.
The total stellar mass of Mrk~177 is estimated to be $M_{*} \sim 10^{8.55}~M_{\odot}$ \citep{koss14}.
The disturbed morphology of Mrk~177 indicates that Mrk~177 is a post-merger galaxy \citep[see][for details]{koss14}.
The gas-phase oxygen abundance is estimated to be $12+\log(\text{O/H})=8.58$ from the SDSS optical spectroscopy of the Mrk~177's nucleus.
The low metallicity galaxies are preferred sites for massive star formation, thus the low metallicity in Mrk~177 may be consistent with the idea that the progenitor of SDSS1133 is a massive star like other LBVs.

Under the LBV scenario, SDSS1133 is a massive star and the local environment of SDSS1133 should be an active star-forming region.
The local environment of SDSS1133 cannot be examined by current observations, and further high spatial resolution observations for the emission lines and dust emission around SDSS1133 are needed to judge the validity of the LBV scenario based on the host galaxy's local environment.

\cite{koss14} estimate a star-formation rate of Mrk~177 as $0.05$~$M_{\odot}$~yr${}^{-1}$ using $GALEX$ UV photometry.
$WISE$ magnitudes of Mrk~177 are 13.25, 13.20, 10.03, and 7.68~Vega mag at W1, W2, W3, and W4 band, respectively, and the WISE colors W1$-$W2=0.05~Vega mag and W2$-$W3=3.17~Vega mag are outside of empirical AGN selection windows \citep{hai16}.
The SDSS optical spectrum of the nucleus of Mrk~177 exhibits star-forming galaxy-like emission line ratio of [NII]/H$\alpha$ and [OIII]/H$\beta$, indicating that Mrk~177 does not harbor a strong AGN at the galactic center \citep{koss14,tob14}.
Also, the weak X-ray luminosity of $L_{0.3-10~\text{keV}} \sim 3 \times 10^{38}$~erg~s${}^{-1}$ at the nucleus of Mrk~177 observed by {\it Chandra} (Section~\ref{sec:chandra}) is well below the empirical threshold luminosity to identify AGNs ($\sim 3 \times 10^{42}~\text{erg}~\text{s}^{-1}$), and can be explained by integrated luminosities from X-ray binary populations in the galaxy without the need for AGN \citep[e.g.,][]{leh16,bir20,sch21}.
The recoiling AGN scenario requires that SDSS1133 is a merged SMBH that was ejected from the center of Mrk~177 at the time of the SMBHB coalescence, and it requires that the Mrk~177's nucleus currently possesses no SMBH.
The absence of an AGN in the nucleus of Mrk~177 is not inconsistent with the recoiling AGN scenario, though the observations do not exclude the possibility of the presence of an inactive SMBH in Mrk~177.

The narrow line redshift of SDSS1133 is consistent with that of Mrk~177's nucleus, indicating that the line-of-sight velocity of SDSS1133 relative to the systemic velocity is currently much less than 100~km~s${}^{-1}$ (Section~\ref{sec:spectral_decomposition}).
Then it is safe to assume that the projected velocity is also smaller than 100~km~s${}^{-1}$.
Assuming the recoiling AGN scenario for SDSS1133, in order to explain the current projected galactocentric distance of 0.81~kpc in terms of the recoil kick from the galactic center, the required life-time of the AGN activity of SDSS1133 is at least $0.81~\text{kpc}/100~\text{km}~\text{s}{}^{-1} \sim 10^{7}~\text{yrs}$.  
This time scale of $10^{7}~\text{yrs}$ is too long for a recoiling AGN to maintain the AGN activity \citep[e.g.,][]{ble08,war21}, and we conclude that the observed small velocity of SDSS1133 relative to the systemic velocity does not fit the recoiling AGN scenario.

\subsection{Light curve evolution}
\label{sec:longterm}

\begin{figure*}
\center{
\includegraphics[clip, width=6.8in]{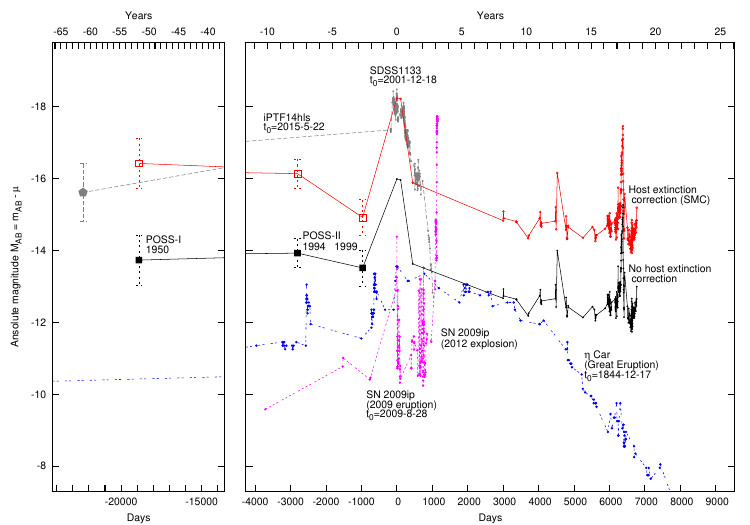}
}
 \caption{
 The historical synthetic $g$-band magnitude light curve of SDSS1133 from 1950 to 2020.
 The horizontal axis is the observed-frame days since the peak epoch ($t_{0}$ = 2001 December 18). 
 The upper light curve (open symbols) is corrected for the host galaxy extinction assuming SMC-like extinction (Section~\ref{sec:balmer_decrement}). 
 Square symbols indicate the POSS-I (in 1950) and POSS-II (in 1994 and 1999) photographic plate photometries, and circle symbols indicate the synthetic $g$ band light curve $M_{g}$ calculated from the $g$, $r$, and $i$-band measurements with the assumption of $g-r=0.29$~mag and $g-i=0.18$~mag. The long-term linear decline rates between 1994 and 2020 are 0.058~mag~yr${}^{-1}$ and 0.051~mag~yr${}^{-1}$ for host extinction uncorrected and corrected light curves, respectively (Section~\ref{sec:longterm_trend}). 
 The $V$ band light curve for the $\eta$ Carinae's Great Eruption (blue circle symbol; peaks in $t_{0}$ = 1844 December 17), pseudo $R$ band light curve of pre-SN eruptions (peaks on $t_{0}$ = 2009 August 28) and subsequent terminal SN explosion of SN~2009ip in September 2012 (magenta circle symbol), and $g$ band light curve of iPTF14hls (gray circle and pentagon symbols; peaks on $t_{0}$ = 2015 May 22) are shown for comparison.}
 \label{fig:historic_lightcurve}
\end{figure*}

\subsubsection{Long-term declining trend}
\label{sec:longterm_trend}

Figure~\ref{fig:historic_lightcurve} shows the 70~years worth of absolute magnitude light curve of SDSS1133, which include the POSS-I and POSS-II photographic plate measurements in 1950, 1994, and 1999 (see Section~\ref{sec:koss_photometry}).
Except for the POSS data, Figure~\ref{fig:historic_lightcurve} shows $g$ band absolute magnitudes of SDSS1133; in addition to the $g$ band measurements between 2001 and 2021, $r$ and $i$ band measurements between 2010 and 2018 are used to calculate the synthetic $g$ band magnitudes assuming constant colors of $g-r=0.29$~mag and $g-i=0.18$~mag from the ZTF data in 2018 (Figure~\ref{fig:ztf_lc_zoom}).
Two light curves are shown; one is uncorrected and the other is corrected for the possible extinction in the host galaxy estimated from the Balmer decrement (Section~\ref{sec:balmer_decrement}).

SDSS1133 has been a bright variable object at least since 1950, as first noticed by \cite{koss14}.
In addition to the outbursts, SDSS1133 also exhibits small amplitude flux variations even in the non-outbursting phase.
After 1994, SDSS1133 light curve probably shows a long-term declining trend since 1994.
We calculated the long-term linear trend by fitting a straight line for the data points in Figure~\ref{fig:historic_lightcurve} between 1994 April 14 and 2020 June 29 (49456 $\leq$ MJD $<$ 59030), excluding the outbursting epochs between 
2001 December 18 and 2002 April 1 (52261 $\leq$ MJD $<$ 52366),
2014 March 16 and 2014 May 10 (56732 $\leq$ MJD $<$ 56788), and
2018 October 31 and 2019 July 22 (58422 $\leq$ MJD $<$ 58687).
This linear declining model provides a reasonable fitting to the long-term light curve in the non-outbursting phase, and we obtained the best-fitting decline rates of 0.058~mag~yr${}^{-1}$ and 0.051~mag~yr${}^{-1}$ for host extinction uncorrected and corrected light curves, respectively.

As reference models, consider simple SN light curve models.
Late-phase SN luminosity can be powered either by radioactive decays in cobalt or SN ejecta-CSM shock interactions.
The former light curve is approximately an exponential form of $L(t) \propto e^{-t/\tau_{\text{Co}}}$ where $\tau_\text{Co}= 111.3$~days is the decay time of ${}^{56}$Co, and the latter can be approximately a power-law form of $L(t) \propto t^{-3/(n-2)}$ (assuming a steady stellar wind CSM) where $n=7-12$ depends on the radial density structure of the ejecta \citep[$\rho_{\text{ej}} \propto r^{-n}$; e.g.,][]{mor13}.
In either case, a single SN exploded somewhere in the 20th century cannot explain the very slow decline trend of the long-term light curve of SDSS1133.
For the same reasoning, we can conclude that the non-outbursting phase light curve of SDSS1133 cannot be attributed to a tidal disruption of a star by a SMBH \citep[$L(t) \propto t^{-5/3}$;][]{phi89b}, or in general  to any other known one-off explosive events.

AGNs are known to show decades-long optical variability at amplitudes of a few 0.1~mag \cite[e.g.,][]{mac12}, thus the long-term light curve in the non-outbursting phase of SDSS1133 alone cannot exclude the possibility of the AGN origin.
However, the co-existence of the decades-long trend and short time-scale large-amplitude outbursts is rarely seen in AGNs.
As shown below, some giant eruption LBVs are observed to show long-lasting light curves with sporadic outbursts, and the observed light variations of SDSS1133 are not inconsistent with these LBVs.

\subsubsection{Outbursts}
\label{sec:outbursts}

As shown in Figures~\ref{fig:ztf_lc} and \ref{fig:historic_lightcurve}, we find that SDSS1133 experienced at least four optical outbursts since 2001; these four outbursts are referred to as 2001 outburst, 2014 outburst, 2019 outburst, and 2021 outburst, respectively.

SDSS1133 was in a very bright outbursting phase during the two SDSS imaging observations on 2001 December 18 and 2002 April 1 \citep{koss14}. In both observations (separated by 104 days), the brightness level was similar ($g = 16.3$~mag).
As pointed out by \cite{koss14}, SDSS1133 was not detected on 2MASS images on 2000 January 10 (MJD=51553), suggesting that the first brightening event happened sometime between 2000 January 10 (2MASS non-detection) and 2001 December 18 (SDSS detection).
Also, the SDSS spectrum taken on 2003 March 9 (MJD=52707) provides a rough estimate of the SDSS1133 brightness at this epoch; $g = 18.70 \pm 0.18$, $r=18.28 \pm 0.06$, $i=18.65 \pm 0.05$, and $z=18.49 \pm 0.08$~mag \citep{koss14}, which are $\sim 2$~mag fainter than the that during the 2001 outburst.
Therefore, the duration of the 2001 outburst is constrained to be less than $52707 - 51553 = 1154$~days ($3.2$~yrs) and longer than 104~days ($0.3$~yrs).
The absolute $g$-band magnitude peaks at $M_{g} = -16.30$~mag ($M_{g} = -18.26$~mag after correcting for the possible host galaxy extinction), thus the 2001 outburst event can be regarded as a SN impostor \citep{pas13,pas19,per20}.

The PTF and PS1 photometry reveals that SDSS1133 experienced a brief optical outburst between 2014 March 8 (MJD=56738.5; PTF $R=18.99$~mag) and 2014 July 8 (MJD=56846.3; PS1 $y=18.65$~mag).
The observed peak magnitude is PS1 $i=18.12 \pm 0.01$~mag on 2014 May 10 (MJD=56787.3).
The duration of the 2014 outburst is roughly constrained to be $56846.3 - 56738.5 = 108$~days ($0.3$~yrs), suggesting that this was shorter and fainter than the 2001 outburst.
This event is also noted by \cite{war21}, who report the small flare detection detected by unpublished data from the Catalina Real-time Transient Survey (CRTS) between 2014 April 28 and 2014 June 5.

%56738.4549 18.9870 0.0500 PTF R
%56740.5256 18.9527 0.0115 PS1 r
%56754.3968 19.6829 0.0242 PS1 i
%56787.2822 18.1170 0.0050 PS1 i
%56835.2491 18.5579 0.0573 PS1 y
%56846.2512 18.6480 0.0700 PS1 y

The ZTF $g$, $r$, and $i$ band observations provide the densely-sampled multi-band light curves for the 2019 outburst (Figure~\ref{fig:ztf_lc_zoom}).
The $g$ and $r$ band light curves peak on 2019 June 5 (MJD=58639.2) at $g=17.03$ and $r=16.81$~mag, about 3~mag brighter than the non-outbursting phase.
The absolute $g$-band magnitude at the peak is $M_{g} =-15.27$~mag ($M_{g} =-17.46$~mag after correcting for the possible host galaxy extinction), thus the 2019 outburst should be regarded as another SN impostor event in SDSS1133 after the 2001 outburst.
We can see from Figure~\ref{fig:ztf_lc_zoom} that the duration of the entire 2019 outburst is about $\sim 100$~days \citep[see also][]{war21}, thus the 2019 outburst is at least $\sim 0.6-0.7$~mag fainter and is probably shorter-lived compared to the 2001 outburst.

The {\it Gaia} photometry in 2021 reveals the fourth outburst event occurred between 2021 April 25 (MJD=59329.0; $G=19.07$~mag) and 2021 August 14 (MJD=59440.3; $G=20.22$~mag).
The observed peak magnitude is $G=17.87$~mag on 2021 June 21 (MJD=59386.3).
The duration of the $2021$ outburst is roughly constrained to be less than $59440.3 - 59329.0 = 111$~days ($0.3$~yrs), which is similar to the 2014 and 2019 outbursts.
We can say that the three outbursts in 2014, 2019, and 2021 had similar observational properties, and they were fainter and had shorter durations compared to the 2001 outburst.

The latter three outbursts have similar observational characteristics, but the outburst in 2001 was relatively brighter and had a longer duration.

Among the four outbursts, only the 2019 outburst is temporally resolved thanks to the high cadence ZTF observations.
The outburst light curve is more complex than a simple rise and fall, exhibiting small bumps on the outburst profile (see Figure~\ref{fig:ztf_lc_zoom}).
Following the definition in \cite{per20}, the rise time and fade time of the 2019 outburst are calculated as the time durations between the peak and the point where the light curve drops to 0.75~mag below peak (equivalent to half the peak flux); by using the linearly-interpolated ZTF $g$-band light curve, we obtain $t_{\text{rise}} = 23.9$~days and $t_{\text{fade}} = 3.6$~days, and the duration is $t_{\text{rise}} + t_{\text{fade}} = 27.5$~days.
The long rise and short fall light curve behavior of the 2019 outburst is very different from normal SNe or Galactic variable stars \citep[see Fig.~3 of][]{per20}.

The multi-band optical light curves during the 2019 outburst (and also the 2001 outburst) do not shows significant color evolution, and there is little difference in the color between the outbursting and non-outbursting phases \citep[Figure~\ref{fig:ztf_lc_zoom}; see also][]{war21}.
This suggests that the effective temperature of SDSS1133 is kept constant (see Section~\ref{sec:sed} for further details), and that the outburst event cannot be explained by an adiabatically-cooling expanding ejecta as observed in SNe.

The long-term and outburst light curve behaviors of SDSS1133 are suggested to be consistent with LBV outbursts \citep{koss14,war21}, but the constant temperature may indicate that the outbursts are due not to normal LBV eruptions.
As pointed out by \cite{hum99}, LBV eruptions can largely be classified into two subclasses; S~Doradus (S~Dor)-type ``normal'' eruptions and $\eta$ Carinae analog giant eruptions \citep[see also][and references therein]{smi11,smi17}.
The optical outbursts in the S~Dor-type LBV eruptions are associated with the decrease of the temperature so that the bolometric luminosity is unchanged while the photospheric radius increases \citep[Figure~\ref{fig:hrd};][and references therein]{wol89,vin11,smi11,pas13,smi14,smi17,kil18}, which is not the case for SDSS1133's outbursts.
Instead, we suggest that the flux variations in SDSS1133 are due to the intrinsic increase in the bolometric luminosity as observed in some of $\eta$~Carinae analog giant eruption LBVs = SN impostor, e.g., pre-SN outbursts of SN~2009ip \citep[e.g.,][and references therein]{hum99,smi11,pas13,dav20,wei20}.
We will return to this point in Section~\ref{sec:sed}.

As mentioned in Section~\ref{sec:intro}, the large variability amplitude of $\sim$2 mag may be achieved by a rare population of extremely variable AGNs.
Further quantitative evaluation of the light curve properties is needed to discuss the differences/similarities with the AGN variability, which is beyond the scope of this paper, but at least we can say that multiple short-duration outbursts observed in SDSS1133 are likely to be unusual for an AGN.

\subsubsection{Comparisons with other LBV giant eruptions}
\label{sec:comparison_with_other_lbvs}

In Figure~\ref{fig:historic_lightcurve}, the absolute $g$-band magnitude light curve of SDSS1133 is compared with the historical $V$ band light curve of the 19th century Great Eruption of $\eta$ Carinae's compiled by \cite{fre04} and \cite{smi11b}.
The original light curve is shifted by 0.02~mag to match the AB magnitude system, and the absolute magnitude is calculated by assuming a distance to $\eta$ Carinae as 2.3~kpc ($\mu = 11.81$~mag), and $A_{V}=1.4$~mag \citep{dav97,smi06,smi11b}.
The absolute $V$-band magnitude of $\eta$ Carinae had been in a range between $-9$ and $-11$~mag before 1822 and reached $\sim -14$~mag during the Great Eruption during $1822-1864$ (peaked at $-14.2$~mag in 1845).
The short durations of the 2019 outburst of SDSS1133 is comparable to $\eta$ Carinae's brief precursor eruptions in 1838 and 1843 just before the final brightening in 1844 December \citep[see][]{smi11b}.
Figure~\ref{fig:historic_lightcurve} suggests that SDSS1133 remains to be brighter than the $\eta$ Carinae's Great Eruption over the period since 1950.
Since the total mass-loss from $\eta$ Carinae during the Great Eruption is estimated to be about $\sim 10$~$M_{\odot}$ \citep{smi11b}, the mass-loss from SDSS1133 should be much larger than $10$~$M_{\odot}$ if we assume the mass-loss rate ($\propto$ kinetic energy) is roughly in proportion to the radiation energy.
Also this may possibly mean that the SDSS1133's current mass is more massive than $\eta$ Carinae, i.e., $>100~M_{\odot}$ \citep[e.g.,][]{smi08}.

We also show a $R$ band absolute magnitude light curve of SN~2009ip \citep{smi10,mau13,pas13} in Figure~\ref{fig:historic_lightcurve}.
It is constructed from $R$ band, F606W band, and unfiltered data between 1999 June 29 and 2009 August 28 in \cite{smi10} and the $R$ band data between 2009 August 30 and 2012 October 16 in \cite{pas13}.
The absolute magnitude is calculated by assuming a distance modulus to SN~2009ip as $\mu = 31.55$~mag, and $A_{R}=0.05$~mag \citep{mau13,pas13}, and the light curve is shifted by 0.21~mag to match the AB magnitude system.
SN~2009ip is believed to have exploded as a terminal SN (namely SN~IIn) in 2012 September \citep[but see][]{pas13}, and before that there were multiple precursor eruptions.
The first identified eruption was in 2009 August-September whose peak was on 2009 August 28 \citep[MJD=55071.75; $m_{R} = 17$~mag;][]{smi10}.
The second eruption was observed in 2011 May-October, and then in 2012 there were two large eruptions \citep[2012a event from August 8 to September 24, and 2012b event from September 24 to December;][]{smi10,pas13,margutti14}.
The 2012b event is considered to be the terminal SN explosion of SN~2009ip, and the pre-SN eruptions (observed as SN impostor) are suggested to be originated from the LBV giant eruptions of the progenitor star \citep{mau13,smi14b,graham14}.
The peak absolute magnitude of SN~2009ip during the eruptions (2009 and 2012a events) is similar to that of $\eta$ Carinae's giant eruptions, and thus very close to the brightness of SDSS1133 (Figure~\ref{fig:historic_lightcurve}).

Another interesting object to be compared with SDSS1133 is iPTF14hls, which is a hydrogen-rich optical transient at $z= 0.0344$ ($d_{L} = 156$~Mpc), exhibiting multiple peaks before and after the maximum epoch, and lasting for 2 years \citep{arc17,yal19,sol19}.
The spectrum shows broad hydrogen Balmer P-Cygni absorption features at $-6000$~km~s${}^{-1}$ but no signs of narrow P-Cygni features \citep{arc17}, similar to SDSS1133.
Although the absolute magnitude reaches over $-18$~mag, the peculiar light curve shape of iPTF14hls is inconsistent with normal SNe.
There is a suggestion that iPTF14hls is not a terminal SN explosion; specifically, \cite{mor20} suggest that iPTF14hls is a non-terminal eruption event like $\eta$ Carinae's Great Eruption.
\cite{arc17} report that iPTF14hls was detected on the POSS-I image on 1954 February 23 at $20.4 \pm 0.8$~mag and not detected on the POSS-II image on 1993 January 2, suggesting that iPTF14hls experienced an optical outburst in 1954.
Figure~\ref{fig:historic_lightcurve} shows the absolute $g$ band magnitude light curve of iPTF14hls, where the Galactic extinction coefficient is assumed to be $A_{g} = 0.05$~mag.
The brightness of iPTF14hls is consistent with SDSS1133 if the large host galaxy extinction of $E(B-V)_{\text{host}} = 0.628$ (SMC-like; Equation~\ref{eqn:eb_v_balmer_decrement}) of SDSS1133 is taken at its face value.
The time scale of the light curve evolution of iPTF14hls is similar to the 2001 outburst of SDSS1133.

Figure~\ref{fig:historic_lightcurve} shows that there exists a vast 
diversity in the light curve properties of the LBV giant eruptions (and giant eruption candidates), and at least we can say that the long-term and outburst properties of SDSS1133 are within the diversity in the LBV giant eruptions.

\subsubsection{Periodicity}
\label{sec:periodicity}

The similar brightness and variability time scales between SDSS1133 and other giant eruption LBVs suggest a common physical processes behind these objects.
The best studied object among the giant eruption LBVs is $\eta$ Carinae, which is suggested to be a binary system composed of a $\sim 100~M_{\odot}$ LBV star and O-type main sequence star, with a orbital period of 5.5~yrs \citep{dam96,par09,smi11b}.
The giant eruptions of $\eta$ Carinae are suggested to have occurred at the times of pericenter passage of the companion star \citep[][]{smi11b}.

The brightness and durations of each of the outbursts in SDSS1133 are different with each other (Section~\ref{sec:outbursts}), which may imply that the outbursts in SDSS1133 cannot be explained by the simple binary model.
Nevertheless, if SDSS1133 is a binary system, the binary orbital period may be inferred from the recurrence timescale of the observed outbursts.

We observed four large outbursts from SDSS1133 in 2001, 2014, 2019, and 2021; specifically, the observed peak epochs are 2001 December 18 (SDSS, MJD=52261.4), 2014 May 10 (PS1 $i$ band, MJD=56787.3), 2019 June 5 (ZTF, MJD=58639.2), and 2021 June 21 ({\it Gaia}, MJD=59386.3), respectively.
The observations before 2014 were scarce, and we may have missed other outbursts between 2003 and 2014.
The time interval between the 2001 and 2014 outbursts is $\Delta t_{2001-2014, \text{obs}} = 4525.9$~days, 2014 and 2019 outbursts is $\Delta t_{2014-2019, \text{obs}} = 1851.9$~days, and 2019 and 2021 outbursts is $\Delta t_{2019-2021, \text{obs}} = 747.1$~days.
Interestingly, the outburst interval looks decreasing as a function of time, which may be related to an orbital evolution due to significant mass-loss episodes, if the binary scenario is true for SDSS1133 \citep[e.g.,][]{smi11b,smi11c}.
Alternatively, the decreasing outburst interval could be related to the increasing pair-instability toward the end of the stellar evolution \citep[e.g.,][]{yos16,woo17}.

The linear regression relationship that predicts the time of the next outburst ($\text{MJD}_{i}$) from the time of the previous outburst ($\text{MJD}_{i-1}$) can be expressed by a linear recurrence relation as
\begin{equation}
\text{MJD}_{i} - \text{MJD}_{i-1} =  -0.59 \times (\text{MJD}_{i-1} - 60000) - 55.36.
\label{eqn:pred_outburst}
\end{equation}
Based on the well-determined observed peak of the 2019 outburst ($\text{MJD} = 58639.2$), Equation~\ref{eqn:pred_outburst} predicts outburst epochs as $\text{MJD}_{\text{pred}} = 52369.2$ (in 2002), $56816.0$ (in 2014), 59386.7 (in 2021), which are close to the observed outburst epochs.
Equation~\ref{eqn:pred_outburst} also predicts multiple outbursts in 2022 and converges to $\text{MJD}_{\text{pred}} = 59906.2$ (2022 November 23).
Further intensive monitoring observations are definitely needed to examine the possible temporal pattern.

\if0

python
import numpy as np
from sklearn.linear_model import LinearRegression
import matplotlib.pyplot as plt

MJD    = np.array( [ 52261.4, 56787.3, 58639.2, 59386.3 ] )
dMJD   = ( MJD[1:]+MJD[:-1] )/2.0
delta = np.diff( MJD )

model = LinearRegression().fit(dMJD.reshape(-1, 1), delta)
print("intercept:", model.intercept_)
print("slope:", model.coef_)

model = LinearRegression().fit(MJD[1:].reshape(-1, 1), delta)
print("intercept:", model.intercept_)
print("slope:", model.coef_)

model = LinearRegression().fit(MJD[0:-1].reshape(-1, 1), delta)
print("intercept:", model.intercept_)
print("slope:", model.coef_)
print("intercept - slope*60000:", model.intercept_ + model.coef_*60000)

plt.plot(  MJD[0:-1], delta, "."  )
plt.plot(  MJD[0:-1], delta, "-"  )

\fi

\subsection{Spectroscopic features}
\label{sec:spec_properties}

\begin{figure*}
\center{
\includegraphics[clip, width=6.8in]{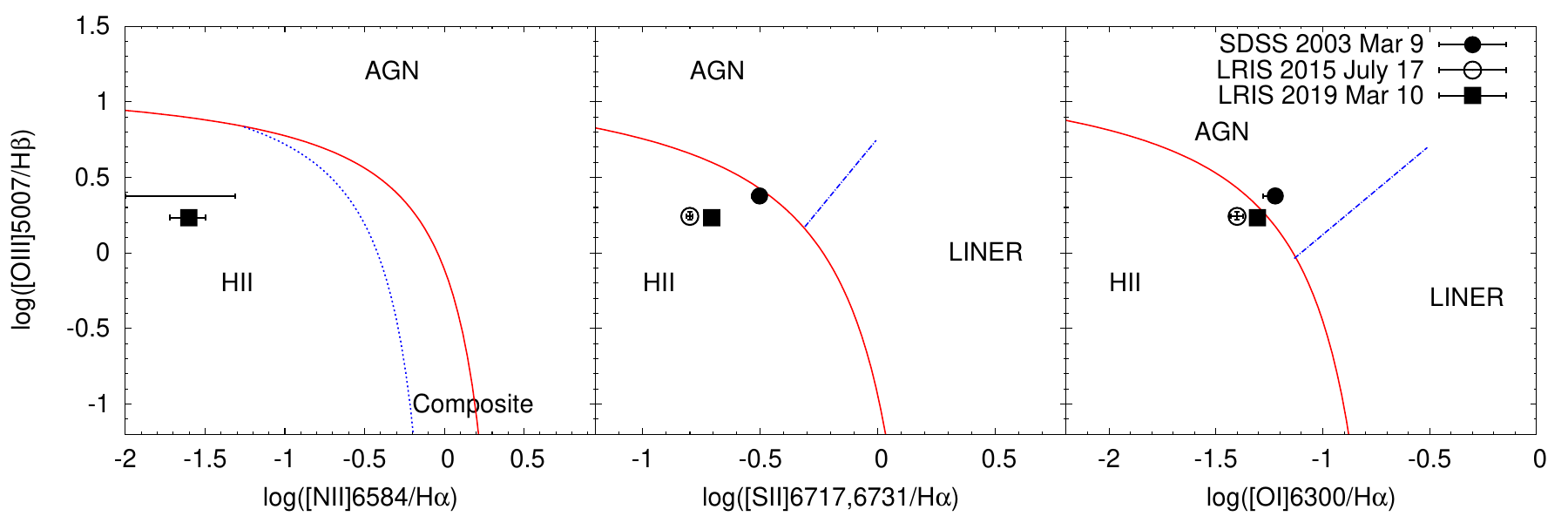}
}
 \caption{BPT classification diagrams for the narrow emission line ratios of SDSS1133.
 The SDSS and two Keck/LRIS measurements are indicated by filled circles, open circles, and filled squares, respectively. The extreme starburst lines (solid lines; Kewley et al. 2001), pure star-formation line (dotted line; Kauffmann et al. 2003), and  Seyfert-LINER (low-ionization nuclear emission-line region) classification lines (dot-dashed lines; Kewley et al. 2006) are shown. 
 }
 \label{fig:bpt_diagram}
\end{figure*}

\begin{figure}
\center{
\includegraphics[clip, width=3.4in]{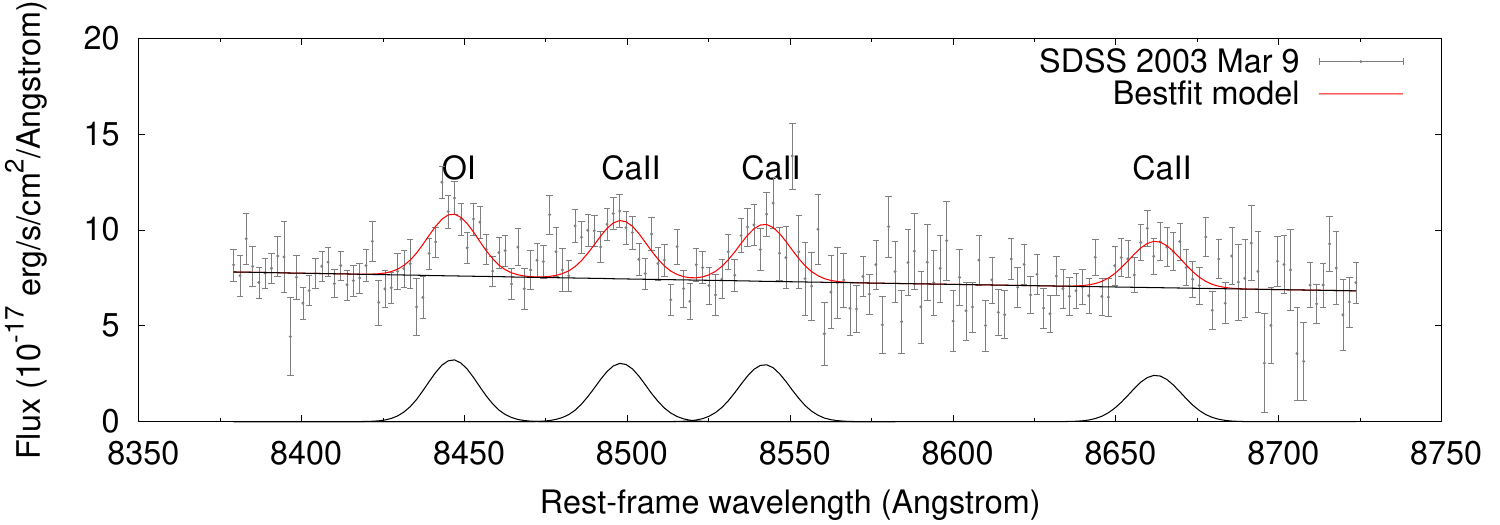}
\includegraphics[clip, width=3.4in]{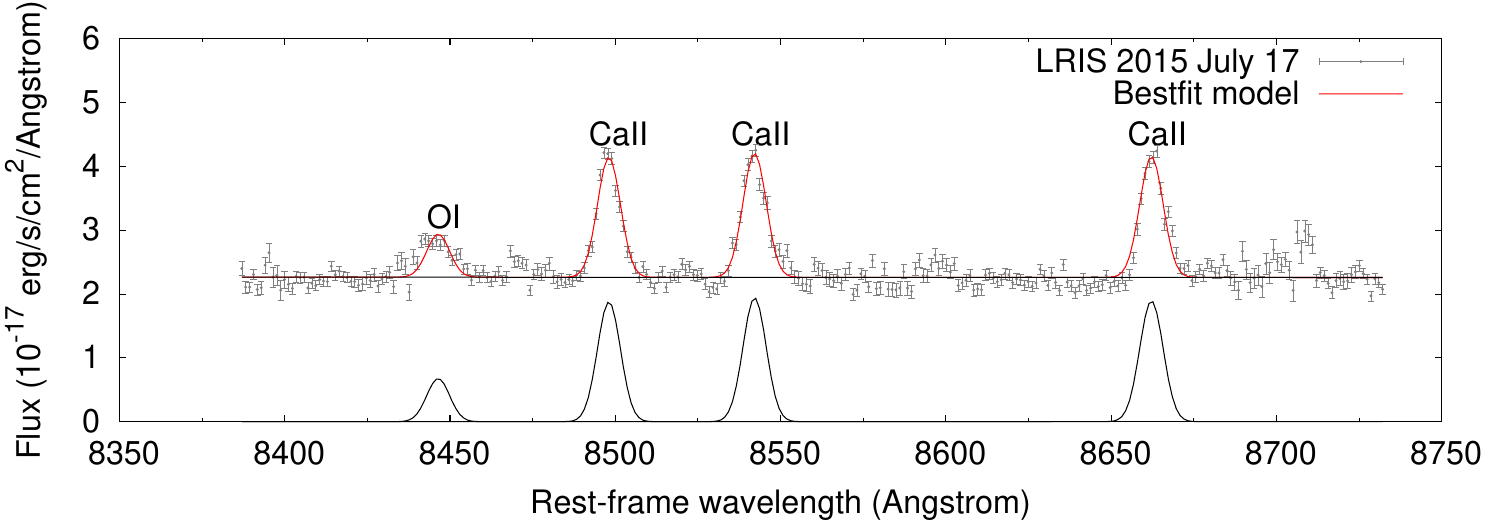}
\includegraphics[clip, width=3.4in]{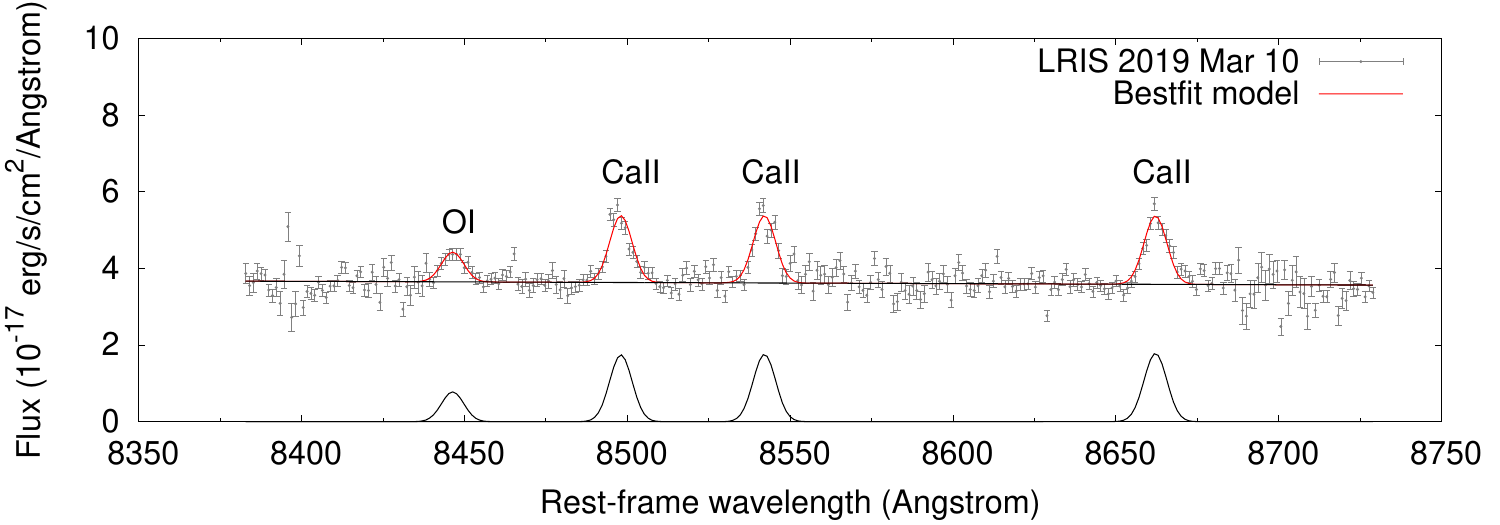}
}
 \caption{The SDSS and Keck spectra at the Ca~II IR $\lambda \lambda 8498, 8542, 8662$ triplet spectral region. A power-law + Gaussian emission line model is fitted.
 }
 \label{fig:catriplet}
\end{figure}

\subsubsection{Narrow emission lines and BPT diagnostic diagrams}
\label{sec:metal_lines}

The results of the spectral decomposition analysis in Table~\ref{tab:decomposition} indicate that the low-ionization nebular narrow lines ([OIII],[NII], and [SII]) seem to display flux variations over the spectroscopic periods \cite[see also][]{pur19b}.
The narrow [SII] doublet ratio [SII]6717/[SII]6731 is consistent at $\sim 1.4$, indicating that the electron density of the narrow emission line emission region is consistent with typical electron densities for local galaxies, $n_{e} = 10-100$~cm${}^{-3}$ \citep{ost89}.
Figure~\ref{fig:bpt_diagram} presents the Baldwin-Phillips-Terlevich (BPT) narrow line diagnostic diagrams for SDSS1133 \citep{bal81,kew01,kew06,kau03}.
The [NII] narrow emission lines of SDSS1133 are much weaker compared to the H$\alpha$ narrow line in SDSS1133, while the narrow line ratio of $\log$[OIII]5007/H$\beta$ is relatively high compared to typical star-forming galaxies.
Also, the [OI]6300/H$\alpha$ and [SII]6717,6731/H$\alpha$ narrow line ratios are much less than 1.

AGN narrow line flux variability is rarely observed since the narrow line region in AGNs is extended over $\sim$~kpc scales, thus it can be taken as evidence suggesting that SDSS1133 is not an AGN.
In terms of the LBV scenario, the narrow line variability may imply that these lines in SDSS1133 are originated from circumstellar materials in close vicinity of the central hot star.
But we should note that careful spectroscopic analyses (especially host galaxy subtraction and slit-loss corrections) are needed to examine the significance of the narrow line flux variability of SDSS1133, which is beyond the scope of this paper.

According to the BPT diagrams, the narrow line ratios of SDSS1133 are consistent with low-mass star-forming galaxies, and suggest that the narrow emission line region of SDSS1133 has a high ionization parameter and low-metallicity \citep[e.g.,][]{kew13,simmonds16}.
The BPT diagnostics for SDSS1133 do not support the standard AGN ionization and thus the recoiling AGN scenario, as already pointed out by \cite{koss14} and \cite{simmonds16}.
However, we should note that, since the narrow emission lines are not spectrally-resolved, they are probably originated from multiple regions, such as the host galaxy H~II region and photo-ionized circumstellar region.
If the AGN narrow emission lines are unusually weak, the emission lines from the star-forming region can dominate the observed emission. Actually, a fraction of X-ray bright AGNs is observed to exhibit BPT line ratios similar to SDSS1133 \citep[e.g.,][]{smith14,ago19}.
In that case, the interpretation of the BPT diagnostics is ambiguous.

% foiii5007 = [ 242.94, 68.74, 110.44 ] 
% fhbeta    = [ 102.09, 39.43, 64.55  ]
% foiii5007 = np.array(foiii5007)
% fhbeta = np.array(fhbeta)
% foiii5007/fhbeta
% array([2.379665  , 1.74334263, 1.71092177])
% np.log10(foiii5007/fhbeta)
% array([0.37651582, 0.24138275, 0.23323015])
%
% np.log10(0.18 / 531.08)
% -3.4698874414802305
%>>> 
% np.log10(8.39/335.58)
% -1.6020341088831915
%>>> 

\subsubsection{Calcium and iron emission lines}
\label{sec:ca_lines}

As shown in Figure~\ref{fig:sdss_decomposition}, the optical spectra of SDSS1133 are heavily contaminated by iron emission line forests.
In addition to the iron emission lines, another interesting spectroscopic feature of SDSS1133 is intermediate-width or narrow calcium emission lines clearly detected in the optical spectra, as already reported by \cite{koss14} and \cite{war21}.
In the LRIS spectra, [Fe~II]$\lambda$7155, [Ca~II]$\lambda \lambda 7291, 7324$, O~I$\lambda 8446$ and Ca~II IR triplet $\lambda \lambda 8498, 8542, 8662$ are clearly detected, while adjacent Paschen lines (e.g., P14 at 8598\AA) are much weaker.
The Ca~II H+K emission lines at 3968 and 3933~\AA\ are not detected.
The [Fe~II]$\lambda$7155 and [Ca~II]$\lambda \lambda 7291, 7324$ forbidden lines are unobserved in the SDSS spectrum while they are clearly detected in the LRIS spectra \citep[see Figure~10 of][for the Fe~II, Ca~II, and O~I line profile fitting to the SDSS and Keck/DEIMOS spectra of SDSS1133]{koss14}, probably implying that the electron density of circumstellar materials exhibits time-evolution around $n_{e} \sim 10^{7}~\text{cm}^{-3}$ \citep[e.g.,][]{hum13}.

Figure~\ref{fig:catriplet} shows a power-law + Gaussian emission line model fitting to the SDSS and LRIS spectra at the Ca triplet spectral region.
The rest-frame velocity widths are 386.45, 172.67, and 170.43~km~s${}^{-1}$ for the SDSS and two LRIS spectra (the later two are spectrally-unresolved), respectively, suggesting that the line widths of these lines are variable.
The Ca~II triplet line ratios are roughly 1:1:1 rather than the theoretically predicted value of 1:9:5, suggesting that the line-forming region is optically-thick \citep[e.g.,][]{app88,ban21}.
The resemblance between the Ca~II and Fe~II line widths in the SDSS spectrum implies that Ca~II and Fe~II emission lines are originated from the same, high density photo-ionization region.
The broader line widths of Ca~II and Fe~II compared to the other H, N, S, and O nebular narrow lines in the SDSS spectrum (Table~\ref{tab:decomposition}) suggest that the Ca~II and Fe~II emission region is separated from the low density, low velocity dispersion nebular responsible for the nebular narrow lines, and is located in closer vicinity to SDSS1133.
The narrow Ca~II lines are occasionally seen in stellar explosion events and LBVs \citep[e.g.,][]{smi10,fol11,war21}, while they are rarely seen in AGNs \citep[though not exclusively;][]{per88,marinello16}.
The presence of the temporally variable intermediate-width Ca~II lines makes SDSS1133 even more unusual for an AGN \citep[e.g.,][]{koss14,war21}.

\subsubsection{The broad P-Cygni absorption feature due to high velocity ejecta}
\label{sec:broad_pcygni}

As shown in Section~\ref{sec:spectral_decomposition}, the hydrogen Balmer lines in SDSS1133 exhibit multi-component line profiles, composed of an unresolved narrow component, FWHM $= 1,000 - 2,000$~km~s${}^{-1}$ broad component, and FWHM $\sim 4,000$~km~s${}^{-1}$ very broad component. Moreover, the broad and very broad components exhibit blue-shifted broad P-Cygni absorption feature.
The spectral changes in the broad line profile can be reproduced by the model where large variations in the absorption component are allowed.
As shown in Figure~\ref{fig:spectra}, the spectroscopic properties of the H$\beta$ and H$\alpha$ profiles of SDSS1133 remain essentially unchanged during the epochs of the spectroscopic observations since 2003.
The P-Cygni absorption feature is clearly detected with the absorption peak at $\sim -2,000$~km~s${}^{-1}$, and the absorption feature continuously extends from $\sim 0$~km~s${}^{-1}$ up to $\sim -5,000$~km~s${}^{-1}$ (Figure~\ref{fig:sdss_decomposition_velocity}).

The optical spectra of SDSS1133 resemble those of type II SNe \citep{pur19b}, but the long-duration light curve and multiple outbursts observed in SDSS1133 rule out the possibility that SDSS1133 is an one-off event, and are in favor of the LBV scenario \citep[see also][]{war21}.
Under the assumption that SDSS1133 is a LBV star, the persistent broad absorption feature implies that the ejected materials are moving at a high velocity of $\sim -2,000$~km~s${}^{-1}$ (up to $-5,000$~km~s${}^{-1}$).
Observations for giant eruption LBVs (namely $\eta$ Carinae and SN~2009ip's pre-SN eruptions) have revealed that at least a fraction of outflowing materials ejected during the LBV giant eruptions can have a velocity higher than $5,000$~km~s${}^{-1}$ \citep[see Figure~\ref{fig:keck_spectra}; e.g.,][]{smi08,fol11,pas13,mau13}.
Moreover, the time variations of the absorption minimum observed in SDSS1133 (Figure~\ref{fig:spectra}) resemble those observed in the SN~2009ip's pre-SN eruptions \citep[see Figure~8 of][]{pas13}.
The persistent broad P-Cygni absorption seen in SDSS1133 suggests that the mass-loss in SDSS1133 is one of the most energetic among the known examples of the LBV giant eruptions \citep{pas13}.
Such high velocity mass ejections (moving faster than the star's escape velocity) are hard to be explained solely by the radiation-driven winds of the star, and they imply the existence of non-terminal explosions in the stellar core inducing outward blast waves into the envelope, which are observed like scaled-down SNe \citep[e.g.,][]{smi08,pas13,smi13,smi14,smi17,smi14b,smi18,str21}; see Section~\ref{sec:broadlines}.

About 10\% of broad line AGNs are known to show blue-shifted broad absorption line (BAL) features in the UV wavelengths due to line absorption by high velocity disk winds \citep[e.g.,][]{hal02}.
However, only a few examples of hydrogen Balmer BAL AGNs are known to date, because the disk winds are generally too low-density and too highly-ionized to keep neutral hydrogen atoms populated in the $n=2$ shell \citep[e.g.,][]{hall07,zha15,kok17,bur21}.
Considering the intrinsic rarity of the Balmer BAL AGNs, it is unreasonable to assume that the P-Cygni features of SDSS1133 (selected as a rare offset AGN) are due to the Balmer BAL nature.
We should note that the discovery of SDSS1133 suggests that a fraction of Balmer BAL AGN candidates known to date \citep[e.g., SDSS~J1025 in][]{bur21} can actually be SDSS1133-like LBVs.

\subsubsection{Interpretations of the broad emission line component}
\label{sec:broadlines}

Most LBVs generally exhibit narrow hydrogen Balmer emission lines from the stellar winds with the velocity of a few $100$~km~s${}^{-1}$, reflecting the low escape speed of blue supergiants stars \citep[][]{smi14,dav20}.
In contrast to the normal LBVs, the line width of the broad emission lines in SDSS1133 is much broader than 1,000~km~s${}^{-1}$; the FWHM of the broad component is $\text{FWHM}_{\text{b}} = 2\sqrt{2\ln2} \sigma_{\text{b}} \sim 1,500-2,000$~km~s${}^{-1}$, and very broad component is $\text{FWHM}_{\text{vb}} = 5,000-10,000$~km~s${}^{-1}$, which are comparable to the line widths observed in SNe.

As with the broad P-Cygni absorption, the broad Balmer emission lines of SDSS1133 are in common with the pre-SN outbursts of SN~2009ip \citep[e.g.,][]{smi08,fol11,pas13,mau13,war21}.
The broad P-Cygni absorption feature implies the presence of the high velocity outflowing materials (Section~\ref{sec:broad_pcygni}), and the broad emission lines are presumably emitted from the same high velocity materials.
As already mentioned in Section~\ref{sec:broad_pcygni}, although the relatively narrow lines ($\lesssim 1,000$~km~s${}^{-1}$) observed in typical LBV eruptions can be attributed to stellar winds \citep[e.g.,][]{smi14,dav20}, the broad lines of $> 1,000$~km~s${}^{-1}$ observed in SDSS1133 and pre-SN outbursts of SN~2009ip are too broad to be explained by the same wind mechanism \citep[e.g.,][]{smi08,pas13,smi14}, suggesting that the broad line-emitting fast materials are originated from the non-terminal explosions \citep{smi08,smi13,smi18,fra22}.

It is known that SNe~IIn exhibit broad/very broad emission lines and absorption features originated from ejecta-CSM interaction regions, where the ejecta kinematics and electron scattering broaden the emission lines up to $\sim$ 10,000~km~s${}^{-1}$ \citep[e.g.,][]{chu01,kok19}.
The blue optical continuum and broad and narrow emission lines produced by the ejecta-CSM interactions in SNe~IIn are known to closely resemble those of type~1 AGNs \citep{fil89,baldassare16}, as observed in SDSS1133.
We can expect that a similar mechanism can produce broad emission and absorption features in the case of mass-loss ejecta-CSM interactions or collisions between shells of matter ejected from a LBV \citep[e.g.,][]{woosley07,des09,pas10,yos16,woo17}.
\cite{smi13} suggests a model for the 19th century eruption of $\eta$ Carinae where a strong wind blowing for 30~yrs and subsequent non-terminal $\sim 10^{50}$~erg explosion occurring in 1844 result in ejecta-CSM interactions like a scaled-down SN~IIn \citep[see also][]{smi08,smi18}.
If SDSS1133 is an event similar to these $\eta$ Carinae analog giant eruptions, the persistent blue UV-optical continuum and broad emission lines from SDSS1133 can naturally be explained by these non-terminal explosions and subsequent shock interactions.
The shock interactions also produce X-ray photons \citep{che82,che06,des09,dwa12,smi13}, and X-ray emission escaped from the shock region can naturally explain the observed X-ray luminosity of SDSS1133 (see Section~\ref{sec:multiwavelength_sed}).
We will discuss the energetics of this non-terminal explosion scenario in detail in Sections~\ref{sec:emission_mechanism}.

SDSS1133 shows no signs of narrow P-Cygni absorption features.
SNe~IIn with strong ejecta-CSM interactions typically show the narrow P-Cygni absorption, which is attributed to unshocked pre-SN stellar wind components \citep[e.g.,][]{tad13}.
But observationally, some SNe~IIn lack narrow P-Cygni absorption features \citep[e.g.,][]{kok19}, which is probably due to viewing angle effects caused by asphericity in the CSM matter distribution.
Therefore, under the scaled-down SN~IIn scenario, the absence of narrow P-Cygni absorption may imply the aspherical CSM interactions in SDSS1133.

\subsection{Spectral Energy Distribution}
\label{sec:sed}

\begin{figure}
\center{
\includegraphics[clip, width=3.4in]{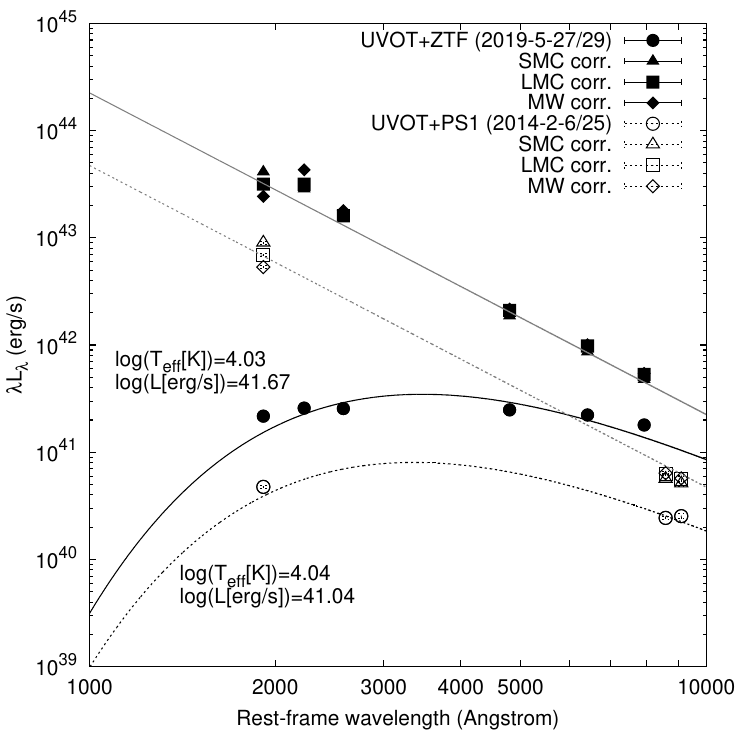}
}
 \caption{UV-optical SEDs of SDSS1133 on 2014 February 6-25 (non-outbursting phase) and 2019 May 27-29 (outbursting phase). The best-fitting black body spectra are overplotted. The host galaxy extinction-corrected SEDs assuming the extinction curves determined by the Balmer decrement (Figure~\ref{fig:balmer_decrement}) are also shown; the extinction-corrected SEDs are consistent with Rayleigh-Jeans power-law ($\lambda L_{\lambda} \propto \lambda^{-3}$), for which the black body fitting was not available.
 }
 \label{fig:BBfit}
\end{figure}

\begin{figure}
\center{
\includegraphics[clip, width=3.4in]{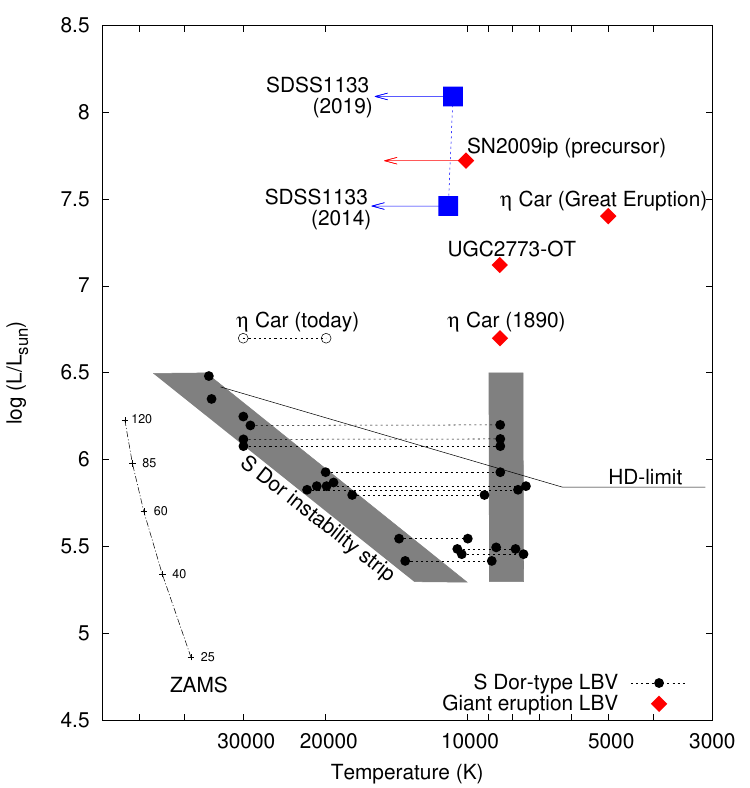}
}
 \caption{The H-R diagram for SDSS1133 in 2014 (non-outbursting phase) and 2019 (outbursting phase), compared with known S~Dor-type LBVs and giant eruption LBVs taken from \citet{res12}. Except for $\eta$ Carinae, the same objects are connected with dotted lines. The S~Dor instability strip (LBVs in quiescence) and constant temperature eruptive LBV strip (LBVs in outburst) are indicated by gray bands, and the empirical Humphreys$-$Davidson (HD) limit is indicated by a solid line. The dashed-dotted line indicates the Zero-Age Main Sequence (ZAMS) for stars with $M=25-120~M_{\odot}$ \citep[no rotation and $Z=Z_{\odot}$;][]{eks12}. The temperatures and luminosities for SDSS1133 and SN~2009ip's pre-SN eruption in 2009 are lower limits because of the possible effects of host galaxy reddening.}
 \label{fig:hrd}
\end{figure}

\subsubsection{Black body model for the UV-optical emission}
\label{sec:BBmodel_fitting}

The UVOT data obtained on 2014 February 6-25 (non-outbursting phase) and 2019 May 27-29 (during the 2019 outburst) are useful to constrain the UV-to-optical SED of SDSS1133.
While the 2019 data cover the full UV wavelength range with the UVW2, UVM2, and UVW1 bands, the 2014 data are obtained only with the UVW2 band.
Since simultaneous optical broad band photometries were not available at the epochs of the UVOT observations, ZTF $g$, $r$, and $i$ band magnitudes in 2019 and PS1 $z$ and $y$ band magnitudes in 2014 at the epochs of the UVOT observations were estimated by linearly-interpolating the light curves; the time differences between the {\it Swift}/UVOT and the closest optical measurements are 1.78, 0.75, and 5.84 days for the ZTF $g$, $r$, $i$ band measurements in 2019, and 24.38 and 24.37 days for the PS1 $z$ and $y$ and measurements in 2014, respectively.

Assuming a spherical black body radiator with a radius $R$, the observed flux is modelled as
\begin{equation}
f_{\nu, \text{obs}} = \left(\frac{r}{d_{L}}\right)^2 (1+z) \pi B_{\nu, \text{rest}}(T_{\text{eff}}).
\end{equation}
There are two free parameters; $T_{\text{eff}}$ and overall scaling factor $C \equiv (r/d_{L})^{2}$.
The bolometric luminosity $L$ can be evaluated from $T_{\text{eff}}$ and $C$:
\begin{equation}
L = 4\pi d_{L}^2 C \sigma_{\text{SB}} T_{\text{eff}}^{4}, 
\end{equation}
from the Stefan-Boltzmann law, $\int \pi B_{\nu}(T_{\text{eff}}) d\nu = \sigma_{\text{SB}} T_{\text{eff}}^4$, where $\sigma_{\text{SB}}$ is the Stefan-Boltzmann constant and $B_{\nu}$ is the black body intensity.
Therefore, there are two fitting parameters, $T_{\text{eff}}$ and $L$. 
The SED fitting was performed with the $\chi^2$ minimization, where the model broad band magnitudes calculated by the filter convolution were compared to the observed magnitudes.

Figure~\ref{fig:BBfit} shows the UV-optical SEDs of SDSS1133 on 2014 February 6-25 (non-outbursting phase) and 2019 May 27-29 (outbursting phase).
The SEDs corrected for possible host galaxy extinction assuming the extinction curves determined by the Balmer decrement (Figure~\ref{fig:balmer_decrement}; Section~\ref{sec:balmer_decrement}) are also shown. 
The best-fitting black body spectra are overplotted.
The best-fitting black body parameters for the SEDs with no host galaxy extinction correction are
$T_{\text{eff}}, L = 
 10^{4.04}~\text{K}, 10^{7.46}~L_{\odot}$ and 
$10^{4.03}~\text{K}, 10^{8.09}~L_{\odot}$
for the 2014 and 2019 data, respectively.
The corresponding black body radius is $R = 10^{3.17}~R_{\odot}$ and $10^{3.52}~R_{\odot}$, respectively.
This fitting may imply that the effective temperature of SDSS1133 did not vary significantly between the outbursting and non-outbursting phases.
Even considering the uncertainty on the intrinsic dust extinction this fitting may imply that the effective temperature of SDSS1133 did not vary significantly between the outbursting and non-outbursting phases, specifically the effective temperature never drops well below $\sim 10,000$~K.

%$T_{\text{eff}}, L = 
%10^{4.04}~\text{K}, 10^{41.04}$~erg~s${}^{-1}$ and 
%$10^{4.03}~\text{K}, 10^{41.67}$~erg~s${}^{-1}$
%for the 2014 and 2019 data, respectively.
%The corresponding black body radius is $R = 10^{3.17}~R_{\odot}$ and $10^{3.52}~R_{\odot}$, respectively.

%Since the 2019 outburst of SDSS1133 occurred on a time scale of $\sim 50-100$~days,  (Figure~\ref{fig:ztf_lc_zoom}) thus the the expansion velocity of the photosphere should be higher than $\sim 150-300$~km~s${}^{-1}$, which is consistent with the high wind velocity inferred from the broad P-Cygni absorption feature (Section~\ref{sec:broad_pcygni}).

As already discussed in Section~\ref{sec:outbursts}, the constant temperature is in contrast to normal LBV eruptions or S~Dor-type variables.
The eruptions in the normal LBVs can be explained by the outward motion of the photosphere either due to a true increase in the star's hydrostatic radius or the radius of a pseudo-photosphere in the opaque wind; the optical outbursts are accompanied with lowering of effective temperature so that the bolometric luminosity remains nearly constant \citep[see Figure~\ref{fig:hrd};][]{vin11,smi11c,smi11,van12,hum14}.
The constant effective temperature during the eruptions in SDSS1133 is more similar to LBV giant eruptions, like the 19th century Great Eruption of $\eta$ Carinae \citep[e.g.,][]{dav97,smi11,smi13,smi14,smi17}.

In terms of the recoiling AGN scenario, the UV-optical luminosity can be related to the Eddington luminosity of the recoiling SMBH.
Considering the host galaxy stellar mass of $10^{8.55}~M_{\odot}$ (Section~\ref{sec:hostgalaxy}), the SMBH mass may be $\sim 10^{6}$~$M_{\odot}$ \citep{rei15}, and the corresponding Eddington luminosity is $L_{\text{Edd}} \sim 10^{10.5} L_{\odot}$.
Therefore, the observed UV-optical luminosity of SDSS1133 is not inconsistent with the recoiling AGN scenario if the accretion rate is sub-Eddington ($\sim 0.1-1$\%).

Although the effective temperature and bolometric luminosity from the fitting for the host galaxy extinction-uncorrected SEDs are physically reasonable, the observed SED shape is much redder than the black body spectra, probably being consistent with the large Balmer decrements described in Section~\ref{sec:balmer_decrement}.
Actually, as shown in Figure~\ref{fig:BBfit}, the host galaxy extinction correction make the SEDs consistent with Rayleigh-Jeans power-law ($\lambda L_{\lambda} \propto \lambda^{-3}$).
If the host galaxy extinction coefficients are taken at their face values, the weak UV turnover (if any) in the SEDs in Figure~\ref{fig:BBfit} prevents us from determining the effective temperature.
This implies that the true temperature of SDSS1133 is much higher than $10^{4}$~K.
In this sense, $T_{\text{eff}}$ and $L$ derived for the host galaxy extinction-uncorrected SEDs should be taken as the lower limits, as indicated in Figure~\ref{fig:hrd}.
Also, taking into account the uncertainties in the host galaxy extinction estimates (Section~\ref{sec:balmer_decrement}), the host galaxy extinction-corrected SEDs shown in Figure~\ref{fig:BBfit} should be regarded as the upper limits.

We should note that the artificial excess emission in the UVM2 band seen in the host galaxy extinction-corrected SEDs in Figure~\ref{fig:BBfit} is due to the 2175\AA\ bump in the extinction curves.
Assuming that the intrinsic continuum SEDs should be smooth, the 2175\AA\ extinction bump toward SDSS1133 should be suppressed than the MW, SMC, LMC extinction curves.
Since the 2175\AA\ extinction bump is considered to be caused by Polycyclic Aromatic Hydrocarbons (PAHs) and/or small graphite dust grains \citep[e.g.,][]{dra84,taz20}, the weak 2175\AA\ extinction bump may indicate that these small grains are eliminated toward the line of sight, probably due to the strong emission of SDSS1133.

\if0

Conversely, the multi-band UV photometry on 2019 May 27-29 enabled to evaluate the host galaxy dust extinction.
The model spectrum is fixed to the Rayleigh-Jeans spectrum.
We adopt the \cite{pei92}'s SMC-like extinction curve with $R_V = 2.93$ as the host galaxy dust extinction model, where an additional fitting parameter $E(B-V)_{\text{host}}$ was introduced.
By using $\chi^2$ minimization, the best-fitting color excess is 
\begin{equation}
E(B-V)_{\text{host}}=0.576~\text{mag}.
\end{equation}
This is slightly smaller than the $E(B-V)_{\text{host}}$ estimated from the narrow line Balmer decrement (Equation~\ref{eqn:eb_v_balmer_decrement}).

\fi

\subsubsection{H-R~diagram and the energy source of SDSS1133}
\label{sec:hrd}

Figure~\ref{fig:hrd} shows the Hertzsprung-Russell (H-R) diagram for SDSS1133 in 2014 (non-outbursting phase) and 2019 (outbursting phase), compared with known S~Dor-type LBVs and giant eruption LBVs taken from \citet{res12}.
The luminosity and temperature of SDSS1133 are obtained from the black body fitting without any host galaxy extinction corrections, described in Section~\ref{sec:BBmodel_fitting}.
Normal S Dor-type LBV eruptions are characterized by transitions from a hot quiescent state in the S Dor instability strip (late O type or early B type) to cool state as an F-type supergiant with $T \sim 7000-8000$~K, as shown in Figure~\ref{fig:hrd} \citep[e.g.,][]{hum94,smi11c,wei20}.
On the other hand, giant eruptions or $\eta$ Carinae analogs distribute over the H-R diagram, typically above the empirical \cite{hum79} (HD) limit \citep[e.g.,][]{dav20}.
The HD-limit is interpreted as the Eddington limit for massive stars, thus the giant eruption LBVs are considered to be in an unstable super-Eddington state \citep[e.g.,][]{smi06,smi14}.

SDSS1133 locates well above the HD limit in the H-R diagram even in its relatively faint phase in 2014, placing it in the parameter space consistent with other giant eruption LBVs \citep[e.g.,][]{dav20}.
This indicates that SDSS1133 remains to be in an eruption phase at least since 1950, resembling the decades-long 19th century Great Eruption of $\eta$ Carinae \citep[Figure~\ref{fig:historic_lightcurve};][]{smi11b}.

% references for SN 2000ch added; Smith+11 and Pastorello+10

It is suggested that the giant eruption LBVs may be subdivided into two classes; `hot LBV' subclasses exemplified by the SN impostors SN 2009ip and SN 2000ch, and `cool LBVs' exemplified by UGC 2773-OT \citep[e.g.,][]{pas10,smi10,fol11,smi11,smi13,war21}.
The hot LBVs have high temperature of $>10,000$~K, high luminosity, and broad line profiles, while the cool LBVs have cooler $7,000-8,000$~K spectra of F-type supergiants superposed with narrower emission/absorption features \citep[Figure~\ref{fig:hrd}; e.g.,][]{smi10}.
SDSS1133 resembles SN 2009ip’s precursor (pre-SN) eruption (Figure~\ref{fig:hrd}), and thus belongs to the hot LBV subclass.
Combined with the other observational evidence (high velocity P-Cygni profile and X-ray emission; Sections~\ref{sec:broad_pcygni} and \ref{sec:broadlines}), we can speculate \citep[as pointed out by][]{smi10} that the hot LBVs (including SDSS1133) may be powered by ejecta-CSM or ejecta-ejecta interactions by a large fraction, while the cool LBVs are dominated by the photospheric emission of opaque radiation-driven outflows \citep[e.g.,][and references therein]{dav20}.

\subsubsection{X-ray emission}
\label{sec:xray_emission}

\begin{figure}
\center{
\includegraphics[clip, width=3.4in]{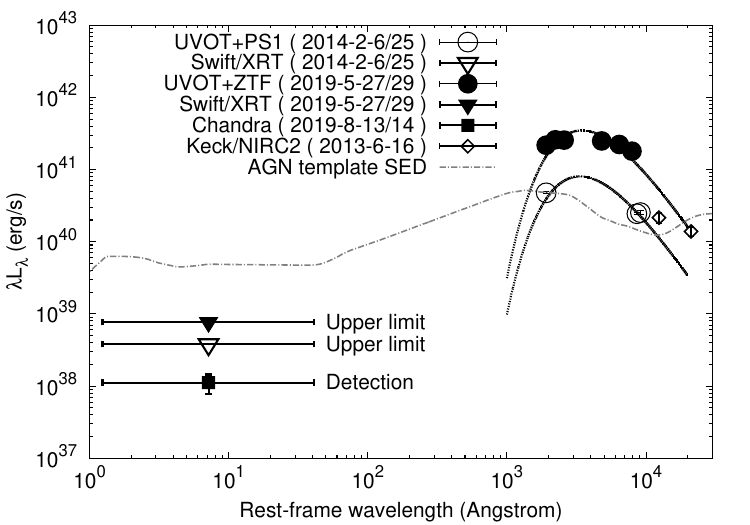}
\includegraphics[clip, width=3.4in]{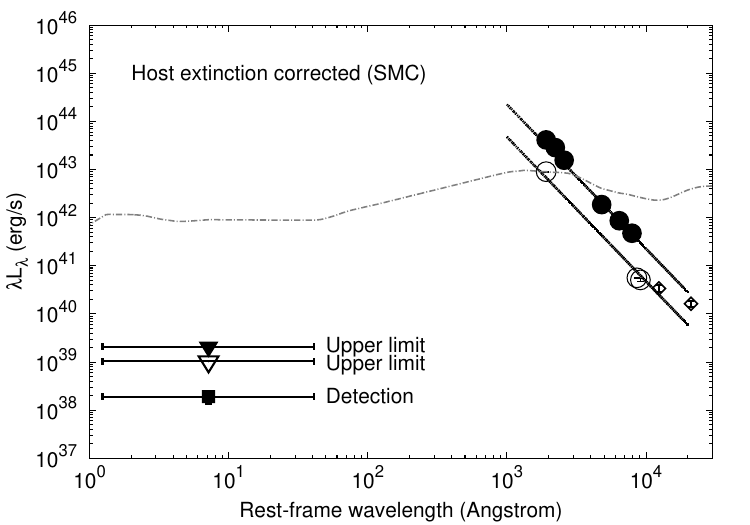}
}
 \caption{The same as Figure~\ref{fig:BBfit}, but including X-ray and NIR data of SDSS1133.
 The X-ray and NIR data were obtained in the non-outbursting phase.
 %The band center of the $0.3-10$~keV ($1.2 - 41.3$~\AA) X-ray measurements are set to $7.2$~\AA.
 Top: the X-ray-to-NIR SED uncorrected for the host galaxy extinction.
 Bottom: the X-ray-to-NIR SED corrected for the host galaxy extinction assuming the SMC-like extinction (Section~\ref{sec:multiwavelength_sed}).
 The X-ray SED is calculated as $\lambda L_{\lambda} = L_{0.3-10~\text{keV}}/\ln(10~\text{keV}/0.3~\text{keV})$, i.e., $\Gamma=2$.
 The triangle symbols indicate the Swift/XRT X-ray upper limits, and the square symbol indicates the Chandra X-ray detection on 2019 August 13-14.
 The dotted lines indicate the best-fit black body SEDs.
 The dashed-dotted line indicates the template SED for radio-quiet AGNs \citep{elv94}, scaled to match the UVW2 luminosity in 2014 (non-outbursting phase).
 }
 \label{fig:fullSED_corrected}
\end{figure}

\begin{figure}
\center{
\includegraphics[clip, width=3.2in]{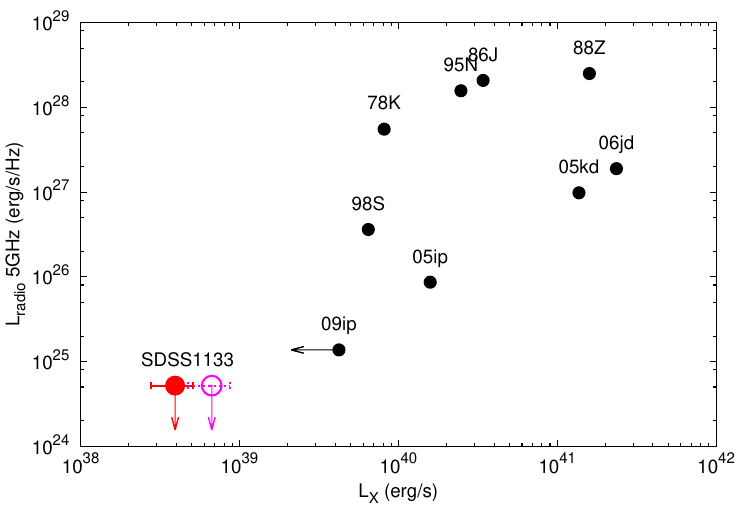}
}
 \caption{The radio and X-ray luminosity of SDSS1133 observed in the non-outbursting phase, compared with the radio and X-ray data of ejecta-CSM interacting SNe IIn at the radio peak (black dots) taken from \citet{margutti14}. The two data points for SDSS1133 indicate the X-ray luminosity uncorrected (filled circle) or corrected (open circle) for the possible host galaxy extinction assuming the SMC-like extinction (Table~\ref{tab:xray_table}). 
 The data for SN~2009ip are for the 2012b event (Section~\ref{sec:longterm}). 
 Note that the LBVs known to date have the X-ray luminosity much less than $10^{34}$~erg~s${}^{-1}$ \citep{naz12}.}
 \label{fig:radio_xray}
\end{figure}

The dust extinction in the host galaxy of SDSS1133 inferred by the Balmer decrement also implies the presence of X-ray absorption.
As mentioned in Section~\ref{sec:balmer_decrement}, the hydrogen column density in the host galaxy inferred from the color excess is $N_{\text{H}, \text{host}} = (2.84, 1.46, 0.31) \times 10^{22}$~cm${}^{-2}$ under the assumption of the SMC-like, LMC-like, and MW-like extinction, respectively.
The X-ray absorption due to the gas of the columns density of $N_{H, \text{host}}$ is estimated by adding {\tt zvphabs} ({\tt xszvphabs}) component in the {\tt XSPEC} ({\tt CIAO}) X-ray spectral modelling, where the heavy-element abundances relative to MW are assumed to be 1/8, 1/3 for SMC and LMC, respectively \citep{pei92}, the absorber's redshift is fixed to $0.0079$.
Table~\ref{tab:xray_table} lists the X-ray luminosity corrected for the X-ray absorption in the host galaxy.
By comparing the values in Table~\ref{tab:xray_table}, we can see that the correction factor due to the X-ray absorption in the host galaxy is at most $\sim 3$.
This mean that the uncertainty on the X-ray absorption strength in the host galaxy have only little influence on our order-of-magnitude discussion about the X-ray luminosity of SDSS1133.

Figure~\ref{fig:fullSED_corrected} shows the X-ray-to-NIR SED of SDSS1133 in the non-outbursting phase in 2014 and outbursting phase in 2019.
The \cite{elv94}'s template SED for luminous ($M_{V} < -22$~mag) radio-quiet AGNs is overplotted in Figure~\ref{fig:fullSED_corrected} for comparison.
The {\it Swift}/XRT X-ray upper limits obtained simultaneously with the Swift/UVOT UV measurements reveal that the X-ray emission of SDSS1133 is much less than the X-ray emission expected from the typical X-ray-to-UV/optical luminosity ratio in AGNs \citep[see also][]{simmonds16}.
Moreover, the {\it Chandra} X-ray detection in the non-outbursting phase in 2019 (but this is just after the 2019 outburst) suggests that the X-ray luminosity is at least a thousand times smaller than the UV-optical luminosity.
Such an low X-ray-to-optical luminosity ratio is unusual for an AGN \citep[e.g.,][]{has08,gib08,liu18b}.
As shown in Figure~\ref{fig:fullSED_corrected}, the corrections for the X-ray and UV-optical host galaxy absorption/extinction further decrease the X-ray-to-optical luminosity ratio.
The weak X-ray emission of SDSS1133 is suggestive of the non-AGN nature.
Actually, the X-ray luminosity of SDSS1133 is well below an empirical threshold luminosity to identify AGNs \citep[$\sim 10^{42}~\text{erg}~\text{s}^{-1}$;][]{leh16,bir20,sch21}, thus SDSS1133 cannot be classified as an AGN based on the X-ray luminosity.

%The radio luminosity at 5~GHz expected from the AGN SED template is about $10^{24}$~erg~s${}^{-1}$~Hz${}^{-1}$, which is not inconsistent with the eEVN upper limit of $5.2 \times 10^{24}$~erg~s${}^{-1}$~Hz${}^{-1}$ \citep{perez15}.

The UV-optical black body modelling for the host extinction-uncorrected SED in the non-outbursting phase in 2014 suggests that the UV-optical luminosity of SDSS1133 is $L \sim 10^{41}$~erg~s${}^{-1}$ (Section~\ref{sec:BBmodel_fitting}).
By using the {\it Chandra} X-ray luminosity ($L_{X} = 4 \times 10^{38}$~erg~s${}^{-1}$), the X-ray-to-optical luminosity ratio is $\log(L_{X}/L) \sim -2.4$.
On the other hand, if the host galaxy extinction correction shown in Figure~\ref{fig:fullSED_corrected} is taken at its face value, the UV-optical luminosity of SDSS1133 can be as large as $L \sim 10^{43}$~erg~s${}^{-1}$, thus $\log(L_{X}/L) \sim -4.4$.
\cite{naz12} show that the X-ray-to-optical luminosity ratio ($\log(L_{X}/L)$) of known LBVs is at most $\sim -5$ and generally less than $-6$.
We should note that the X-ray luminosity of SDSS1133 is at least 4 orders of magnitude larger than that observed in Galactic LBVs \citep[$L_{0.5-8~\text{keV}} < 10^{34}$~erg~s${}^{-1}$;][]{naz12,koss14}.
Therefore, we can conclude that SDSS1133 is extremely X-ray bright compared to normal LBVs.
In summary, SDSS1133 is too faint to be a ``typical'' AGN, but too bright to be a ``typical'' LBV.

\subsubsection{Multi-wavelength SED}
\label{sec:multiwavelength_sed}

The large discrepancy between the X-ray luminosities of SDSS1133 and other LBVs known to date \citep{naz12} is probably because the current LBV sample observed at X-ray is limited to normal S~Dor-type LBVs, and essentially no detailed X-ray data is available for giant eruption LBVs or SN impostors.

As discussed in Sections~\ref{sec:broad_pcygni} and \ref{sec:broadlines}, a population of LBV giant eruptions can be powered by shock interactions between ejecta materials and CSM \citep[e.g.,][]{smi08,smi13,smi10,smi18}.
As observationally confirmed in ejecta-CSM interacting SNe~IIn, the interaction region can produce strong UV-optical, X-ray, and radio emission \citep{smi13,smi17b}.
It is naturally expected that the observational properties of the interaction-powered LBVs resemble the SNe~IIn, and observed as scaled-down SNe~IIn \citep{smi13}.

In Figure~\ref{fig:radio_xray}, the X-ray and radio luminosities of SDSS1133 are compared with those of a sample of ejecta-CSM interacting SNe IIn compiled by \cite{margutti14}.
We should note that the {\it Chandra} X-ray observation was carried out just after the 2019 outburst (Figure~\ref{fig:ztf_lc_zoom}), thus the X-ray luminosity may be an ``afterglow'' emission of the 2019 outburst event, while the radio luminosity reflects the emission in the non-outbursting phase before the 2014 outburst.
In this figure, we can see that SDSS1133 may locate at the faint end of the X-ray vs. radio correlation space of the interacting SNe.
The X-ray luminosity of SDSS1133 is about 2 orders of magnitude lower than the typical values of the interacting SNe~IIn, suggesting that SDSS1133 is a scaled-down version of these objects.
In terms of the multi-wavelength energetics, we can say that SDSS1133 is in between LBVs and interacting SNe.
Also, the radio luminosity of SDSS1133 is at least two orders of magnitude below compared to the X-ray/radio correlation of low-luminosity AGNs at the given X-ray luminosity \citep[e.g.,][]{mer03,gro22}.

\cite{koss14} point out that the Keck/NIRC2 $K$ band measurement may indicate the presence of an IR excess emission at $\lambda \sim 20,000$~\AA (see Figure~\ref{fig:fullSED_corrected}).
Since the hot dust emission of various strength is one of the observational features of interacting SN \citep[e.g.,][]{sza19,kok19}, the IR excess in SDSS1133 can also be naturally explained by the above mentioned scaled-down SN~IIn model.
Future optical-NIR simultaneous observations are needed to confirm this weak NIR excess emission.

%, the $K$ band excess in SDSS1133 is, if any, much weaker than the typical AGN SED.
%We should note that the weak NIR excess emission alone cannot rule out the possibility that SDSS1133 is AGN-origin, since the matter in the dust torus could be unbounded due the high kick velocity at the time of the gravitational wave recoil \citep[e.g.,][]{kom08,liu12}.
%Future optical-NIR simultaneous observations are needed to confirm this NIR excess emission.

\subsection{  Energetics: LBV with CSM interactions?  }
\label{sec:emission_mechanism}

In this Section we discuss the energetics of the UV-optical emission in the non-outbursting and outbursting phases of SDSS1133.
Constructing models to reproduce the detailed structure of the multi-wavelength light curves is beyond the scope of this paper and will be presented elsewhere.
Here we show that the required radiation energy to explain SDSS1133 can be generated within existing observational and theoretical frameworks; namely, some form of non-terminal explosions and subsequent efficient kinetic-to-radiation energy conversion by ejecta-CSM or ejecta-ejecta interactions.

The UV-optical luminosity of SDSS1133 is at least $L_{\text{outburst}}=10^{8.09}~L_{\odot} = 10^{41.67}~\text{erg}~\text{s}^{-1}$ during the 2019 outburst (no host galaxy extinction correction; Section~\ref{sec:BBmodel_fitting}), that continued for more than 27.5~days (see Section~\ref{sec:outbursts}). 
Thus, the total energy required to explain the 2019 outburst is at least 
\begin{equation}
E_{\text{rad}, \text{outburst}} \sim 10^{48}~\text{erg} \times \left(  \frac{L_{\text{outburst}}}{10^{8.09}~L_{\odot}}  \right) \left(\frac{\Delta t}{27.5~\text{days}} \right).
\end{equation}
SDSS1133 experienced multiple outbursts of similar radiation energy at least four times since 2001.
On the other hand, the persistent continuum luminosity in the non-outbursting phase since 1950 is roughly $L_{\text{non-outburst}}=10^{7.49}~L_{\odot} = 10^{41.07}~\text{erg}~\text{s}^{-1}$, thus the the total radiation energy released so far is 
\begin{equation}
E_{\text{rad}, \text{non-outburst}} \sim 10^{50}~\text{erg} \times \left(  \frac{L_{\text{non-outburst}}}{10^{7.49}~L_{\odot}}  \right) \left(\frac{\Delta t}{70~\text{yrs}} \right).
\end{equation}
We should note that these estimates for $E_{\text{rad}}$ are lower-limit, as the possible host galaxy extinction is uncorrected correction in the above calculations.

The Eddington ratio or Eddington parameter $\Gamma$ given the luminosity $L$ and stellar mass $M$ is
\begin{equation}
\Gamma = \frac{\kappa_{e} L}{4 \pi G M c} = 2.6 \times \left( \frac{L}{10^{7}L_{\odot}} \right)  \left(\frac{M}{100~M_{\odot}}\right)^{-1},
\end{equation}
where $\kappa_{e}=0.34~\text{cm}^{2}~\text{g}^{-1}$ is the opacity due to Thomson scattering, and we assume $M=100~M_{\odot}$ \citep[the current mass of $\eta$ Carinae; e.g.,][]{smi08} as a reference value.
The luminosity of SDSS1133 ($L > 10^{7}~L_{\odot}$) implies that it exceeds the classical Eddington limit of $\Gamma = 1$ even in the non-outbursting phase at least for 70 yrs.
This situation is reminiscent of the $\eta$ Carinae's Great Eruption, that had been luminous at $\Gamma = 5$ for over decades.
Like the $\eta$ Carinae's Great Eruption, SDSS1133 is probably continuously launching inhomogeneous, porous super-Eddington (SE) continuum-driven winds \citep[e.g.,][]{smi06b,owo12,dot12}.
A dense CSM around SDSS1133 can be produced by these radiation-driven winds \citep[e.g.,][]{smi14}.
The radiation-driven wind mechanisms, however, cannot produce high velocity outflows of $>1,000$~km~s${}^{-1}$ (well beyond the star's escape velocity) observed in SDSS1133 as the broad absorption, thus we need some form of non-terminal explosions and subsequent enhanced mass ejections (like scaled-down SN) to explain the high velocity materials observed in SDSS1133 (Sections~\ref{sec:broad_pcygni} and \ref{sec:broadlines}).

The radiated energy of SDSS1133 is comparable to that of a single core-collapse SN ($E_{\text{rad}} \sim 10^{49}~\text{erg}$).
As suggested in Sections~\ref{sec:broad_pcygni} and \ref{sec:broadlines}, the large radiation energy of SDSS1133 can be naturally explained by mass ejections due to non-terminal explosions ($E_\text{kin} \sim 10^{49}-10^{51}~\text{erg}$, where $E_\text{kin}$ is the kinetic energy of the explosion) and subsequent CSM interactions.
Because the CSM interactions can have high efficiency of converting kinetic energy into radiation \citep[a few 10\%; e.g.,][]{mor13b,smi14b}, the shocks induced by the non-terminal explosions and propagating into the dense CSM can produce enough radiation energy to account for both the outbursting and non-outbursting phases of SDSS1133.
In the same way as the CSM interactions, if massive shells ejected by the multiple explosion events have different velocities, collisions between the shells may produce distinct outbursts \citep[e.g.,][]{heg03,woosley07,des09,mor13b,yos16,arc17}.

Consider a simple ejecta-CSM shock interaction model \citep[e.g.,][]{smi13,mor13}.
The CSM is assumed to be formed via a steady wind with a velocity of $v_{w}$.
The kinetic energy of the ejected massive shell is converted into radiation energy at the shocked region moving at a velocity of $v_{s}$.
The shocked shell velocity $v_{s}$ is time-dependent, but here we replace it by the initial velocity of the ejecta: $v_{s} \sim \sqrt{2E_{\text{kin}}/M_{\text{ej}}} \simeq 1,000~\text{km}~\text{s}^{-1} \times ( E_{\text{kin}}/10^{50}~\text{erg} )^{1/2} ( M_{\text{ej}}/10~M_{\odot} )^{-1/2}$, where $M_{\text{ej}}$ is the ejected mass.
A kinetic-to-radiation conversion efficiency, $\epsilon$, is usually assumed to be in a range of $0.1-0.5$ \citep[e.g.,][]{mor13b,mor13}.
The wind mass-loss rate $\dot{M}$ can be related to the bolometric luminosity via $\epsilon$ as:
\begin{eqnarray}
L &=& \frac{1}{2}\epsilon\frac{\dot{M}}{v_{w}}v_{s}^3 \nonumber \\
&=& 10^{7.92}~L_{\odot} \times \left(\frac{\epsilon}{0.1}\right) \left(\frac{\dot{M}}{M_{\odot}~\text{yr}^{-1}}\right) \nonumber \\
& & \times \left(\frac{v_{w}}{100~\text{km}~\text{s}^{-1}}\right)^{-1} \left(\frac{E_{\text{kin}}}{10^{50}~\text{erg}}\right)^{3/2} \left(\frac{M_{\text{ej}}}{10~M_{\odot}}\right)^{-3/2}.
\end{eqnarray}
Although most of the parameters are uncertain in the case of SDSS1133, this estimate suggests that the luminosities in the non-outbursting and outburst phases can be produced by the CSM interactions if the persistent mass-loss of the order of $\dot{M} \sim 1~M_{\odot}~\text{yr}^{-1}$ and non-terminal explosions of $E_{\text{kin}} \sim 10^{50}~\text{erg}$ and $M_{\text{ej}} \sim 10~M_{\odot}$.
%SDSS1133 has kept a mass-loss of the order of $\dot{M} \sim 0.1~M_{\odot}~\text{yr}^{-1}$ at least for 70~years since 1950, and occasionally experienced giant mass-loss episodes of the order of $\dot{M} \sim 1~M_{\odot}~\text{yr}^{-1}$.
Such a large mass-loss rate is suggested to present in giant eruption LBVs \citep{smi14}, and actually the mass-loss rate at the time of the $\eta$ Carinae's Great Eruption is estimated to be a few $M_{\odot}~\text{yr}^{-1}$ from observations for the Homunculus nebula \citep[e.g.,][]{hum05,smi13}.
The large mass-loss rate also indicates that SDSS1133 is a massive star of the mass of $100~M_{\odot}$ like $\eta$ Carinae.

While the dense CSM may be naturally formed via the SE winds \citep[e.g.,][]{smi14}, it is unclear what mechanisms can trigger (multiple) non-terminal explosions in LBVs \citep[e.g.,][]{smi14b}.
From the viewpoint of energy deposition, it is suggested that the non-terminal explosions of $E_\text{kin} \sim 10^{50}~\text{erg}$ may not be explained by the envelope instability, and require to transfer a large amount of extra energy from the stellar core into the outer envelope \citep[e.g.,][]{smi14,smi14b,str21}.

Many mechanisms to generate the mass-loss episodes in the giant eruption LBVs or SN impostors have been suggested in the literature, including pulsational pair instability \citep[PPI; e.g.,][]{woosley07,woo17,arc17}, wave-driven mass loss \citep[e.g.,][]{qua12,shi14}, and stellar collisions or mergers in a binary system \citep[e.g.,][]{kas10,smi11c,hir21} \citep[see][for a review]{smi11,smi14}.
Although $\eta$ Carinae is known to be a binary (or triple) system, currently there is no reason to consider that SDSS1133 is in a binary system and its eruptions are caused by binary interactions.
Moreover, the binary interactions alone may not be enough to explain the highly-luminous multiple outbursts observed in SDSS1133.
Among these scenarios, the PPI-driven mass ejection is of particular interest since this mechanism may possibly provide a natural explanation for the multiple giant eruptions (=non-terminal explosions) every few years with the explosion energy of $\sim 10^{50}~\text{erg}$, ejected mass of $\sim 10~M_{\odot}$, and ejecta velocity of $\sim 2,000$~km~s${}^{-1}$, as observed in SDSS1133 and other extreme LBV giant eruptions \citep[e.g.,][]{smi08,yos16,woo17,arc17,fra22}.
An interesting possibility is that, since the PPI (as well as the wave-driven mass loss) is predicted to occur at the very late stage of massive star evolution, we may be able to observe the terminal SN explosion of SDSS1133 after a few additional pre-SN eruptions, like other precursor eruptions observed prior to SNe~IIn \citep[e.g.,][]{gal09,smi11,lan12,mau13,moriya14,ofe14,arc17,smi17,str21,fra22}.
Another possibility is that SDSS1133 may end up with a collapse to a massive BH without a bright SN \citep[e.g.,][]{heg03,all20}.
Further multi-wavelength follow-up observations are encouraged to trace the evolution of SDSS1133.

\if0

python

from astropy import units as u
from astropy import constants as const
import numpy as np

Mdot    = 1.0 * u.M_sun/u.yr
vw      = 100.0*u.km/u.s

Mej  = 10.0*u.M_sun
Ekin = 1.0e50*u.erg
vs   = np.sqrt(2.0*Ekin/Mej)

epsilon = 0.1

Lbol = (0.5 * epsilon * Mdot / vw * vs**3).to(u.erg/u.s)
Lbol_sun = (Lbol/const.L_sun).cgs

print( np.log10(Lbol.value), np.log10(Lbol_sun.value) )

%41.50215091778503 7.919178988680224

\fi

\if0

python

from astropy import units as u
from astropy import constants as const
import numpy as np

M       = 100.0 * u.M_sun
kappa_e = 0.34 * u.cm**2/u.g
L       = 1.0e7 * u.L_sun

Gamma = (kappa_e*L)/(4.0*np.pi*const.G*M*const.c)

print( Gamma.cgs.value )

%41.498386263174254 7.91541433406945

Mdot    = 1.0 * u.M_sun/u.yr
vw      = 100.0*u.km/u.s
R_ph    = (kappa_e*Mdot)/(4.0*np.pi*vw)

print( R_ph.to(u.R_sun) )

\fi

\subsection{Volumetric rate of SDSS1133-like objects}

The volumetric rate of the SDSS1133-like LBV objects may roughly be estimated by the fact that there are only two objects found to be offset from the host galaxy nucleus among 3,579 broad H$\alpha$ emission lines objects at $z<0.31$ targeted by the 8,032~deg${}^{2}$ SDSS-I/II spectroscopic survey \citep[$>1,000$~km~s${}^{-1}$;][]{aba09,ste12,koss14}.
Among the two objects, one is SDSS1133, and another is Mrk~739, but Mrk~739 is confirmed to be a dual AGN \citep{kos11}.
The comoving volume within redshift $z=0.31$ is 7.8 Gpc${}^{3} \times (8,032/41,253)$ = 1.5~Gpc${}^{3}$ \footnote{$\Lambda$CDM cosmology of $\Omega_{\Lambda}=0.7$, $\Omega_{\text{m}}=0.3$, and $H_{0}=70$~km~s${}^{-1}$~Mpc${}^{-1}$ is assumed}, thus the volumetric rate of SDSS1133-like objects may be estimated as $\sim 0.7$~Gpc${}^{-3}$.
The estimated rate of the SDSS1133-like LBV objects is much less than the core-collapse SN rate in the local Universe \citep[$\sim 100,000$~Gpc${}^{-3}$~yr${}^{-1}$;][]{per20}.
It may indicate that at most a few percent of core-collapse SN progenitors experience the SDSS1133-like long-lasting ($> 60$~years) bright LBV eruption phase.
This may be consistent with the scenario that LBVs with enhanced mass-loss end up with (super-luminous) SNe IIn, i.e., a rare population of core-collapse SNe \citep{ofe14,smi17,str21}.

The above volumetric rate estimate ignores several objects similar to SDSS1133 individually reported in the literature; for example, PHL~293B, which is found in a local blue compact dwarf galaxy at $z=0.00517$, is suggested to be another example belonging to the same class of extragalactic transient/variable as SDSS1133 \citep[e.g.,][]{izo09,ter14,bur20,all20}.
Another similar object is recently reported by ZTF \citep[AT2018kle at $z=0.013$ with the peak absolute magnitude of $M_{g} = -16.34$;][]{per20}, which is classified as an eruption of a LBV.
%The volumetric rate estimated above is for the extremely bright extragalactic blue variables with offsets from the host galaxy centers.
It is worth noticing that the spectral resemblance between SDSS1133 and broad line AGNs suggests that many SDSS1133-like objects are missed by the current surveys; for example, a fraction of nuclear/off-nuclear AGN-like objects can actually be LBVs \citep[e.g.,][]{izo07,izo08,baldassare16,simmonds16,bur21,war21}.
In this sense, the volumetric rate of the SDSS1133-like objects could be higher than the above estimate, and more comprehensive survey is needed to quantify the true occurrence rate of the SDSS1133-like objects.

\section{Summary and Conclusions}
\label{sec:conclusion}

In this paper we have presented the comprehensive analysis of the multi-wavelength data of SDSS1133.
As summarized in Table~\ref{tbl:scenario}, our analysis reveals that the observational properties of SDSS1133 are unusual for an AGN, while consistent with several well-known giant eruption LBVs.
SDSS1133 is an off-nuclear object, thus if it is an AGN it is already a very rare ``recoiling AGN'', and it seems unreasonable to accept that such a special object simultaneously has many unusual observational properties for an AGN as discussed in this paper.
Therefore, we conclude that SDSS1133 is not a recoiling AGN but is more likely to be an extragalactic giant eruption LBV.

The observed peak absolute magnitude of the outbursts reaches $M_{g} \sim -16$~mag ($-18$~mag after correcting for the possible host galaxy extinction), which places SDSS1133 as one of the brightest LBV giant eruptions of known Galactic and extragalactic LBVs and SN impostors.
The observational properties of SDSS1133 (multiple outbursts, broad P-Cygni profile, UV-optical luminosity, high black body temperature of $>10,000$~K) is reminiscent of pre-SN eruptions of SN~2009ip (Sections~\ref{sec:outbursts}, \ref{sec:broad_pcygni}, \ref{sec:broadlines}, and \ref{sec:BBmodel_fitting}), suggesting that SDSS1133 belongs to the hot LBV subclass (Section~\ref{sec:hrd}).
The persistent P-Cygni absorption feature seen in the hydrogen Balmer lines extends up to $-5,000$~km~s${}^{-1}$.
The fast ejected materials responsible for the P-Cygni absorption in SDSS1133 likely originate from multiple non-terminal explosive outbursts, rather than radiation-driven stellar winds (Section~\ref{sec:broad_pcygni}).
We suggest that both of the bright persistent emission and UV-optical outbursts in SDSS1133 are produced by interactions of the ejected shell with CSM and/or different shells (Sections~\ref{sec:broadlines} and \ref{sec:emission_mechanism}).
The interaction model naturally explains the long-duration SN~IIn-like multi-wavelength photometric and spectroscopic properties of SDSS1133 including X-ray luminosity, persistent broad emission/absorption lines, and blue UV-optical continuum (Sections~\ref{sec:broadlines}, \ref{sec:multiwavelength_sed}, and \ref{sec:emission_mechanism}).

The origin of the non-terminal explosions is highly debated.
Among the many theoretical possibilities, we suggest that pulsational pair-instability may provide a viable explanation for the multiple energetic eruptions in SDSS1133, in terms of the energetics and occurence rate of the eruptions (Sections~\ref{sec:outbursts} and \ref{sec:emission_mechanism}).
If the current activity of SDSS1133 is a bona-fide precursor of a terminal SN explosion, we may be able to observe a few additional giant eruptions and then the terminal SN explosion or collapse to a massive BH in future observations (in a few years to a few thousand years).

\section*{Acknowledgements}

\small{


M.~K. acknowledges the JSPS overseas research fellowship.
M.~K. acknowledges the referee for his/her constructive comments to improve this paper.
M.~K. thanks Wenbin Lu, Masaomi Tanaka and Hirofumi Noda for a fruitful discussion.
A part of this work was initiated during my stay at Caltech under the Caltech-Japan visiting program in Astronomy, and M.~K. is grateful to Matthew J. Graham for his support and hospitality during the visit.

This research made use of {\tt Astropy}, a community-developed core Python package for Astronomy \citep{ast13}.
We used {\tt numpy} \citep{num20}, {\tt scipy} \citep{sci20}, {\tt extinction} \citep{bar16}, and {\tt speclite}\footnote{\href{https://github.com/desihub/speclite}{https://github.com/desihub/speclite}} in the data analysis.

%KOA

This research has made use of the Keck Observatory Archive (KOA), which is operated by the W. M. Keck Observatory and the NASA Exoplanet Science Institute(NExScI), under contract with the National Aeronautics and Space Administration.
The Keck archival data (KOAID = LB.20150717.21554, LB.20150717.22197, LB.20190310.44548, LR.20150717.21568, LR.20150717.22257, LR.20190310.44557) obtained through the programs C280LA and C300 (PI: F. Harrison) were used.

%Gaia

This work has made use of data from the European Space Agency (ESA) mission
{\it Gaia} (\url{https://www.cosmos.esa.int/gaia}), processed by the {\it Gaia}
Data Processing and Analysis Consortium (DPAC,
\url{https://www.cosmos.esa.int/web/gaia/dpac/consortium}). Funding for the DPAC
has been provided by national institutions, in particular the institutions
participating in the {\it Gaia} Multilateral Agreement.

We acknowledge ESA Gaia, DPAC and the Photometric Science Alerts Team (http://gsaweb.ast.cam.ac.uk/alerts).

%IPAC

This research has made use of the NASA/IPAC Infrared Science Archive, which is operated by the Jet Propulsion Laboratory, California Institute of Technology, under contract with the National Aeronautics and Space Administration.

%Chandra

This research has made use of data obtained from the Chandra Data Archive and the Chandra Source Catalog, and software provided by the Chandra X-ray Center (CXC) in the application packages CIAO, ChIPS, and Sherpa.

%XMM

Based on observations obtained with {\it XMM-Newton}, an ESA science mission with instruments and contributions directly funded by ESA Member States and NASA.

%Swift

We acknowledge the use of public data from the Swift data archive.

%ZTF

Based on observations obtained with the Samuel Oschin 48-inch Telescope at the Palomar Observatory as part of the Zwicky Transient Facility project. ZTF is supported by the National Science Foundation under Grant No. AST-1440341 and a collaboration including Caltech, IPAC, the Weizmann Institute for Science, the Oskar Klein Center at Stockholm University, the University of Maryland, the University of Washington, Deutsches Elektronen-Synchrotron and Humboldt University, Los Alamos National Laboratories, the TANGO Consortium of Taiwan, the University of Wisconsin at Milwaukee, and Lawrence Berkeley National Laboratories. Operations are conducted by COO, IPAC, and UW.

% PS1

The Pan-STARRS1 Surveys (PS1) and the PS1 public science archive have been made possible through contributions by the Institute for Astronomy, the University of Hawaii, the Pan-STARRS Project Office, the Max-Planck Society and its participating institutes, the Max Planck Institute for Astronomy, Heidelberg and the Max Planck Institute for Extraterrestrial Physics, Garching, The Johns Hopkins University, Durham University, the University of Edinburgh, the Queen's University Belfast, the Harvard-Smithsonian Center for Astrophysics, the Las Cumbres Observatory Global Telescope Network Incorporated, the National Central University of Taiwan, the Space Telescope Science Institute, the National Aeronautics and Space Administration under Grant No. NNX08AR22G issued through the Planetary Science Division of the NASA Science Mission Directorate, the National Science Foundation Grant No. AST-1238877, the University of Maryland, Eotvos Lorand University (ELTE), the Los Alamos National Laboratory, and the Gordon and Betty Moore Foundation.

%Legacy Survey

The Legacy Surveys consist of three individual and complementary projects: the Dark Energy Camera Legacy Survey (DECaLS; NOAO Proposal ID \# 2014B-0404; PIs: David Schlegel and Arjun Dey), the Beijing-Arizona Sky Survey (BASS; NOAO Proposal ID \# 2015A-0801; PIs: Zhou Xu and Xiaohui Fan), and the Mayall z band Legacy Survey (MzLS; NOAO Proposal ID \# 2016A-0453; PI: Arjun Dey). DECaLS, BASS and MzLS together include data obtained, respectively, at the Blanco telescope, Cerro Tololo Inter-American Observatory, National Optical Astronomy Observatory (NOAO); the Bok telescope, Steward Observatory, University of Arizona; and the Mayall telescope, Kitt Peak National Observatory, NOAO. The Legacy Surveys project is honored to be permitted to conduct astronomical research on Iolkam Du'ag (Kitt Peak), a mountain with particular significance to the Tohono O'odham Nation.

This research has made use of the services of the ESO ScienceArchive Facility. This research is partly based on observations col-lected at the European Organisation for Astronomical Research inthe Southern Hemisphere under ESO programme 087.D-0693(B).

NOAO is operated by the Association of Universities for Research in Astronomy (AURA) under a cooperative agreement with the National Science Foundation.

This project used data obtained with the Dark Energy Camera (DECam), which was constructed by the Dark Energy Survey (DES) collaboration. Funding for the DES Projects has been provided by the U.S. Department of Energy, the U.S. National Science Foundation, the Ministry of Science and Education of Spain, the Science and Technology Facilities Council of the United Kingdom, the Higher Education Funding Council for England, the National Center for Supercomputing Applications at the University of Illinois at Urbana-Champaign, the Kavli Institute of Cosmological Physics at the University of Chicago, Center for Cosmology and Astro-Particle Physics at the Ohio State University, the Mitchell Institute for Fundamental Physics and Astronomy at Texas A\&M University, Financiadora de Estudos e Projetos, Fundacao Carlos Chagas Filho de Amparo, Financiadora de Estudos e Projetos, Fundacao Carlos Chagas Filho de Amparo a Pesquisa do Estado do Rio de Janeiro, Conselho Nacional de Desenvolvimento Cientifico e Tecnologico and the Ministerio da Ciencia, Tecnologia e Inovacao, the Deutsche Forschungsgemeinschaft and the Collaborating Institutions in the Dark Energy Survey. The Collaborating Institutions are Argonne National Laboratory, the University of California at Santa Cruz, the University of Cambridge, Centro de Investigaciones Energeticas, Medioambientales y Tecnologicas-Madrid, the University of Chicago, University College London, the DES-Brazil Consortium, the University of Edinburgh, the Eidgenossische Technische Hochschule (ETH) Zurich, Fermi National Accelerator Laboratory, the University of Illinois at Urbana-Champaign, the Institut de Ciencies de l'Espai (IEEC/CSIC), the Institut de Fisica d'Altes Energies, Lawrence Berkeley National Laboratory, the Ludwig-Maximilians Universitat Munchen and the associated Excellence Cluster Universe, the University of Michigan, the National Optical Astronomy Observatory, the University of Nottingham, the Ohio State University, the University of Pennsylvania, the University of Portsmouth, SLAC National Accelerator Laboratory, Stanford University, the University of Sussex, and Texas A\&M University.

BASS is a key project of the Telescope Access Program (TAP), which has been funded by the National Astronomical Observatories of China, the Chinese Academy of Sciences (the Strategic Priority Research Program "The Emergence of Cosmological Structures" Grant \# XDB09000000), and the Special Fund for Astronomy from the Ministry of Finance. The BASS is also supported by the External Cooperation Program of Chinese Academy of Sciences (Grant \# 114A11KYSB20160057), and Chinese National Natural Science Foundation (Grant \# 11433005).

The Legacy Survey team makes use of data products from the Near-Earth Object Wide-field Infrared Survey Explorer (NEOWISE), which is a project of the Jet Propulsion Laboratory/California Institute of Technology. NEOWISE is funded by the National Aeronautics and Space Administration.

The Legacy Surveys imaging of the DESI footprint is supported by the Director, Office of Science, Office of High Energy Physics of the U.S. Department of Energy under Contract No. DE-AC02-05CH1123, by the National Energy Research Scientific Computing Center, a DOE Office of Science User Facility under the same contract; and by the U.S. National Science Foundation, Division of Astronomical Sciences under Contract No. AST-0950945 to NOAO.

}

%%%%%%%%%%%%%%%%%%%%%%%%%%%%%%%%%%%%%%%%%%%%%%%%%%

\section*{Data Availability}

The data underlying this article will be shared on reasonable request to the corresponding author.

%%%%%%%%%%%%%%%%% APPENDICES %%%%%%%%%%%%%%%%%%%%%

%%%%%%%%%%%%%%%%%%%%%%%%%%%%%%%%%%%%%%%%%%%%%%%%%%


\bibliography{sdss1133.bib}

\begin{thebibliography}{}
\makeatletter
\relax
\def\mn@urlcharsother{\let\do\@makeother \do\$\do\&\do\#\do\^\do\_\do\%\do\~}
\def\mn@doi{\begingroup\mn@urlcharsother \@ifnextchar [ {\mn@doi@}
  {\mn@doi@[]}}
\def\mn@doi@[#1]#2{\def\@tempa{#1}\ifx\@tempa\@empty \href
  {http://dx.doi.org/#2} {doi:#2}\else \href {http://dx.doi.org/#2} {#1}\fi
  \endgroup}
\def\mn@eprint#1#2{\mn@eprint@#1:#2::\@nil}
\def\mn@eprint@arXiv#1{\href {http://arxiv.org/abs/#1} {{\tt arXiv:#1}}}
\def\mn@eprint@dblp#1{\href {http://dblp.uni-trier.de/rec/bibtex/#1.xml}
  {dblp:#1}}
\def\mn@eprint@#1:#2:#3:#4\@nil{\def\@tempa {#1}\def\@tempb {#2}\def\@tempc
  {#3}\ifx \@tempc \@empty \let \@tempc \@tempb \let \@tempb \@tempa \fi \ifx
  \@tempb \@empty \def\@tempb {arXiv}\fi \@ifundefined
  {mn@eprint@\@tempb}{\@tempb:\@tempc}{\expandafter \expandafter \csname
  mn@eprint@\@tempb\endcsname \expandafter{\@tempc}}}

\bibitem[\protect\citeauthoryear{{Abazajian} et~al.,}{{Abazajian}
  et~al.}{2009}]{aba09}
{Abazajian} K.~N.,  et~al., 2009, \mn@doi [\apjs]
  {10.1088/0067-0049/182/2/543}, \href
  {https://ui.adsabs.harvard.edu/abs/2009ApJS..182..543A} {182, 543}

\bibitem[\protect\citeauthoryear{{Agostino} \& {Salim}}{{Agostino} \&
  {Salim}}{2019}]{ago19}
{Agostino} C.~J.,  {Salim} S.,  2019, \mn@doi [\apj]
  {10.3847/1538-4357/ab1094}, \href
  {https://ui.adsabs.harvard.edu/abs/2019ApJ...876...12A} {876, 12}

\bibitem[\protect\citeauthoryear{{Allan}, {Groh}, {Mehner}, {Smith}, {Boian},
  {Farrell}  \& {Andrews}}{{Allan} et~al.}{2020}]{all20}
{Allan} A.~P.,  {Groh} J.~H.,  {Mehner} A.,  {Smith} N.,  {Boian} I.,
  {Farrell} E.~J.,   {Andrews} J.~E.,  2020, \mn@doi [\mnras]
  {10.1093/mnras/staa1629}, \href
  {https://ui.adsabs.harvard.edu/abs/2020MNRAS.496.1902A} {496, 1902}

\bibitem[\protect\citeauthoryear{{Apparao} \& {Tarafdar}}{{Apparao} \&
  {Tarafdar}}{1988}]{app88}
{Apparao} K.~M.~V.,  {Tarafdar} S.~P.,  1988, \aap, \href
  {https://ui.adsabs.harvard.edu/abs/1988A&A...192..255A} {192, 255}

\bibitem[\protect\citeauthoryear{{Arcavi} et~al.,}{{Arcavi}
  et~al.}{2017}]{arc17}
{Arcavi} I.,  et~al., 2017, \mn@doi [\nat] {10.1038/nature24030}, \href
  {https://ui.adsabs.harvard.edu/abs/2017Natur.551..210A} {551, 210}

\bibitem[\protect\citeauthoryear{{Astropy Collaboration} et~al.,}{{Astropy
  Collaboration} et~al.}{2013}]{ast13}
{Astropy Collaboration} et~al., 2013, \mn@doi [\aap]
  {10.1051/0004-6361/201322068}, \href
  {http://adsabs.harvard.edu/abs/2013A%26A...558A..33A} {558, A33}

\bibitem[\protect\citeauthoryear{{Baldassare} et~al.,}{{Baldassare}
  et~al.}{2016}]{baldassare16}
{Baldassare} V.~F.,  et~al., 2016, \mn@doi [\apj] {10.3847/0004-637X/829/1/57},
  \href {https://ui.adsabs.harvard.edu/abs/2016ApJ...829...57B} {829, 57}

\bibitem[\protect\citeauthoryear{{Baldwin}, {Phillips}  \&
  {Terlevich}}{{Baldwin} et~al.}{1981}]{bal81}
{Baldwin} J.~A.,  {Phillips} M.~M.,   {Terlevich} R.,  1981, \mn@doi [\pasp]
  {10.1086/130766}, \href
  {https://ui.adsabs.harvard.edu/abs/1981PASP...93....5B} {93, 5}

\bibitem[\protect\citeauthoryear{{Banerjee}, {Mathew}, {Paul}, {Subramaniam},
  {Bhattacharyya}  \& {Anusha}}{{Banerjee} et~al.}{2021}]{ban21}
{Banerjee} G.,  {Mathew} B.,  {Paul} K.~T.,  {Subramaniam} A.,  {Bhattacharyya}
  S.,   {Anusha} R.,  2021, \mn@doi [\mnras] {10.1093/mnras/staa3469}, \href
  {https://ui.adsabs.harvard.edu/abs/2021MNRAS.500.3926B} {500, 3926}

\bibitem[\protect\citeauthoryear{Barbary}{Barbary}{2016}]{bar16}
Barbary K.,  2016, extinction v0.3.0, \mn@doi{10.5281/zenodo.804967}, \url
  {https://doi.org/10.5281/zenodo.804967}

\bibitem[\protect\citeauthoryear{{Bellm} et~al.,}{{Bellm} et~al.}{2019}]{bel19}
{Bellm} E.~C.,  et~al., 2019, \mn@doi [\pasp] {10.1088/1538-3873/aaecbe}, \href
  {https://ui.adsabs.harvard.edu/abs/2019PASP..131a8002B} {131, 018002}

\bibitem[\protect\citeauthoryear{{Bertin}}{{Bertin}}{2011}]{ber11}
{Bertin} E.,  2011, in {Evans} I.~N.,  {Accomazzi} A.,  {Mink} D.~J.,   {Rots}
  A.~H.,  eds,  Astronomical Society of the Pacific Conference Series Vol. 442,
  Astronomical Data Analysis Software and Systems XX. p.~435

\bibitem[\protect\citeauthoryear{{Bertin} \& {Arnouts}}{{Bertin} \&
  {Arnouts}}{1996}]{ber96}
{Bertin} E.,  {Arnouts} S.,  1996, \mn@doi [\aaps] {10.1051/aas:1996164}, \href
  {http://ads.nao.ac.jp/abs/1996A%26AS..117..393B} {117, 393}

\bibitem[\protect\citeauthoryear{{Birchall}, {Watson}  \& {Aird}}{{Birchall}
  et~al.}{2020}]{bir20}
{Birchall} K.~L.,  {Watson} M.~G.,   {Aird} J.,  2020, \mn@doi [\mnras]
  {10.1093/mnras/staa040}, \href
  {https://ui.adsabs.harvard.edu/abs/2020MNRAS.492.2268B} {492, 2268}

\bibitem[\protect\citeauthoryear{{Blecha} \& {Loeb}}{{Blecha} \&
  {Loeb}}{2008}]{ble08}
{Blecha} L.,  {Loeb} A.,  2008, \mn@doi [\mnras]
  {10.1111/j.1365-2966.2008.13790.x}, \href
  {https://ui.adsabs.harvard.edu/abs/2008MNRAS.390.1311B} {390, 1311}

\bibitem[\protect\citeauthoryear{{Boroson} \& {Green}}{{Boroson} \&
  {Green}}{1992}]{bor92}
{Boroson} T.~A.,  {Green} R.~F.,  1992, \mn@doi [\apjs] {10.1086/191661}, \href
  {http://ads.nao.ac.jp/abs/1992ApJS...80..109B} {80, 109}

\bibitem[\protect\citeauthoryear{{Breeveld}, {Landsman}, {Holland}, {Roming},
  {Kuin}  \& {Page}}{{Breeveld} et~al.}{2011}]{bre11}
{Breeveld} A.~A.,  {Landsman} W.,  {Holland} S.~T.,  {Roming} P.,  {Kuin}
  N.~P.~M.,   {Page} M.~J.,  2011, in {McEnery} J.~E.,  {Racusin} J.~L.,
  {Gehrels} N.,  eds,  American Institute of Physics Conference Series Vol.
  1358, American Institute of Physics Conference Series. pp 373--376
  (\mn@eprint {arXiv} {1102.4717}), \mn@doi{10.1063/1.3621807}

\bibitem[\protect\citeauthoryear{{Burke} et~al.,}{{Burke} et~al.}{2020}]{bur20}
{Burke} C.~J.,  et~al., 2020, \mn@doi [\apjl] {10.3847/2041-8213/ab88de}, \href
  {https://ui.adsabs.harvard.edu/abs/2020ApJ...894L...5B} {894, L5}

\bibitem[\protect\citeauthoryear{{Burke}, {Liu}, {Chen}, {Shen}  \&
  {Guo}}{{Burke} et~al.}{2021}]{bur21}
{Burke} C.~J.,  {Liu} X.,  {Chen} Y.-C.,  {Shen} Y.,   {Guo} H.,  2021, \mn@doi
  [\mnras] {10.1093/mnras/stab912}, \href
  {https://ui.adsabs.harvard.edu/abs/2021MNRAS.504..543B} {504, 543}

\bibitem[\protect\citeauthoryear{{Chambers} et~al.,}{{Chambers}
  et~al.}{2016}]{chambers16}
{Chambers} K.~C.,  et~al., 2016, arXiv e-prints, \href
  {https://ui.adsabs.harvard.edu/abs/2016arXiv161205560C} {p. arXiv:1612.05560}

\bibitem[\protect\citeauthoryear{{Chevalier}}{{Chevalier}}{1982}]{che82}
{Chevalier} R.~A.,  1982, \mn@doi [\apj] {10.1086/160126}, \href
  {https://ui.adsabs.harvard.edu/abs/1982ApJ...258..790C} {258, 790}

\bibitem[\protect\citeauthoryear{{Chevalier}, {Fransson}  \&
  {Nymark}}{{Chevalier} et~al.}{2006}]{che06}
{Chevalier} R.~A.,  {Fransson} C.,   {Nymark} T.~K.,  2006, \mn@doi [\apj]
  {10.1086/500528}, \href
  {https://ui.adsabs.harvard.edu/abs/2006ApJ...641.1029C} {641, 1029}

\bibitem[\protect\citeauthoryear{{Chugai}}{{Chugai}}{2001}]{chu01}
{Chugai} N.~N.,  2001, \mn@doi [\mnras] {10.1111/j.1365-2966.2001.04717.x},
  \href {https://ui.adsabs.harvard.edu/abs/2001MNRAS.326.1448C} {326, 1448}

\bibitem[\protect\citeauthoryear{{Damineli}}{{Damineli}}{1996}]{dam96}
{Damineli} A.,  1996, \mn@doi [\apjl] {10.1086/309961}, \href
  {https://ui.adsabs.harvard.edu/abs/1996ApJ...460L..49D} {460, L49}

\bibitem[\protect\citeauthoryear{{Davidson}}{{Davidson}}{2020}]{dav20}
{Davidson} K.,  2020, \mn@doi [Galaxies] {10.3390/galaxies8010010}, \href
  {https://ui.adsabs.harvard.edu/abs/2020Galax...8...10D} {8, 10}

\bibitem[\protect\citeauthoryear{{Davidson} \& {Humphreys}}{{Davidson} \&
  {Humphreys}}{1997}]{dav97}
{Davidson} K.,  {Humphreys} R.~M.,  1997, \mn@doi [\araa]
  {10.1146/annurev.astro.35.1.1}, \href
  {https://ui.adsabs.harvard.edu/abs/1997ARA&A..35....1D} {35, 1}

\bibitem[\protect\citeauthoryear{{Dessart}, {Hillier}, {Gezari}, {Basa}  \&
  {Matheson}}{{Dessart} et~al.}{2009}]{des09}
{Dessart} L.,  {Hillier} D.~J.,  {Gezari} S.,  {Basa} S.,   {Matheson} T.,
  2009, \mn@doi [\mnras] {10.1111/j.1365-2966.2008.14042.x}, \href
  {https://ui.adsabs.harvard.edu/abs/2009MNRAS.394...21D} {394, 21}

\bibitem[\protect\citeauthoryear{{Dey} et~al.,}{{Dey} et~al.}{2019}]{dey19}
{Dey} A.,  et~al., 2019, \mn@doi [\aj] {10.3847/1538-3881/ab089d}, \href
  {https://ui.adsabs.harvard.edu/abs/2019AJ....157..168D} {157, 168}

\bibitem[\protect\citeauthoryear{{Dom{\'\i}nguez} et~al.,}{{Dom{\'\i}nguez}
  et~al.}{2013}]{dom13}
{Dom{\'\i}nguez} A.,  et~al., 2013, \mn@doi [\apj]
  {10.1088/0004-637X/763/2/145}, \href
  {https://ui.adsabs.harvard.edu/abs/2013ApJ...763..145D} {763, 145}

\bibitem[\protect\citeauthoryear{{Dotan} \& {Shaviv}}{{Dotan} \&
  {Shaviv}}{2012}]{dot12}
{Dotan} C.,  {Shaviv} N.~J.,  2012, \mn@doi [\mnras]
  {10.1111/j.1365-2966.2012.22020.x}, \href
  {https://ui.adsabs.harvard.edu/abs/2012MNRAS.427.3071D} {427, 3071}

\bibitem[\protect\citeauthoryear{{Draine} \& {Lee}}{{Draine} \&
  {Lee}}{1984}]{dra84}
{Draine} B.~T.,  {Lee} H.~M.,  1984, \mn@doi [\apj] {10.1086/162480}, \href
  {http://adsabs.harvard.edu/abs/1984ApJ...285...89D} {285, 89}

\bibitem[\protect\citeauthoryear{{Dwarkadas} \& {Gruszko}}{{Dwarkadas} \&
  {Gruszko}}{2012}]{dwa12}
{Dwarkadas} V.~V.,  {Gruszko} J.,  2012, \mn@doi [\mnras]
  {10.1111/j.1365-2966.2011.19808.x}, \href
  {https://ui.adsabs.harvard.edu/abs/2012MNRAS.419.1515D} {419, 1515}

\bibitem[\protect\citeauthoryear{{Ekstr{\"o}m} et~al.,}{{Ekstr{\"o}m}
  et~al.}{2012}]{eks12}
{Ekstr{\"o}m} S.,  et~al., 2012, \mn@doi [\aap] {10.1051/0004-6361/201117751},
  \href {https://ui.adsabs.harvard.edu/abs/2012A&A...537A.146E} {537, A146}

\bibitem[\protect\citeauthoryear{{Elvis} et~al.,}{{Elvis} et~al.}{1994}]{elv94}
{Elvis} M.,  et~al., 1994, \mn@doi [\apjs] {10.1086/192093}, \href
  {https://ui.adsabs.harvard.edu/abs/1994ApJS...95....1E} {95, 1}

\bibitem[\protect\citeauthoryear{{Filippenko}}{{Filippenko}}{1989}]{fil89}
{Filippenko} A.~V.,  1989, \mn@doi [\aj] {10.1086/115018}, \href
  {https://ui.adsabs.harvard.edu/abs/1989AJ.....97..726F} {97, 726}

\bibitem[\protect\citeauthoryear{{Fitzpatrick}}{{Fitzpatrick}}{1999}]{fit99}
{Fitzpatrick} E.~L.,  1999, \mn@doi [\pasp] {10.1086/316293}, \href
  {http://adsabs.harvard.edu/abs/1999PASP..111...63F} {111, 63}

\bibitem[\protect\citeauthoryear{{Foley}, {Berger}, {Fox}, {Levesque},
  {Challis}, {Ivans}, {Rhoads}  \& {Soderberg}}{{Foley} et~al.}{2011}]{fol11}
{Foley} R.~J.,  {Berger} E.,  {Fox} O.,  {Levesque} E.~M.,  {Challis} P.~J.,
  {Ivans} I.~I.,  {Rhoads} J.~E.,   {Soderberg} A.~M.,  2011, \mn@doi [\apj]
  {10.1088/0004-637X/732/1/32}, \href
  {https://ui.adsabs.harvard.edu/abs/2011ApJ...732...32F} {732, 32}

\bibitem[\protect\citeauthoryear{{Fransson} et~al.,}{{Fransson}
  et~al.}{2022}]{fra22}
{Fransson} C.,  et~al., 2022, arXiv e-prints, \href
  {https://ui.adsabs.harvard.edu/abs/2022arXiv220606497F} {p. arXiv:2206.06497}

\bibitem[\protect\citeauthoryear{{Frew}}{{Frew}}{2004}]{fre04}
{Frew} D.~J.,  2004, Journal of Astronomical Data, \href
  {https://ui.adsabs.harvard.edu/abs/2004JAD....10....6F} {10, 6}

\bibitem[\protect\citeauthoryear{{Fruscione} et~al.,}{{Fruscione}
  et~al.}{2006}]{fru06}
{Fruscione} A.,  et~al., 2006, in {Silva} D.~R.,  {Doxsey} R.~E.,  eds,
  Society of Photo-Optical Instrumentation Engineers (SPIE) Conference Series
  Vol. 6270, Society of Photo-Optical Instrumentation Engineers (SPIE)
  Conference Series. p. 62701V, \mn@doi{10.1117/12.671760}

\bibitem[\protect\citeauthoryear{{Gabriel} et~al.,}{{Gabriel}
  et~al.}{2004}]{gab04}
{Gabriel} C.,  et~al., 2004, in {Ochsenbein} F.,  {Allen} M.~G.,   {Egret} D.,
  eds,  Astronomical Society of the Pacific Conference Series Vol. 314,
  Astronomical Data Analysis Software and Systems (ADASS) XIII. p.~759

\bibitem[\protect\citeauthoryear{{Gaia Collaboration} et~al.,}{{Gaia
  Collaboration} et~al.}{2016}]{gai16}
{Gaia Collaboration} et~al., 2016, \mn@doi [\aap]
  {10.1051/0004-6361/201629272}, \href
  {https://ui.adsabs.harvard.edu/abs/2016A&A...595A...1G} {595, A1}

\bibitem[\protect\citeauthoryear{{Gal-Yam} \& {Leonard}}{{Gal-Yam} \&
  {Leonard}}{2009}]{gal09}
{Gal-Yam} A.,  {Leonard} D.~C.,  2009, \mn@doi [\nat] {10.1038/nature07934},
  \href {https://ui.adsabs.harvard.edu/abs/2009Natur.458..865G} {458, 865}

\bibitem[\protect\citeauthoryear{{Gehrels} et~al.,}{{Gehrels}
  et~al.}{2004}]{geh04}
{Gehrels} N.,  et~al., 2004, \mn@doi [\apj] {10.1086/422091}, \href
  {https://ui.adsabs.harvard.edu/abs/2004ApJ...611.1005G} {611, 1005}

\bibitem[\protect\citeauthoryear{{Gibson}, {Brandt}  \& {Schneider}}{{Gibson}
  et~al.}{2008}]{gib08}
{Gibson} R.~R.,  {Brandt} W.~N.,   {Schneider} D.~P.,  2008, \mn@doi [\apj]
  {10.1086/590403}, \href
  {https://ui.adsabs.harvard.edu/abs/2008ApJ...685..773G} {685, 773}

\bibitem[\protect\citeauthoryear{{Graham} et~al.,}{{Graham}
  et~al.}{2014}]{graham14}
{Graham} M.~L.,  et~al., 2014, \mn@doi [\apj] {10.1088/0004-637X/787/2/163},
  \href {https://ui.adsabs.harvard.edu/abs/2014ApJ...787..163G} {787, 163}

\bibitem[\protect\citeauthoryear{{Graham}, {Djorgovski}, {Drake}, {Stern},
  {Mahabal}, {Glikman}, {Larson}  \& {Christensen}}{{Graham}
  et~al.}{2017}]{gra17}
{Graham} M.~J.,  {Djorgovski} S.~G.,  {Drake} A.~J.,  {Stern} D.,  {Mahabal}
  A.~A.,  {Glikman} E.,  {Larson} S.,   {Christensen} E.,  2017, \mn@doi
  [\mnras] {10.1093/mnras/stx1456}, \href
  {https://ui.adsabs.harvard.edu/abs/2017MNRAS.470.4112G} {470, 4112}

\bibitem[\protect\citeauthoryear{{Graham} et~al.,}{{Graham}
  et~al.}{2019}]{gra19}
{Graham} M.~J.,  et~al., 2019, \mn@doi [\pasp] {10.1088/1538-3873/ab006c},
  \href {https://ui.adsabs.harvard.edu/abs/2019PASP..131g8001G} {131, 078001}

\bibitem[\protect\citeauthoryear{{Greene} \& {Ho}}{{Greene} \&
  {Ho}}{2007}]{gre07}
{Greene} J.~E.,  {Ho} L.~C.,  2007, \mn@doi [\apj] {10.1086/522082}, \href
  {https://ui.adsabs.harvard.edu/abs/2007ApJ...670...92G} {670, 92}

\bibitem[\protect\citeauthoryear{{Grossov{\'a}} et~al.,}{{Grossov{\'a}}
  et~al.}{2022}]{gro22}
{Grossov{\'a}} R.,  et~al., 2022, \mn@doi [\apjs] {10.3847/1538-4365/ac366c},
  \href {https://ui.adsabs.harvard.edu/abs/2022ApJS..258...30G} {258, 30}

\bibitem[\protect\citeauthoryear{HEASARC}{HEASARC}{2014}]{nasa14}
HEASARC 2014, {HEAsoft: Unified Release of FTOOLS and XANADU} (\mn@eprint
  {ascl} {1408.004})

\bibitem[\protect\citeauthoryear{{Hainline}, {Reines}, {Greene}  \&
  {Stern}}{{Hainline} et~al.}{2016}]{hai16}
{Hainline} K.~N.,  {Reines} A.~E.,  {Greene} J.~E.,   {Stern} D.,  2016,
  \mn@doi [\apj] {10.3847/0004-637X/832/2/119}, \href
  {https://ui.adsabs.harvard.edu/abs/2016ApJ...832..119H} {832, 119}

\bibitem[\protect\citeauthoryear{{Hall}}{{Hall}}{2007}]{hall07}
{Hall} P.~B.,  2007, \mn@doi [\aj] {10.1086/511272}, \href
  {https://ui.adsabs.harvard.edu/abs/2007AJ....133.1271H} {133, 1271}

\bibitem[\protect\citeauthoryear{{Hall} et~al.,}{{Hall} et~al.}{2002}]{hal02}
{Hall} P.~B.,  et~al., 2002, \mn@doi [\apjs] {10.1086/340546}, \href
  {https://ui.adsabs.harvard.edu/abs/2002ApJS..141..267H} {141, 267}

\bibitem[\protect\citeauthoryear{Harris et~al.,}{Harris et~al.}{2020}]{num20}
Harris C.~R.,  et~al., 2020, \mn@doi [Nature] {10.1038/s41586-020-2649-2}, 585,
  357–362

\bibitem[\protect\citeauthoryear{{Harutyunyan} et~al.,}{{Harutyunyan}
  et~al.}{2008}]{harutyunyan08}
{Harutyunyan} A.~H.,  et~al., 2008, \mn@doi [\aap]
  {10.1051/0004-6361:20078859}, \href
  {https://ui.adsabs.harvard.edu/abs/2008A&A...488..383H} {488, 383}

\bibitem[\protect\citeauthoryear{{Hasinger}}{{Hasinger}}{2008}]{has08}
{Hasinger} G.,  2008, \mn@doi [\aap] {10.1051/0004-6361:200809839}, \href
  {https://ui.adsabs.harvard.edu/abs/2008A%26A...490..905H} {490, 905}

\bibitem[\protect\citeauthoryear{{Heger}, {Fryer}, {Woosley}, {Langer}  \&
  {Hartmann}}{{Heger} et~al.}{2003}]{heg03}
{Heger} A.,  {Fryer} C.~L.,  {Woosley} S.~E.,  {Langer} N.,   {Hartmann} D.~H.,
   2003, \mn@doi [\apj] {10.1086/375341}, \href
  {https://ui.adsabs.harvard.edu/abs/2003ApJ...591..288H} {591, 288}

\bibitem[\protect\citeauthoryear{{Hirai}, {Podsiadlowski}, {Owocki},
  {Schneider}  \& {Smith}}{{Hirai} et~al.}{2021}]{hir21}
{Hirai} R.,  {Podsiadlowski} P.,  {Owocki} S.~P.,  {Schneider} F. R.~N.,
  {Smith} N.,  2021, \mn@doi [\mnras] {10.1093/mnras/stab571}, \href
  {https://ui.adsabs.harvard.edu/abs/2021MNRAS.503.4276H} {503, 4276}

\bibitem[\protect\citeauthoryear{{Ho}, {Filippenko}  \& {Sargent}}{{Ho}
  et~al.}{1997}]{ho97b}
{Ho} L.~C.,  {Filippenko} A.~V.,   {Sargent} W. L.~W.,  1997, \mn@doi [\apj]
  {10.1086/304638}, \href
  {https://ui.adsabs.harvard.edu/abs/1997ApJ...487..568H} {487, 568}

\bibitem[\protect\citeauthoryear{{Humphreys}}{{Humphreys}}{2005}]{hum05}
{Humphreys} R.~M.,  2005, in {Humphreys} R.,  {Stanek} K.,  eds,  Astronomical
  Society of the Pacific Conference Series Vol. 332, The Fate of the Most
  Massive Stars. p.~14

\bibitem[\protect\citeauthoryear{{Humphreys} \& {Davidson}}{{Humphreys} \&
  {Davidson}}{1979}]{hum79}
{Humphreys} R.~M.,  {Davidson} K.,  1979, \mn@doi [\apj] {10.1086/157301},
  \href {https://ui.adsabs.harvard.edu/abs/1979ApJ...232..409H} {232, 409}

\bibitem[\protect\citeauthoryear{{Humphreys} \& {Davidson}}{{Humphreys} \&
  {Davidson}}{1994}]{hum94}
{Humphreys} R.~M.,  {Davidson} K.,  1994, \mn@doi [\pasp] {10.1086/133478},
  \href {https://ui.adsabs.harvard.edu/abs/1994PASP..106.1025H} {106, 1025}

\bibitem[\protect\citeauthoryear{{Humphreys}, {Davidson}  \&
  {Smith}}{{Humphreys} et~al.}{1999}]{hum99}
{Humphreys} R.~M.,  {Davidson} K.,   {Smith} N.,  1999, \mn@doi [\pasp]
  {10.1086/316420}, \href
  {https://ui.adsabs.harvard.edu/abs/1999PASP..111.1124H} {111, 1124}

\bibitem[\protect\citeauthoryear{{Humphreys}, {Davidson}, {Grammer},
  {Kneeland}, {Martin}, {Weis}  \& {Burggraf}}{{Humphreys}
  et~al.}{2013}]{hum13}
{Humphreys} R.~M.,  {Davidson} K.,  {Grammer} S.,  {Kneeland} N.,  {Martin}
  J.~C.,  {Weis} K.,   {Burggraf} B.,  2013, \mn@doi [\apj]
  {10.1088/0004-637X/773/1/46}, \href
  {https://ui.adsabs.harvard.edu/abs/2013ApJ...773...46H} {773, 46}

\bibitem[\protect\citeauthoryear{{Humphreys}, {Weis}, {Davidson}, {Bomans}  \&
  {Burggraf}}{{Humphreys} et~al.}{2014}]{hum14}
{Humphreys} R.~M.,  {Weis} K.,  {Davidson} K.,  {Bomans} D.~J.,   {Burggraf}
  B.,  2014, \mn@doi [\apj] {10.1088/0004-637X/790/1/48}, \href
  {https://ui.adsabs.harvard.edu/abs/2014ApJ...790...48H} {790, 48}

\bibitem[\protect\citeauthoryear{{Izotov} \& {Thuan}}{{Izotov} \&
  {Thuan}}{2008}]{izo08}
{Izotov} Y.~I.,  {Thuan} T.~X.,  2008, \mn@doi [\apj] {10.1086/591660}, \href
  {https://ui.adsabs.harvard.edu/abs/2008ApJ...687..133I} {687, 133}

\bibitem[\protect\citeauthoryear{{Izotov} \& {Thuan}}{{Izotov} \&
  {Thuan}}{2009}]{izo09}
{Izotov} Y.~I.,  {Thuan} T.~X.,  2009, \mn@doi [\apj]
  {10.1088/0004-637X/690/2/1797}, \href
  {https://ui.adsabs.harvard.edu/abs/2009ApJ...690.1797I} {690, 1797}

\bibitem[\protect\citeauthoryear{{Izotov}, {Thuan}  \& {Guseva}}{{Izotov}
  et~al.}{2007}]{izo07}
{Izotov} Y.~I.,  {Thuan} T.~X.,   {Guseva} N.~G.,  2007, \mn@doi [\apj]
  {10.1086/522923}, \href
  {https://ui.adsabs.harvard.edu/abs/2007ApJ...671.1297I} {671, 1297}

\bibitem[\protect\citeauthoryear{{Jansen} et~al.,}{{Jansen}
  et~al.}{2001}]{jan01}
{Jansen} F.,  et~al., 2001, \mn@doi [\aap] {10.1051/0004-6361:20000036}, \href
  {https://ui.adsabs.harvard.edu/abs/2001A&A...365L...1J} {365, L1}

\bibitem[\protect\citeauthoryear{{Kashi} \& {Soker}}{{Kashi} \&
  {Soker}}{2010}]{kas10}
{Kashi} A.,  {Soker} N.,  2010, \mn@doi [\apj] {10.1088/0004-637X/723/1/602},
  \href {https://ui.adsabs.harvard.edu/abs/2010ApJ...723..602K} {723, 602}

\bibitem[\protect\citeauthoryear{{Kauffmann} et~al.,}{{Kauffmann}
  et~al.}{2003}]{kau03}
{Kauffmann} G.,  et~al., 2003, \mn@doi [\mnras]
  {10.1111/j.1365-2966.2003.07154.x}, \href
  {https://ui.adsabs.harvard.edu/abs/2003MNRAS.346.1055K} {346, 1055}

\bibitem[\protect\citeauthoryear{{Keel} et~al.,}{{Keel} et~al.}{2012}]{kee12}
{Keel} W.~C.,  et~al., 2012, \mn@doi [\mnras]
  {10.1111/j.1365-2966.2011.20101.x}, \href
  {https://ui.adsabs.harvard.edu/abs/2012MNRAS.420..878K} {420, 878}

\bibitem[\protect\citeauthoryear{{Kewley}, {Dopita}, {Sutherland}, {Heisler}
  \& {Trevena}}{{Kewley} et~al.}{2001}]{kew01}
{Kewley} L.~J.,  {Dopita} M.~A.,  {Sutherland} R.~S.,  {Heisler} C.~A.,
  {Trevena} J.,  2001, \mn@doi [\apj] {10.1086/321545}, \href
  {https://ui.adsabs.harvard.edu/abs/2001ApJ...556..121K} {556, 121}

\bibitem[\protect\citeauthoryear{{Kewley}, {Groves}, {Kauffmann}  \&
  {Heckman}}{{Kewley} et~al.}{2006}]{kew06}
{Kewley} L.~J.,  {Groves} B.,  {Kauffmann} G.,   {Heckman} T.,  2006, \mn@doi
  [\mnras] {10.1111/j.1365-2966.2006.10859.x}, \href
  {https://ui.adsabs.harvard.edu/abs/2006MNRAS.372..961K} {372, 961}

\bibitem[\protect\citeauthoryear{{Kewley}, {Dopita}, {Leitherer}, {Dav{\'e}},
  {Yuan}, {Allen}, {Groves}  \& {Sutherland}}{{Kewley} et~al.}{2013}]{kew13}
{Kewley} L.~J.,  {Dopita} M.~A.,  {Leitherer} C.,  {Dav{\'e}} R.,  {Yuan} T.,
  {Allen} M.,  {Groves} B.,   {Sutherland} R.,  2013, \mn@doi [\apj]
  {10.1088/0004-637X/774/2/100}, \href
  {https://ui.adsabs.harvard.edu/abs/2013ApJ...774..100K} {774, 100}

\bibitem[\protect\citeauthoryear{{Kilpatrick} et~al.,}{{Kilpatrick}
  et~al.}{2018}]{kil18}
{Kilpatrick} C.~D.,  et~al., 2018, \mn@doi [\mnras] {10.1093/mnras/stx2675},
  \href {https://ui.adsabs.harvard.edu/abs/2018MNRAS.473.4805K} {473, 4805}

\bibitem[\protect\citeauthoryear{{Kokubo}}{{Kokubo}}{2017}]{kok17}
{Kokubo} M.,  2017, \mn@doi [\mnras] {10.1093/mnras/stx080}, \href
  {http://ads.nao.ac.jp/abs/2017MNRAS.467.3723K} {467, 3723}

\bibitem[\protect\citeauthoryear{{Kokubo} et~al.,}{{Kokubo}
  et~al.}{2019}]{kok19}
{Kokubo} M.,  et~al., 2019, \mn@doi [\apj] {10.3847/1538-4357/aaff6b}, \href
  {https://ui.adsabs.harvard.edu/abs/2019ApJ...872..135K} {872, 135}

\bibitem[\protect\citeauthoryear{{Komossa}}{{Komossa}}{2012}]{kom12}
{Komossa} S.,  2012, \mn@doi [Advances in Astronomy] {10.1155/2012/364973},
  \href {https://ui.adsabs.harvard.edu/abs/2012AdAst2012E..14K} {2012, 364973}

\bibitem[\protect\citeauthoryear{{Koss} et~al.,}{{Koss} et~al.}{2011}]{kos11}
{Koss} M.,  et~al., 2011, \mn@doi [\apjl] {10.1088/2041-8205/735/2/L42}, \href
  {https://ui.adsabs.harvard.edu/abs/2011ApJ...735L..42K} {735, L42}

\bibitem[\protect\citeauthoryear{{Koss} et~al.,}{{Koss} et~al.}{2014}]{koss14}
{Koss} M.,  et~al., 2014, \mn@doi [\mnras] {10.1093/mnras/stu1673}, \href
  {https://ui.adsabs.harvard.edu/abs/2014MNRAS.445..515K} {445, 515}

\bibitem[\protect\citeauthoryear{{Laher} et~al.,}{{Laher} et~al.}{2014}]{lah14}
{Laher} R.~R.,  et~al., 2014, \mn@doi [\pasp] {10.1086/677351}, \href
  {https://ui.adsabs.harvard.edu/abs/2014PASP..126..674L} {126, 674}

\bibitem[\protect\citeauthoryear{{Langer}}{{Langer}}{2012}]{lan12}
{Langer} N.,  2012, \mn@doi [\araa] {10.1146/annurev-astro-081811-125534},
  \href {https://ui.adsabs.harvard.edu/abs/2012ARA&A..50..107L} {50, 107}

\bibitem[\protect\citeauthoryear{{Law} et~al.,}{{Law} et~al.}{2009}]{law09}
{Law} N.~M.,  et~al., 2009, \mn@doi [\pasp] {10.1086/648598}, \href
  {https://ui.adsabs.harvard.edu/abs/2009PASP..121.1395L} {121, 1395}

\bibitem[\protect\citeauthoryear{{Lehmer} et~al.,}{{Lehmer}
  et~al.}{2016}]{leh16}
{Lehmer} B.~D.,  et~al., 2016, \mn@doi [\apj] {10.3847/0004-637X/825/1/7},
  \href {https://ui.adsabs.harvard.edu/abs/2016ApJ...825....7L} {825, 7}

\bibitem[\protect\citeauthoryear{{Liu}, {Luo}, {Brandt}, {Gallagher}  \&
  {Garmire}}{{Liu} et~al.}{2018}]{liu18b}
{Liu} H.,  {Luo} B.,  {Brandt} W.~N.,  {Gallagher} S.~C.,   {Garmire} G.~P.,
  2018, \mn@doi [\apj] {10.3847/1538-4357/aabe8d}, \href
  {https://ui.adsabs.harvard.edu/abs/2018ApJ...859..113L} {859, 113}

\bibitem[\protect\citeauthoryear{{L{\'o}pez-Corredoira} \&
  {Guti{\'e}rrez}}{{L{\'o}pez-Corredoira} \& {Guti{\'e}rrez}}{2006}]{lop06}
{L{\'o}pez-Corredoira} M.,  {Guti{\'e}rrez} C.~M.,  2006, in {Lerner} E.~J.,
  {Almeida} J.~B.,  eds,  American Institute of Physics Conference Series Vol.
  822, First Crisis in Cosmology Conference. pp 75--92 (\mn@eprint {arXiv}
  {astro-ph/0509630}), \mn@doi{10.1063/1.2189124}

\bibitem[\protect\citeauthoryear{{MacLeod} et~al.,}{{MacLeod}
  et~al.}{2010}]{mac10}
{MacLeod} C.~L.,  et~al., 2010, \mn@doi [\apj] {10.1088/0004-637X/721/2/1014},
  \href {http://adsabs.harvard.edu/abs/2010ApJ...721.1014M} {721, 1014}

\bibitem[\protect\citeauthoryear{{MacLeod} et~al.,}{{MacLeod}
  et~al.}{2012}]{mac12}
{MacLeod} C.~L.,  et~al., 2012, \mn@doi [\apj] {10.1088/0004-637X/753/2/106},
  \href {http://adsabs.harvard.edu/abs/2012ApJ...753..106M} {753, 106}

\bibitem[\protect\citeauthoryear{{Magnier} et~al.,}{{Magnier}
  et~al.}{2020}]{magnier20}
{Magnier} E.~A.,  et~al., 2020, \mn@doi [\apjs] {10.3847/1538-4365/abb82a},
  \href {https://ui.adsabs.harvard.edu/abs/2020ApJS..251....6M} {251, 6}

\bibitem[\protect\citeauthoryear{{Margutti} et~al.,}{{Margutti}
  et~al.}{2014}]{margutti14}
{Margutti} R.,  et~al., 2014, \mn@doi [\apj] {10.1088/0004-637X/780/1/21},
  \href {https://ui.adsabs.harvard.edu/abs/2014ApJ...780...21M} {780, 21}

\bibitem[\protect\citeauthoryear{{Marinello}, {Rodr{\'\i}guez-Ardila},
  {Garcia-Rissmann}, {Sigut}  \& {Pradhan}}{{Marinello}
  et~al.}{2016}]{marinello16}
{Marinello} M.,  {Rodr{\'\i}guez-Ardila} A.,  {Garcia-Rissmann} A.,  {Sigut}
  T.~A.~A.,   {Pradhan} A.~K.,  2016, \mn@doi [\apj]
  {10.3847/0004-637X/820/2/116}, \href
  {https://ui.adsabs.harvard.edu/abs/2016ApJ...820..116M} {820, 116}

\bibitem[\protect\citeauthoryear{{Masci} et~al.,}{{Masci} et~al.}{2019}]{mas19}
{Masci} F.~J.,  et~al., 2019, \mn@doi [\pasp] {10.1088/1538-3873/aae8ac}, \href
  {https://ui.adsabs.harvard.edu/abs/2019PASP..131a8003M} {131, 018003}

\bibitem[\protect\citeauthoryear{{Masci et al.}}{{Masci et al.}}{2020}]{mas20}
{Masci et al.} F.~J.,  2020, The ZTF Science Data System (ZSDS) Explanatory
  Supplement, Version 5

\bibitem[\protect\citeauthoryear{{Mason} et~al.,}{{Mason} et~al.}{2001}]{mas01}
{Mason} K.~O.,  et~al., 2001, \mn@doi [\aap] {10.1051/0004-6361:20000044},
  \href {https://ui.adsabs.harvard.edu/abs/2001A&A...365L..36M} {365, L36}

\bibitem[\protect\citeauthoryear{{Mauerhan} et~al.,}{{Mauerhan}
  et~al.}{2013}]{mau13}
{Mauerhan} J.~C.,  et~al., 2013, \mn@doi [\mnras] {10.1093/mnras/stt009}, \href
  {https://ui.adsabs.harvard.edu/abs/2013MNRAS.430.1801M} {430, 1801}

\bibitem[\protect\citeauthoryear{{Merloni}, {Heinz}  \& {di Matteo}}{{Merloni}
  et~al.}{2003}]{mer03}
{Merloni} A.,  {Heinz} S.,   {di Matteo} T.,  2003, \mn@doi [\mnras]
  {10.1046/j.1365-2966.2003.07017.x}, \href
  {https://ui.adsabs.harvard.edu/abs/2003MNRAS.345.1057M} {345, 1057}

\bibitem[\protect\citeauthoryear{{Moriya}, {Blinnikov}, {Tominaga}, {Yoshida},
  {Tanaka}, {Maeda}  \& {Nomoto}}{{Moriya} et~al.}{2013a}]{mor13b}
{Moriya} T.~J.,  {Blinnikov} S.~I.,  {Tominaga} N.,  {Yoshida} N.,  {Tanaka}
  M.,  {Maeda} K.,   {Nomoto} K.,  2013a, \mn@doi [\mnras]
  {10.1093/mnras/sts075}, \href
  {https://ui.adsabs.harvard.edu/abs/2013MNRAS.428.1020M} {428, 1020}

\bibitem[\protect\citeauthoryear{{Moriya}, {Maeda}, {Taddia}, {Sollerman},
  {Blinnikov}  \& {Sorokina}}{{Moriya} et~al.}{2013b}]{mor13}
{Moriya} T.~J.,  {Maeda} K.,  {Taddia} F.,  {Sollerman} J.,  {Blinnikov} S.~I.,
    {Sorokina} E.~I.,  2013b, \mn@doi [\mnras] {10.1093/mnras/stt1392}, \href
  {https://ui.adsabs.harvard.edu/abs/2013MNRAS.435.1520M} {435, 1520}

\bibitem[\protect\citeauthoryear{{Moriya}, {Maeda}, {Taddia}, {Sollerman},
  {Blinnikov}  \& {Sorokina}}{{Moriya} et~al.}{2014}]{moriya14}
{Moriya} T.~J.,  {Maeda} K.,  {Taddia} F.,  {Sollerman} J.,  {Blinnikov} S.~I.,
    {Sorokina} E.~I.,  2014, \mn@doi [\mnras] {10.1093/mnras/stu163}, \href
  {https://ui.adsabs.harvard.edu/abs/2014MNRAS.439.2917M} {439, 2917}

\bibitem[\protect\citeauthoryear{{Moriya}, {Mazzali}  \& {Pian}}{{Moriya}
  et~al.}{2020}]{mor20}
{Moriya} T.~J.,  {Mazzali} P.~A.,   {Pian} E.,  2020, \mn@doi [\mnras]
  {10.1093/mnras/stz3122}, \href
  {https://ui.adsabs.harvard.edu/abs/2020MNRAS.491.1384M} {491, 1384}

\bibitem[\protect\citeauthoryear{{Morris}, {Gull}, {Hillier}, {Barlow},
  {Royer}, {Nielsen}, {Black}  \& {Swinyard}}{{Morris} et~al.}{2017}]{morris17}
{Morris} P.~W.,  {Gull} T.~R.,  {Hillier} D.~J.,  {Barlow} M.~J.,  {Royer} P.,
  {Nielsen} K.,  {Black} J.,   {Swinyard} B.,  2017, \mn@doi [\apj]
  {10.3847/1538-4357/aa71b3}, \href
  {https://ui.adsabs.harvard.edu/abs/2017ApJ...842...79M} {842, 79}

\bibitem[\protect\citeauthoryear{{Naz{\'e}}, {Rauw}  \&
  {Hutsem{\'e}kers}}{{Naz{\'e}} et~al.}{2012}]{naz12}
{Naz{\'e}} Y.,  {Rauw} G.,   {Hutsem{\'e}kers} D.,  2012, \mn@doi [\aap]
  {10.1051/0004-6361/201118040}, \href
  {https://ui.adsabs.harvard.edu/abs/2012A&A...538A..47N} {538, A47}

\bibitem[\protect\citeauthoryear{{Ofek} et~al.,}{{Ofek} et~al.}{2014}]{ofe14}
{Ofek} E.~O.,  et~al., 2014, \mn@doi [\apj] {10.1088/0004-637X/789/2/104},
  \href {https://ui.adsabs.harvard.edu/abs/2014ApJ...789..104O} {789, 104}

\bibitem[\protect\citeauthoryear{{Osterbrock}}{{Osterbrock}}{1989}]{ost89}
{Osterbrock} D.~E.,  1989, {Astrophysics of gaseous nebulae and active galactic
  nuclei}

\bibitem[\protect\citeauthoryear{{Owocki} \& {Shaviv}}{{Owocki} \&
  {Shaviv}}{2012}]{owo12}
{Owocki} S.~P.,  {Shaviv} N.~J.,  2012, {Instability \& Mass Loss near the
  Eddington Limit}.
p.~275, \mn@doi{10.1007/978-1-4614-2275-4_12}

\bibitem[\protect\citeauthoryear{{Parkin}, {Pittard}, {Corcoran}, {Hamaguchi}
  \& {Stevens}}{{Parkin} et~al.}{2009}]{par09}
{Parkin} E.~R.,  {Pittard} J.~M.,  {Corcoran} M.~F.,  {Hamaguchi} K.,
  {Stevens} I.~R.,  2009, \mn@doi [\mnras] {10.1111/j.1365-2966.2009.14475.x},
  \href {https://ui.adsabs.harvard.edu/abs/2009MNRAS.394.1758P} {394, 1758}

\bibitem[\protect\citeauthoryear{{Pastorello} \& {Fraser}}{{Pastorello} \&
  {Fraser}}{2019}]{pas19}
{Pastorello} A.,  {Fraser} M.,  2019, \mn@doi [Nature Astronomy]
  {10.1038/s41550-019-0809-9}, \href
  {https://ui.adsabs.harvard.edu/abs/2019NatAs...3..676P} {3, 676}

\bibitem[\protect\citeauthoryear{{Pastorello} et~al.,}{{Pastorello}
  et~al.}{2002}]{pas02}
{Pastorello} A.,  et~al., 2002, \mn@doi [\mnras]
  {10.1046/j.1365-8711.2002.05366.x}, \href
  {https://ui.adsabs.harvard.edu/abs/2002MNRAS.333...27P} {333, 27}

\bibitem[\protect\citeauthoryear{{Pastorello} et~al.,}{{Pastorello}
  et~al.}{2010}]{pas10}
{Pastorello} A.,  et~al., 2010, \mn@doi [\mnras]
  {10.1111/j.1365-2966.2010.17142.x}, \href
  {https://ui.adsabs.harvard.edu/abs/2010MNRAS.408..181P} {408, 181}

\bibitem[\protect\citeauthoryear{{Pastorello} et~al.,}{{Pastorello}
  et~al.}{2013}]{pas13}
{Pastorello} A.,  et~al., 2013, \mn@doi [\apj] {10.1088/0004-637X/767/1/1},
  \href {https://ui.adsabs.harvard.edu/abs/2013ApJ...767....1P} {767, 1}

\bibitem[\protect\citeauthoryear{{Patterson} et~al.,}{{Patterson}
  et~al.}{2019}]{pat19}
{Patterson} M.~T.,  et~al., 2019, \mn@doi [\pasp] {10.1088/1538-3873/aae904},
  \href {https://ui.adsabs.harvard.edu/abs/2019PASP..131a8001P} {131, 018001}

\bibitem[\protect\citeauthoryear{{Pei}}{{Pei}}{1992}]{pei92}
{Pei} Y.~C.,  1992, \mn@doi [\apj] {10.1086/171637}, \href
  {http://adsabs.harvard.edu/abs/1992ApJ...395..130P} {395, 130}

\bibitem[\protect\citeauthoryear{{Peng}, {Ho}, {Impey}  \& {Rix}}{{Peng}
  et~al.}{2011}]{pen11}
{Peng} C.~Y.,  {Ho} L.~C.,  {Impey} C.~D.,   {Rix} H.-W.,  2011, {GALFIT:
  Detailed Structural Decomposition of Galaxy Images}, Astrophysics Source Code
  Library (\mn@eprint {ascl} {1104.010})

\bibitem[\protect\citeauthoryear{{Perez-Torres}, {Piconcelli}, {Alberdi},
  {Komossa}  \& {Herrero-Illana}}{{Perez-Torres} et~al.}{2015}]{perez15}
{Perez-Torres} M.,  {Piconcelli} N. R.-O.~E.,  {Alberdi} A.,  {Komossa} S.,
  {Herrero-Illana} R.,  2015, The Astronomer's Telegram, \href
  {https://ui.adsabs.harvard.edu/abs/2015ATel.7388....1P} {7388, 1}

\bibitem[\protect\citeauthoryear{{Perley} et~al.,}{{Perley}
  et~al.}{2020}]{per20}
{Perley} D.~A.,  et~al., 2020, \mn@doi [\apj] {10.3847/1538-4357/abbd98}, \href
  {https://ui.adsabs.harvard.edu/abs/2020ApJ...904...35P} {904, 35}

\bibitem[\protect\citeauthoryear{{Persson}}{{Persson}}{1988}]{per88}
{Persson} S.~E.,  1988, \mn@doi [\apj] {10.1086/166509}, \href
  {https://ui.adsabs.harvard.edu/abs/1988ApJ...330..751P} {330, 751}

\bibitem[\protect\citeauthoryear{{Phinney}}{{Phinney}}{1989}]{phi89b}
{Phinney} E.~S.,  1989, in {Morris} M.,  ed.,  Vol. 136, The Center of the
  Galaxy. p.~543

\bibitem[\protect\citeauthoryear{{Poole} et~al.,}{{Poole} et~al.}{2008}]{poo08}
{Poole} T.~S.,  et~al., 2008, \mn@doi [\mnras]
  {10.1111/j.1365-2966.2007.12563.x}, \href
  {https://ui.adsabs.harvard.edu/abs/2008MNRAS.383..627P} {383, 627}

\bibitem[\protect\citeauthoryear{{Pursimo}, {Ighina}, {Ihanec}, {Mandarakas},
  {Skillen}  \& {Terefe}}{{Pursimo} et~al.}{2019a}]{pur19b}
{Pursimo} T.,  {Ighina} L.,  {Ihanec} N.,  {Mandarakas} N.,  {Skillen} K.,
  {Terefe} S.,  2019a, Contributions of the Astronomical Observatory Skalnate
  Pleso, \href {https://ui.adsabs.harvard.edu/abs/2019CoSka..49..539P} {49,
  539}

\bibitem[\protect\citeauthoryear{{Pursimo}, {Galindo-Guil}, {Dennefeld},
  {Ighina}, {Ihanec}, {Mandarakas}, {Skillen}  \& {Terefe}}{{Pursimo}
  et~al.}{2019b}]{pur19}
{Pursimo} T.,  {Galindo-Guil} F.,  {Dennefeld} M.,  {Ighina} L.,  {Ihanec} N.,
  {Mandarakas} N.,  {Skillen} K.,   {Terefe} S.,  2019b, The Astronomer's
  Telegram, \href {https://ui.adsabs.harvard.edu/abs/2019ATel12911....1P}
  {12911, 1}

\bibitem[\protect\citeauthoryear{{Quataert} \& {Shiode}}{{Quataert} \&
  {Shiode}}{2012}]{qua12}
{Quataert} E.,  {Shiode} J.,  2012, \mn@doi [\mnras]
  {10.1111/j.1745-3933.2012.01264.x}, \href
  {https://ui.adsabs.harvard.edu/abs/2012MNRAS.423L..92Q} {423, L92}

\bibitem[\protect\citeauthoryear{{Reines} \& {Volonteri}}{{Reines} \&
  {Volonteri}}{2015}]{rei15}
{Reines} A.~E.,  {Volonteri} M.,  2015, \mn@doi [\apj]
  {10.1088/0004-637X/813/2/82}, \href
  {https://ui.adsabs.harvard.edu/abs/2015ApJ...813...82R} {813, 82}

\bibitem[\protect\citeauthoryear{{Reines}, {Greene}  \& {Geha}}{{Reines}
  et~al.}{2013}]{rei13}
{Reines} A.~E.,  {Greene} J.~E.,   {Geha} M.,  2013, \mn@doi [\apj]
  {10.1088/0004-637X/775/2/116}, \href
  {https://ui.adsabs.harvard.edu/abs/2013ApJ...775..116R} {775, 116}

\bibitem[\protect\citeauthoryear{{Rest} et~al.,}{{Rest} et~al.}{2012}]{res12}
{Rest} A.,  et~al., 2012, \mn@doi [\nat] {10.1038/nature10775}, \href
  {https://ui.adsabs.harvard.edu/abs/2012Natur.482..375R} {482, 375}

\bibitem[\protect\citeauthoryear{{Schirra} et~al.,}{{Schirra}
  et~al.}{2021}]{sch21}
{Schirra} A.~P.,  et~al., 2021, \mn@doi [\mnras] {10.1093/mnras/stab2863},
  \href {https://ui.adsabs.harvard.edu/abs/2021MNRAS.508.4816S} {508, 4816}

\bibitem[\protect\citeauthoryear{{Schlafly} \& {Finkbeiner}}{{Schlafly} \&
  {Finkbeiner}}{2011}]{sch11}
{Schlafly} E.~F.,  {Finkbeiner} D.~P.,  2011, \mn@doi [\apj]
  {10.1088/0004-637X/737/2/103}, \href
  {http://adsabs.harvard.edu/abs/2011ApJ...737..103S} {737, 103}

\bibitem[\protect\citeauthoryear{{Selsing}, {Fynbo}, {Christensen}  \&
  {Krogager}}{{Selsing} et~al.}{2016}]{sel16}
{Selsing} J.,  {Fynbo} J.~P.~U.,  {Christensen} L.,   {Krogager} J.~K.,  2016,
  \mn@doi [\aap] {10.1051/0004-6361/201527096}, \href
  {https://ui.adsabs.harvard.edu/abs/2016A&A...585A..87S} {585, A87}

\bibitem[\protect\citeauthoryear{{Shiode} \& {Quataert}}{{Shiode} \&
  {Quataert}}{2014}]{shi14}
{Shiode} J.~H.,  {Quataert} E.,  2014, \mn@doi [\apj]
  {10.1088/0004-637X/780/1/96}, \href
  {https://ui.adsabs.harvard.edu/abs/2014ApJ...780...96S} {780, 96}

\bibitem[\protect\citeauthoryear{{Simmonds}, {Bauer}, {Thuan}, {Izotov},
  {Stern}  \& {Harrison}}{{Simmonds} et~al.}{2016}]{simmonds16}
{Simmonds} C.,  {Bauer} F.~E.,  {Thuan} T.~X.,  {Izotov} Y.~I.,  {Stern} D.,
  {Harrison} F.~A.,  2016, \mn@doi [\aap] {10.1051/0004-6361/201629310}, \href
  {https://ui.adsabs.harvard.edu/abs/2016A&A...596A..64S} {596, A64}

\bibitem[\protect\citeauthoryear{{Smith}}{{Smith}}{2006}]{smi06}
{Smith} N.,  2006, \mn@doi [\apj] {10.1086/503766}, \href
  {https://ui.adsabs.harvard.edu/abs/2006ApJ...644.1151S} {644, 1151}

\bibitem[\protect\citeauthoryear{{Smith}}{{Smith}}{2008}]{smi08}
{Smith} N.,  2008, \mn@doi [\nat] {10.1038/nature07269}, \href
  {https://ui.adsabs.harvard.edu/abs/2008Natur.455..201S} {455, 201}

\bibitem[\protect\citeauthoryear{{Smith}}{{Smith}}{2011}]{smi11c}
{Smith} N.,  2011, \mn@doi [\mnras] {10.1111/j.1365-2966.2011.18607.x}, \href
  {https://ui.adsabs.harvard.edu/abs/2011MNRAS.415.2020S} {415, 2020}

\bibitem[\protect\citeauthoryear{{Smith}}{{Smith}}{2012}]{smi12}
{Smith} N.,  2012, {All Things Homunculus}.
p.~145, \mn@doi{10.1007/978-1-4614-2275-4_7}

\bibitem[\protect\citeauthoryear{{Smith}}{{Smith}}{2013}]{smi13}
{Smith} N.,  2013, \mn@doi [\mnras] {10.1093/mnras/sts508}, \href
  {https://ui.adsabs.harvard.edu/abs/2013MNRAS.429.2366S} {429, 2366}

\bibitem[\protect\citeauthoryear{{Smith}}{{Smith}}{2014}]{smi14}
{Smith} N.,  2014, \mn@doi [\araa] {10.1146/annurev-astro-081913-040025}, \href
  {https://ui.adsabs.harvard.edu/abs/2014ARA&A..52..487S} {52, 487}

\bibitem[\protect\citeauthoryear{{Smith}}{{Smith}}{2017a}]{smi17b}
{Smith} N.,  2017a, {Interacting Supernovae: Types IIn and Ibn}.
p.~403, \mn@doi{10.1007/978-3-319-21846-5_38}

\bibitem[\protect\citeauthoryear{{Smith}}{{Smith}}{2017b}]{smi17}
{Smith} N.,  2017b, \mn@doi [Philosophical Transactions of the Royal Society of
  London Series A] {10.1098/rsta.2016.0268}, \href
  {https://ui.adsabs.harvard.edu/abs/2017RSPTA.37560268S} {375, 20160268}

\bibitem[\protect\citeauthoryear{{Smith} \& {Frew}}{{Smith} \&
  {Frew}}{2011}]{smi11b}
{Smith} N.,  {Frew} D.~J.,  2011, \mn@doi [\mnras]
  {10.1111/j.1365-2966.2011.18993.x}, \href
  {https://ui.adsabs.harvard.edu/abs/2011MNRAS.415.2009S} {415, 2009}

\bibitem[\protect\citeauthoryear{{Smith} \& {Owocki}}{{Smith} \&
  {Owocki}}{2006}]{smi06b}
{Smith} N.,  {Owocki} S.~P.,  2006, \mn@doi [\apjl] {10.1086/506523}, \href
  {https://ui.adsabs.harvard.edu/abs/2006ApJ...645L..45S} {645, L45}

\bibitem[\protect\citeauthoryear{{Smith}, {Brickhouse}, {Liedahl}  \&
  {Raymond}}{{Smith} et~al.}{2001}]{smi01}
{Smith} R.~K.,  {Brickhouse} N.~S.,  {Liedahl} D.~A.,   {Raymond} J.~C.,  2001,
  \mn@doi [\apjl] {10.1086/322992}, \href
  {https://ui.adsabs.harvard.edu/abs/2001ApJ...556L..91S} {556, L91}

\bibitem[\protect\citeauthoryear{{Smith} et~al.,}{{Smith} et~al.}{2010}]{smi10}
{Smith} N.,  et~al., 2010, \mn@doi [\aj] {10.1088/0004-6256/139/4/1451}, \href
  {https://ui.adsabs.harvard.edu/abs/2010AJ....139.1451S} {139, 1451}

\bibitem[\protect\citeauthoryear{{Smith}, {Li}, {Silverman}, {Ganeshalingam}
  \& {Filippenko}}{{Smith} et~al.}{2011}]{smi11}
{Smith} N.,  {Li} W.,  {Silverman} J.~M.,  {Ganeshalingam} M.,   {Filippenko}
  A.~V.,  2011, \mn@doi [\mnras] {10.1111/j.1365-2966.2011.18763.x}, \href
  {https://ui.adsabs.harvard.edu/abs/2011MNRAS.415..773S} {415, 773}

\bibitem[\protect\citeauthoryear{{Smith}, {Mauerhan}  \& {Prieto}}{{Smith}
  et~al.}{2014a}]{smi14b}
{Smith} N.,  {Mauerhan} J.~C.,   {Prieto} J.~L.,  2014a, \mn@doi [\mnras]
  {10.1093/mnras/stt2269}, \href
  {https://ui.adsabs.harvard.edu/abs/2014MNRAS.438.1191S} {438, 1191}

\bibitem[\protect\citeauthoryear{{Smith}, {Koss}  \& {Mushotzky}}{{Smith}
  et~al.}{2014b}]{smith14}
{Smith} K.~L.,  {Koss} M.,   {Mushotzky} R.~F.,  2014b, \mn@doi [\apj]
  {10.1088/0004-637X/794/2/112}, \href
  {https://ui.adsabs.harvard.edu/abs/2014ApJ...794..112S} {794, 112}

\bibitem[\protect\citeauthoryear{{Smith} et~al.,}{{Smith} et~al.}{2018}]{smi18}
{Smith} N.,  et~al., 2018, \mn@doi [\mnras] {10.1093/mnras/sty1500}, \href
  {https://ui.adsabs.harvard.edu/abs/2018MNRAS.480.1466S} {480, 1466}

\bibitem[\protect\citeauthoryear{{Sollerman} et~al.,}{{Sollerman}
  et~al.}{2019}]{sol19}
{Sollerman} J.,  et~al., 2019, \mn@doi [\aap] {10.1051/0004-6361/201833689},
  \href {https://ui.adsabs.harvard.edu/abs/2019A&A...621A..30S} {621, A30}

\bibitem[\protect\citeauthoryear{{Stanek} et~al.,}{{Stanek}
  et~al.}{2019}]{sta19}
{Stanek} K.~Z.,  et~al., 2019, The Astronomer's Telegram, \href
  {https://ui.adsabs.harvard.edu/abs/2019ATel12794....1S} {12794, 1}

\bibitem[\protect\citeauthoryear{{Stern} \& {Laor}}{{Stern} \&
  {Laor}}{2012}]{ste12}
{Stern} J.,  {Laor} A.,  2012, \mn@doi [\mnras]
  {10.1111/j.1365-2966.2012.20901.x}, \href
  {https://ui.adsabs.harvard.edu/abs/2012MNRAS.423..600S} {423, 600}

\bibitem[\protect\citeauthoryear{{Strotjohann} et~al.,}{{Strotjohann}
  et~al.}{2021}]{str21}
{Strotjohann} N.~L.,  et~al., 2021, \mn@doi [\apj] {10.3847/1538-4357/abd032},
  \href {https://ui.adsabs.harvard.edu/abs/2021ApJ...907...99S} {907, 99}

\bibitem[\protect\citeauthoryear{{Szalai}, {Zs{\'\i}ros}, {Fox}, {Pejcha}  \&
  {M{\"u}ller}}{{Szalai} et~al.}{2019}]{sza19}
{Szalai} T.,  {Zs{\'\i}ros} S.,  {Fox} O.~D.,  {Pejcha} O.,   {M{\"u}ller} T.,
  2019, \mn@doi [\apjs] {10.3847/1538-4365/ab10df}, \href
  {https://ui.adsabs.harvard.edu/abs/2019ApJS..241...38S} {241, 38}

\bibitem[\protect\citeauthoryear{{Taddia} et~al.,}{{Taddia}
  et~al.}{2013}]{tad13}
{Taddia} F.,  et~al., 2013, \mn@doi [\aap] {10.1051/0004-6361/201321180}, \href
  {https://ui.adsabs.harvard.edu/abs/2013A&A...555A..10T} {555, A10}

\bibitem[\protect\citeauthoryear{{Tazaki}, {Ichikawa}  \& {Kokubo}}{{Tazaki}
  et~al.}{2020}]{taz20}
{Tazaki} R.,  {Ichikawa} K.,   {Kokubo} M.,  2020, \mn@doi [\apj]
  {10.3847/1538-4357/ab7822}, \href
  {https://ui.adsabs.harvard.edu/abs/2020ApJ...892...84T} {892, 84}

\bibitem[\protect\citeauthoryear{{Terlevich}, {Terlevich}, {Bosch},
  {D{\'\i}az}, {H{\"a}gele}, {Cardaci}  \& {Firpo}}{{Terlevich}
  et~al.}{2014}]{ter14}
{Terlevich} R.,  {Terlevich} E.,  {Bosch} G.,  {D{\'\i}az} {\'A}.,
  {H{\"a}gele} G.,  {Cardaci} M.,   {Firpo} V.,  2014, \mn@doi [\mnras]
  {10.1093/mnras/stu1806}, \href
  {https://ui.adsabs.harvard.edu/abs/2014MNRAS.445.1449T} {445, 1449}

\bibitem[\protect\citeauthoryear{{Toba} et~al.,}{{Toba} et~al.}{2014}]{tob14}
{Toba} Y.,  et~al., 2014, \mn@doi [\apj] {10.1088/0004-637X/788/1/45}, \href
  {https://ui.adsabs.harvard.edu/abs/2014ApJ...788...45T} {788, 45}

\bibitem[\protect\citeauthoryear{{Tully}}{{Tully}}{1988}]{tul88}
{Tully} R.~B.,  1988, {Nearby galaxies catalog}

\bibitem[\protect\citeauthoryear{{Tully}}{{Tully}}{1994}]{tul94}
{Tully} R.~B.,  1994, VizieR Online Data Catalog, \href
  {https://ui.adsabs.harvard.edu/abs/1994yCat.7145....0T} {p. VII/145}

\bibitem[\protect\citeauthoryear{{Van Dyk} \& {Matheson}}{{Van Dyk} \&
  {Matheson}}{2012}]{van12}
{Van Dyk} S.~D.,  {Matheson} T.,  2012, {The Supernova Impostors}.
p.~249, \mn@doi{10.1007/978-1-4614-2275-4_11}

\bibitem[\protect\citeauthoryear{{Van Dyk}, {Peng}, {King}, {Filippenko},
  {Treffers}, {Li}  \& {Richmond}}{{Van Dyk} et~al.}{2000}]{van00}
{Van Dyk} S.~D.,  {Peng} C.~Y.,  {King} J.~Y.,  {Filippenko} A.~V.,  {Treffers}
  R.~R.,  {Li} W.,   {Richmond} M.~W.,  2000, \mn@doi [\pasp] {10.1086/317727},
  \href {https://ui.adsabs.harvard.edu/abs/2000PASP..112.1532V} {112, 1532}

\bibitem[\protect\citeauthoryear{{Vink}}{{Vink}}{2011}]{vin11}
{Vink} J.~S.,  2011, \mn@doi [\apss] {10.1007/s10509-011-0636-7}, \href
  {https://ui.adsabs.harvard.edu/abs/2011Ap&SS.336..163V} {336, 163}

\bibitem[\protect\citeauthoryear{Virtanen et~al.,}{Virtanen
  et~al.}{2020}]{sci20}
Virtanen P.,  et~al., 2020, \mn@doi [Nature Methods]
  {10.1038/s41592-019-0686-2}, \href {https://rdcu.be/b08Wh} {17, 261}

\bibitem[\protect\citeauthoryear{{Ward} et~al.,}{{Ward} et~al.}{2021}]{war21}
{Ward} C.,  et~al., 2021, \mn@doi [\apj] {10.3847/1538-4357/abf246}, \href
  {https://ui.adsabs.harvard.edu/abs/2021ApJ...913..102W} {913, 102}

\bibitem[\protect\citeauthoryear{{Waters} et~al.,}{{Waters}
  et~al.}{2020}]{wat20}
{Waters} C.~Z.,  et~al., 2020, \mn@doi [\apjs] {10.3847/1538-4365/abb82b},
  \href {https://ui.adsabs.harvard.edu/abs/2020ApJS..251....4W} {251, 4}

\bibitem[\protect\citeauthoryear{{Weiler}}{{Weiler}}{2018}]{wei18}
{Weiler} M.,  2018, \mn@doi [\aap] {10.1051/0004-6361/201833462}, \href
  {https://ui.adsabs.harvard.edu/abs/2018A&A...617A.138W} {617, A138}

\bibitem[\protect\citeauthoryear{{Weis} \& {Bomans}}{{Weis} \&
  {Bomans}}{2020}]{wei20}
{Weis} K.,  {Bomans} D.~J.,  2020, \mn@doi [Galaxies]
  {10.3390/galaxies8010020}, \href
  {https://ui.adsabs.harvard.edu/abs/2020Galax...8...20W} {8, 20}

\bibitem[\protect\citeauthoryear{{Weisskopf}, {Brinkman}, {Canizares},
  {Garmire}, {Murray}  \& {Van Speybroeck}}{{Weisskopf} et~al.}{2002}]{wei02}
{Weisskopf} M.~C.,  {Brinkman} B.,  {Canizares} C.,  {Garmire} G.,  {Murray}
  S.,   {Van Speybroeck} L.~P.,  2002, \mn@doi [\pasp] {10.1086/338108}, \href
  {https://ui.adsabs.harvard.edu/abs/2002PASP..114....1W} {114, 1}

\bibitem[\protect\citeauthoryear{{Wolf}}{{Wolf}}{1989}]{wol89}
{Wolf} B.,  1989, \aap, \href
  {https://ui.adsabs.harvard.edu/abs/1989A&A...217...87W} {217, 87}

\bibitem[\protect\citeauthoryear{{Woosley}}{{Woosley}}{2017}]{woo17}
{Woosley} S.~E.,  2017, \mn@doi [\apj] {10.3847/1538-4357/836/2/244}, \href
  {https://ui.adsabs.harvard.edu/abs/2017ApJ...836..244W} {836, 244}

\bibitem[\protect\citeauthoryear{{Woosley}, {Blinnikov}  \& {Heger}}{{Woosley}
  et~al.}{2007}]{woosley07}
{Woosley} S.~E.,  {Blinnikov} S.,   {Heger} A.,  2007, \mn@doi [\nat]
  {10.1038/nature06333}, \href
  {https://ui.adsabs.harvard.edu/abs/2007Natur.450..390W} {450, 390}

\bibitem[\protect\citeauthoryear{{Yalinewich} \& {Matzner}}{{Yalinewich} \&
  {Matzner}}{2019}]{yal19}
{Yalinewich} A.,  {Matzner} C.~D.,  2019, \mn@doi [\mnras]
  {10.1093/mnras/stz2590}, \href
  {https://ui.adsabs.harvard.edu/abs/2019MNRAS.490..312Y} {490, 312}

\bibitem[\protect\citeauthoryear{{Yoshida}, {Umeda}, {Maeda}  \&
  {Ishii}}{{Yoshida} et~al.}{2016}]{yos16}
{Yoshida} T.,  {Umeda} H.,  {Maeda} K.,   {Ishii} T.,  2016, \mn@doi [\mnras]
  {10.1093/mnras/stv3002}, \href
  {https://ui.adsabs.harvard.edu/abs/2016MNRAS.457..351Y} {457, 351}

\bibitem[\protect\citeauthoryear{{Zhang} et~al.,}{{Zhang} et~al.}{2015}]{zha15}
{Zhang} S.,  et~al., 2015, \mn@doi [\apj] {10.1088/0004-637X/815/2/113}, \href
  {http://adsabs.harvard.edu/abs/2015ApJ...815..113Z} {815, 113}

\bibitem[\protect\citeauthoryear{{Zhou}, {Wang}, {Yuan}, {Lu}, {Dong}, {Wang}
  \& {Lu}}{{Zhou} et~al.}{2006}]{zho06}
{Zhou} H.,  {Wang} T.,  {Yuan} W.,  {Lu} H.,  {Dong} X.,  {Wang} J.,   {Lu} Y.,
   2006, \mn@doi [\apjs] {10.1086/504869}, \href
  {https://ui.adsabs.harvard.edu/abs/2006ApJS..166..128Z} {166, 128}

\bibitem[\protect\citeauthoryear{{Zou} et~al.,}{{Zou} et~al.}{2019}]{zou19}
{Zou} H.,  et~al., 2019, \mn@doi [\apjs] {10.3847/1538-4365/ab48e8}, \href
  {https://ui.adsabs.harvard.edu/abs/2019ApJS..245....4Z} {245, 4}

\bibitem[\protect\citeauthoryear{{van Dokkum}}{{van Dokkum}}{2001}]{dok01}
{van Dokkum} P.~G.,  2001, \mn@doi [\pasp] {10.1086/323894}, \href
  {http://adsabs.harvard.edu/abs/2001PASP..113.1420V} {113, 1420}

\makeatother
\end{thebibliography}

%-------------------------------------------------------------
%                 A figure as large as the width of the column
%-------------------------------------------------------------
%   \begin{figure}
%   \centering
%   \includegraphics[width=\hsize]{empty.eps}
%      \caption{Vibrational stability equation of state
%               $S_{\mathrm{vib}}(\lg e, \lg \rho)$.
%               $>0$ means vibrational stability.
%              }
%         \label{FigVibStab}
%   \end{figure}

% Don't change these lines
\bsp	% typesetting comment
\label{lastpage}
\end{document}